\newcount\mgnf  
\mgnf=2

\ifnum\mgnf=0
\magnification=1000   
\hsize=15truecm\vsize=20.2truecm
   \parindent=0.3cm\baselineskip=0.45cm\fi
\ifnum\mgnf=1
   \magnification=\magstephalf
   \baselineskip=18truept plus0.1pt minus0.1pt \parindent=0.9truecm
\fi

\ifnum\mgnf=2\magnification=1200\fi

\ifnum\mgnf=0
\def\openone{\leavevmode\hbox{\ninerm 1\kern-3.3pt\tenrm1}}%
\def\*{\vskip.7truemm}\fi
\ifnum\mgnf=1
\def\openone{\leavevmode\hbox{\ninerm 1\kern-3.63pt\tenrm1}}%
\def\*{\vglue0.3truecm}\fi
\ifnum\mgnf=2\def\*{\vglue0.7truecm}\fi

\font\titolo=cmbx12\font\titolone=cmbx10 scaled\magstep 2%
\font\cs=cmcsc10\font\sc=cmcsc10\font\css=cmcsc8%
\font\indbf=cmbx10 scaled\magstep2
\font\ottorm=cmr8\font\ninerm=cmr9%
\font\msytw=msbm9 scaled\magstep1%
%
%
%
%
%
\def\st{\scriptstyle}%
%


\font\tenmib=cmmib10 \font\eightmib=cmmib8
\font\sevenmib=cmmib7\font\fivemib=cmmib5 
\font\ottoit=cmti8
\font\fiveit=cmti5\font\sixit=cmti6
\font\fivei=cmmi5\font\sixi=cmmi6\font\ottoi=cmmi8
\font\ottorm=cmr8
\font\ottosy=cmsy8\font\sixsy=cmsy6\font\fivesy=cmsy5
\font\ottobf=cmbx8\font\sixbf=cmbx6\font\fivebf=cmbx5%
\font\ottocss=cmcsc8%

\def\ottopunti{\def\rm{\fam0\ottorm}\def\it{\fam6\ottoit}%
\def\bf{\fam7\ottobf}%
\textfont1=\ottoi\scriptfont1=\sixi\scriptscriptfont1=\fivei%
\textfont2=\ottosy\scriptfont2=\sixsy\scriptscriptfont2=\fivesy%
\textfont4=\ottocss\scriptfont4=\sc\scriptscriptfont4=\sc%
\textfont5=\eightmib\scriptfont5=\sevenmib\scriptscriptfont5=\fivemib%
\textfont6=\ottoit\scriptfont6=\sixit\scriptscriptfont6=\fiveit%
\textfont7=\ottobf\scriptfont7=\sixbf\scriptscriptfont7=\fivebf%
\setbox\strutbox=\hbox{\vrule height7pt depth2pt width0pt}%
\normalbaselineskip=9pt\rm}
\let\nota=\ottopunti%

\textfont5=\tenmib\scriptfont5=\sevenmib\scriptscriptfont5=\fivemib
\mathchardef\Ba   = "050B  
\mathchardef\Bb   = "050C  
\mathchardef\Bg   = "050D  
\mathchardef\Bd   = "050E  
\mathchardef\Be   = "0522  
\mathchardef\Bee  = "050F  
\mathchardef\Bz   = "0510  
\mathchardef\Bh   = "0511  
\mathchardef\Bthh = "0512  
\mathchardef\Bth  = "0523  
\mathchardef\Bi   = "0513  
\mathchardef\Bk   = "0514  
\mathchardef\Bl   = "0515  
\mathchardef\Bm   = "0516  
\mathchardef\Bn   = "0517  
\mathchardef\Bx   = "0518  
\mathchardef\Bom  = "0530  
\mathchardef\Bp   = "0519  
\mathchardef\Br   = "0525  
\mathchardef\Bro  = "051A  
\mathchardef\Bs   = "051B  
\mathchardef\Bsi  = "0526  
\mathchardef\Bt   = "051C  
\mathchardef\Bu   = "051D  
\mathchardef\Bf   = "0527  
\mathchardef\Bff  = "051E  
\mathchardef\Bch  = "051F  
\mathchardef\Bps  = "0520  
\mathchardef\Bo   = "0521  
\mathchardef\Bome = "0524  
\mathchardef\BG   = "0500  
\mathchardef\BD   = "0501  
\mathchardef\BTh  = "0502  
\mathchardef\BL   = "0503  
\mathchardef\BX   = "0504  
\mathchardef\BP   = "0505  
\mathchardef\BS   = "0506  
\mathchardef\BU   = "0507  
\mathchardef\BF   = "0508  
\mathchardef\BPs  = "0509  
\mathchardef\BO   = "050A  
\mathchardef\BDpr = "0540  
\mathchardef\Bstl = "053F  

%
%
%
%
%
%
%

\global\newcount\numsec\global\newcount\numapp
\global\newcount\numfor\global\newcount\numfig
\global\newcount\numsub
\numsec=0\numapp=0\numfig=0
\def\veroparagrafo{\number\numsec}\def\veraformula{\number\numfor}
\def\veraappendice{\number\numapp}\def\verasub{\number\numsub}
\def\verafigura{\number\numfig}

\def\Section(#1,#2){\advance\numsec by 1\numfor=1\numsub=1\numfig=1%
\SIA p,#1,{\veroparagrafo} %
\write15{\string\Fp (#1){\secc(#1)}}%
\write16{ sec. #1 ==> \secc(#1)  }%
\0\hbox
{\titolo\hfill
\number\numsec. #2\hfill%
\expandafter{\hglue-1truecm\alato(sec. #1)}}}

\def\appendix(#1,#2){\advance\numapp by 1\numfor=1\numsub=1\numfig=1%
\SIA p,#1,{A\veraappendice} %
\write15{\string\Fp (#1){\secc(#1)}}%
\write16{ app. #1 ==> \secc(#1)  }%
\hbox to \hsize{\titolo Appendix A\number\numapp. #2\hfill%
\expandafter{\alato(app. #1)}}\*%
}

\def\senondefinito#1{\expandafter\ifx\csname#1\endcsname\relax}

\def\SIA #1,#2,#3 {\senondefinito{#1#2}%
\expandafter\xdef\csname #1#2\endcsname{#3}\else
\write16{???? ma #1#2 e' gia' stato definito !!!!} \fi}

\def \Fe(#1)#2{\SIA fe,#1,#2 }
\def \Fp(#1)#2{\SIA fp,#1,#2 }
\def \Fg(#1)#2{\SIA fg,#1,#2 }

\def\etichetta(#1){(\veroparagrafo.\veraformula)%
\SIA e,#1,(\veroparagrafo.\veraformula) %
\global\advance\numfor by 1%
\write15{\string\Fe (#1){\equ(#1)}}%
\write16{ EQ #1 ==> \equ(#1)  }}

\def\etichettaa(#1){(A\veraappendice.\veraformula)%
\SIA e,#1,(A\veraappendice.\veraformula) %
\global\advance\numfor by 1%
\write15{\string\Fe (#1){\equ(#1)}}%
\write16{ EQ #1 ==> \equ(#1) }}

\def\getichetta(#1){\veroparagrafo.\verafigura%
\SIA g,#1,{\veroparagrafo.\verafigura} %
\global\advance\numfig by 1%
\write15{\string\Fg (#1){\graf(#1)}}%
\write16{ Fig. #1 ==> \graf(#1) }}

\def\etichettap(#1){\veroparagrafo.\verasub%
\SIA p,#1,{\veroparagrafo.\verasub} %
\global\advance\numsub by 1%
\write15{\string\Fp (#1){\secc(#1)}}%
\write16{ par #1 ==> \secc(#1)  }}

\def\Eq(#1){\eqno{\etichetta(#1)\alato(#1)}}
\def\eq(#1){\etichetta(#1)\alato(#1)}
\def\Eqa(#1){\eqno{\etichettaa(#1)\alato(#1)}}
\def\eqa(#1){\etichettaa(#1)\alato(#1)}
\def\eqg(#1){\getichetta(#1)\alato(fig. #1)}
\def\sub(#1){\0\palato(p. #1){\bf \etichettap(#1).}}
\def\asub(#1){\0\palato(p. #1){\bf \etichettapa(#1).}}
\def\apprif(#1){\senondefinito{e#1}%
\eqv(#1)\else\csname e#1\endcsname\fi}

\def\equv(#1){\senondefinito{fe#1}$\clubsuit$#1%
\write16{eq. #1 non e' (ancora) definita}%
\else\csname fe#1\endcsname\fi}
\def\grafv(#1){\senondefinito{fg#1}$\clubsuit$#1%
\write16{fig. #1 non e' (ancora) definito}%
\else\csname fg#1\endcsname\fi}
\def\secv(#1){\senondefinito{fp#1}$\clubsuit$#1%
\write16{par. #1 non e' (ancora) definito}%
\else\csname fp#1\endcsname\fi}

\def\eqo{{\global\advance\numfor by 1}}
\def\equ(#1){\senondefinito{e#1}\equv(#1)\else\csname e#1\endcsname\fi}
\def\graf(#1){\senondefinito{g#1}\grafv(#1)\else\csname g#1\endcsname\fi}
\def\figura(#1){{\css Figura} \getichetta(#1)}
\def\secc(#1){\senondefinito{p#1}\secv(#1)\else\csname p#1\endcsname\fi}
\def\sec(#1){{\secc(#1)}}
\def\refe(#1){{[\secc(#1)]}}

\def\BOZZA{
\def\alato(##1){\rlap{\kern-\hsize\kern-.5truecm{$\scriptstyle##1$}}}
\def\palato(##1){\rlap{\kern-.5truecm{$\scriptstyle##1$}}}
}

\def\alato(#1){}
\def\galato(#1){}
\def\palato(#1){}


{\count255=\time\divide\count255 by 60 \xdef\hourmin{\number\count255}
        \multiply\count255 by-60\advance\count255 by\time
   \xdef\hourmin{\hourmin:\ifnum\count255<10 0\fi\the\count255}}

\def\oramin{\hourmin }

\def\data{\number\day/\ifcase\month\or gennaio \or febbraio \or marzo \or
aprile \or maggio \or giugno \or luglio \or agosto \or settembre
\or ottobre \or novembre \or dicembre \fi/\number\year;\ \oramin}
\setbox200\hbox{$\scriptscriptstyle \data $}

\newdimen\xshift \newdimen\xwidth \newdimen\yshift \newdimen\ywidth

\def\ins#1#2#3{\vbox to0pt{\kern-#2\hbox{\kern#1 #3}\vss}\nointerlineskip}

\def\eqfig#1#2#3#4#5{
\par\xwidth=#1 \xshift=\hsize \advance\xshift
by-\xwidth \divide\xshift by 2
\yshift=#2 \divide\yshift by 2%
{\hglue\xshift \vbox to #2{\vfil
#3 \includegraphics{#4.ps}
}\hfill\raise\yshift\hbox{#5}}}

\def\8{\write12}

\openin13=#1.aux \ifeof13 \relax \else
\input #1.aux \closein13\fi
\openin14=\jobname.aux \ifeof14 \relax \else
\input \jobname.aux \closein14 \fi
\immediate\openout15=\jobname.aux


\let\a=\alpha \let\b=\beta  \let\g=\gamma  \let\d=\delta \let\e=\varepsilon
\let\z=\zeta  \let\h=\eta   \let\th=\theta \let\k=\kappa \let\l=\lambda
\let\m=\mu    \let\n=\nu    \let\x=\xi     \let\p=\pi    \let\r=\rho
\let\s=\sigma \let\t=\tau   \let\f=\varphi 
\let\ch=\chi  \let\ps=\psi   \let\o=\omega
\let\G=\Gamma \let\D=\Delta  \let\L=\Lambda 
     \let\F=\Phi    
\let\O=\Omega 

\def\\{\hfill\break} \let\==\equiv
\let\txt=\textstyle

\let\io=\infty 
\def\Dpr{\BDpr\,}
\def\ap{{\it a priori\ }}
\let\0=\noindent\def\pagina{{\vfill\eject}}

\def\media#1{{\langle#1\rangle}}
\def\ie{\hbox{\it i.e.\ }}\def\eg{\hbox{\it e.g.\ }}
\let\dpr=\partial

\def\tende#1{\,\vtop{\ialign{##\crcr\rightarrowfill\crcr
 \noalign{\kern-1pt\nointerlineskip} \hskip3.pt${\scriptstyle
 #1}$\hskip3.pt\crcr}}\,}
\def\circage{\lower2pt\hbox{$\,\buildrel > \over {\scriptstyle \sim}\,$}}
\def\otto{\,{\kern-1.truept\leftarrow\kern-5.truept\to\kern-1.truept}\,}
\def\fra#1#2{{#1\over#2}}

\def\CC{{\cal C}}\def\FF{{\cal F}} \def\HH{{\cal H}} 
 \def\BB{{\cal B}} 
  
 \def\GG{{\cal G}}

\def\T#1{{#1_{\kern-3pt\lower7pt\hbox{$\widetilde{}$}}\kern3pt}}
\def\VVV#1{{\underline #1}_{\kern-3pt
\lower7pt\hbox{$\widetilde{}$}}\kern3pt\,}
\def\W#1{#1_{\kern-3pt\lower7.5pt\hbox{$\widetilde{}$}}\kern2pt\,}

\def\lis{\overline}\def\tto{\Rightarrow}

\def\indica{\leaders \hbox to 0.5cm{\hss.\hss}\hfill}
\def\guida{\leaders\hbox to 1em{\hss.\hss}\hfill}

\def\qed{\raise1pt\hbox{\vrule height5pt width5pt depth0pt}}

\def\indic{\hbox{\raise-2pt \hbox{\indbf 1}}}

\def\RRR{\hbox{\msytw R}}

 \def\ZZZ{\hbox{\msytw Z}}

\def\defi{\,{\buildrel def\over=}\,}
\def\lhs{{\it l.h.s.}\ }
\def\rhs{{\it r.h.s.}\ }
\def\cfr{{\it c.f.\ }}

\def\sqr#1#2{{\vcenter{\vbox{\hrule height.#2pt%
        \hbox{\vrule width.#2pt height#1pt \kern#1pt%
          \vrule width.#2pt}%
        \hrule height.#2pt}}}}

\def\ig{\int}

\footline={\rlap{\hbox{\copy200}}\tenrm\hss \number\pageno\hss}
\def\V#1{{\bf#1}}

\def\Tr{{\rm \, Tr\,}}

\centerline{\titolone Equilibrium Statistical Mechanics}
\*

\centerline
{\it Giovanni Gallavotti}

\centerline
{\it I.N.F.N. Roma 1, Fisica Roma1}

\*
\Section(1, Foundations: atoms and molecules)

\*
Classical Statistical Mechanics studies properties of macroscopic
aggregates of particles, atoms and molecules, based on the assumption
that they are point masses subject to the laws of
classical mechanics. Distinction between macroscopic and
microscopic is evanescent and in fact the foundations of statistical
mechanics have been laid on properties, proved or assumed, of
few particles systems.

Macroscopic systems are often considered in stationary states: which
means that their microscopic configurations follow each other as time
evolves while looking the same macroscopically. Observing time
evolution is the same as sampling (``not too closely'' time-wise)
independent copies of the system prepared in the same way.

A basic distinction is necessary: a stationary state can be either in
equilibrium or not. The first case arises when the particles are
enclosed in a container $\O$ and are subject only to their mutual
conservative interactions and, possibly, to external conservative
forces: typical example is a gas in a container subject forces due to
the walls of $\O$ and to gravity, besides the internal
interactions. This is a very restricted class of systems and states.

A more general case is when the system is in a stationary
state but it is also subject to non conservative forces: a typical
example is a gas or fluid in which a wheel rotates, as in the Joule
experiment, with some device acting to keep the temperature
constant. The device is called a thermostat and in statistical
mechanics it has to be modeled by forces, non conservative as well,
which prevent an indefinite energy transfer from the external forcing
to the system: such transfer would impede reaching stationary
states. For instance the thermostat could be simply a constant
friction force (as in stirred incompressible liquids or as in electric
wires in which current circulates because of an electromotive force).

A more fundamental approach would be to imagine that
the thermostatting device is not a phenomenologically introduced non
conservative force (like a friction force) but is due to the
interaction with an external infinite system which is in
``equilibrium at infinity''.

In any event non equilibrium stationary states are intrinsically more
complex than equilibrium states. Here attention will be confined to
equilibrium statistical mechanics of systems of $N$ identical point
particles $\V Q=(\V q_1,\ldots,\V q_N)$ enclosed in a cubic box $\O$,
with volume $V$ and side $L$, usually supposed with perfectly
reflecting walls.

Particles of mass $m$ located at $\V q,\V q'$ will be supposed to
interact via a pair potential $\f(\V q-\V q')$. Microscopic motion
will follow the equations

$$m \ddot{\V q}_i=-\sum_{j=1}^N\Dpr_{\V q_i} \f(\V q_i-\V q_j)+\sum_i
W_{wall}(\V q_i)\defi -\Dpr_{\V q_i}\F(\V Q)\Eq(1.1)$$
where the potential $\f$ will be supposed smooth except, possibly, for
$|\V q-\V q'|\le r_0$ where it could be $+\io$ meaning the particles
cannot come closer than $r_0$ and at $r_0$ \equ(1.1) is interpreted by
imagining that they undergo elastic collisions; the potential
$W_{wall}$ models the container and it will be replaced, unless
explicitly stated, by an elastic collision rule.

The time evolution $(\V Q,
\dot{\V Q})\to S_t(\V Q,\dot{\V Q})$  will, therefore, be described
on the space $\widehat\FF(N)$ of the pairs position-velocity of the
$N$ particles or, more conveniently, on {\it phase space}: \ie by a
time evolution $S_t$ on the space $\FF(N)$ of the pairs $(\V P,\V Q)$
momentum--position with $\V P=m\dot\V Q$. The motion being
conservative, the energy $U\defi \sum_i\fra1{2m} \V
p_i^2+\sum_{i<j}\f(\V q_i-\V q_j)+\sum_i W_{wall}(\V q_i)\defi K(\V
P)+\F(\V Q)$ will be a constant of motion; the last term in $\F$ is
missing if walls are perfect. This makes it convenient to regard the
dynamics as associated with two dynamical systems $(\FF(N),S_t)$ on
the $6N$ dimensional phase space, and $(\FF_U(N),S_t)$ on the $6N-1$
dimensional surface of energy $U$.  Since the dynamics \equ(1.1) is
Hamiltonian on phase space, with Hamiltonian $H(\V P,\V Q)\defi
\sum_i\fra1{2m} \V p_i^2+\F(\V Q)
\defi K+\F$, it follows that the volume $d^{3N}\V P d^{3N}\V Q$
is conserved (\ie a region $E$ has the same volume as $S_t E$) and
also the area $\d(H(\V P,\V Q)-U)d^{3N}\V P d^{3N}\V Q$ is conserved.

The above dynamical systems are well defined, \ie $S_t$ is a map on
phase space globally defined for all $t\in (-\io,\io)$, when the
interaction potential is bounded below: this is implied by the \ap bounds
due to energy conservation. For gravitational or Coulomb
interactions a lot more has to be said, assumed and done in order to even
define the key quantities needed for a statistical theory of motion.

Although our world is $3$--dimensional (or {\it at least was} so
believed to be until recent revolutionary theories) it will be useful
to consider also systems of particles in dimension $d\ne3$: in this
case the above $6N$ and $3N$ become, respectively, $2d N$ and $d
N$. Systems with dimension $d=1,2$ are in fact sometimes very good
models for thin filaments or thin films. For the same reason it is
often useful to imagine that space is discrete and particles can
only be located on a lattice, \eg on $\ZZZ^d$, see Sect.\sec(15).
\*
\0{\it Bibliography:} [Ga99].
\*

\Section(2, Pressure, temperature and kinetic energy)
\*

The beginning was {\cs Bernoulli}'s derivation of the perfect gas law
via the identification of the {\it pressure} at numerical
{\it density} $\r$ with the average momentum transferred per unit time
to a surface element of area $dS$ on the walls: \ie the average of the
observable $2m v \,\r v\,dS$ with $v$ the normal component of the
velocity of the particles which undergo collisions with $dS$. If
$f(v)dv$ is the distribution of the normal component velocity and $f(\V v)d^3
\V v\=\prod_i f(v_i) \,d^3 \V v$, $\V v=(v_1,v_2,v_3)$, is the total
velocity distribution the average of $\fra K N$ is $p dS$ given by

$$dS \ig_{v>0}2 m v^2\r f(v)dv=dS \ig m v^2\r f(v)dv=\r \fra23 dS
\ig\fra12  \V v^2 f(\V v)d^3\V v=\r \fra23\media{\fra{K}{N}} dS\Eq(2.1)$$
{\it Furthermore} $\fra23 \media{\fra{K}{N}}$ was identified as
proportional to the {\it absolute temperature}
$\media{\fra{K}{N}}\defi$ $const\fra32T$ which, with present day
notations, is written $\fra23\media{\fra{K}{N}}=k_B T$. The constant
$k_B$ was (later) called {\it Boltzmann's constant} and it is the same
constant for all perfect gases at least. Its independence on the
particular nature of the gas is a consequence of {\it Avogadro's law}
stating that equal volumes of gases at the same conditions of
temperature and pressure contain equal number of molecules.

{\it Proportionality between average kinetic energy and temperature via
the universal constant $k_B$ became, since, a fundamental assumption
extending to all aggregates of particles gaseous or not, never
challenged in all later works} (until quantum mechanics, where this is
no longer true, see Sect. 26).
\*
\0{\it Bibliography:} [Ga99].
\*

\Section(3, Heat and entropy)
\*

After Clausius' discovery of entropy  and to explain it
mechanically, {\cs Boltzmann} introduced the {\it heat theorem}, which
he developed to full generality between 1866 and 1884. Identification
of absolute temperature with average kinetic energy and the heat
theorem can be considered the founding elements of statistical
mechanics.

The theorem makes precise the notion of time average and then states
in great generality that given {\it any} mechanical system one can
associate with its dynamics four quantities $U,V,p,T$, defined as time
averages of suitable mechanical observables (\ie functions on phase
space) so that when the external conditions are infinitesimally varied
and the quantities $U,V$ change by $dU,dV$ respectively, the ratio
$\fra{d\,U\,+\,p\, d\,V}T$ is exact, \ie there is a function $S(U,V)$
whose corresponding variation equals the ratio. It will
be better, for the purpose of considering very large boxes ($V\to\io$)
to write this relation in terms of intensive quantities
$u\defi\fra{U}{N}$ and $v=\fra{V}{N}$ as

$$\fra{d\,u\,+\,p\, d\,v}T\,\hbox{\rm is exact}\Eq(3.1)$$
\ie the ratio equals the variation $d s$ of
$s(\fra{U}{N},\fra{V}N)\=\fra1N S(U,V)$.

The proof originally dealt with {\it monocyclic systems}, \ie systems
in which all motions are periodic. The assumption is clearly much too
restrictive and justification for it developed from the early ``{\it non
periodic motions can be regarded as periodic with infinite period}''
(1866), to the later {\it ergodic hypothesis} and finally to the
realization that, after all, the heat theorem does not really depend
on the ergodic hypothesis (1884).

Although for a one dimensional system the proof of the heat theorem is
a simple check it was a real breakthrough because it led to answering the
general question of under which conditions one could define mechanical
quantities whose variations were constrained to satisfy \equ(3.1) and
therefore could be interpreted as mechanical model of Clausius'
macroscopic thermodynamics. It is reproduced in the next few lines.

Consider a one-dimensional system subject to forces with a confining
potential $\f(x)$ such that $|\f'(x)|>0$ for $|x|>0$, $\f''(0)>0$ and
$\f(x)\tende{x\to\io}+\io$. All motions are periodic so that the
system is {\it monocyclic}. Suppose that the potential $\f(x)$
depends on a parameter $V$ and define {\it a state} to be a motion with
given energy $U$ and given $V$; let

$$\vcenter{\halign{#\ $=$\ & #\hfill\cr
$U$ & total energy of the system $\= K+\F$\cr
$T$ & time average of the kinetic energy $K$, $\media{K}$\cr
$V$ & the parameter on which $\f$ is supposed to depend\cr
$p$ & $-$ time average of $\dpr_V \f$, $-\media{\dpr_V\f}$.\cr}}\Eq(3.2)$$

\0A state is thus parameterized by $U,V$; if such parameters change by
$dU, dV$, respectively, let: $dL\defi -p dV, dQ\defi dU+p dV$. Then \equ(3.1)
holds.  In fact let $x_\pm(U,V)$ be the extremes of the oscillations
of the motion with given $U,V$ and define $S$ as:

$$
S=2\log \ig_{x_-(U,V)}^{x_+(U,V)}
\sqrt{(U-\f(x))}dx\ \tto\ dS=
\fra{\ig \big(dU-\dpr_V\f(x) dV\big)\, \fra{dx}{\sqrt{K}}}{
\ig K\fra{dx}{\sqrt{K}}}
\Eq(3.3)$$
Noting that $\fra{dx}{\sqrt K} =\sqrt{\fra2m} dt$, \equ(3.1) follows
because time averages are given by integrating with respect to $\fra{dx}{\sqrt
K}$ and dividing by the integral of $\fra{1}{\sqrt K}$.
\*
\0{\it Bibliography:} [Bo84], [Ga99].
\*

\Section(4, Heat theorem and ergodic hypothesis)
\*

Boltzmann tried to extend the result beyond the one dimensional
systems: for instance to Keplerian motions: which are not monocyclic
unless only motions with a fixed eccentricity are considered.  However
the early statement that ``aperiodic motions can be regarded as
periodic with infinite period'' is really the heart of the application
of the heat theorem for monocyclic systems to the far more complex gas
in a box.

Imagine that the gas container $\O$ is closed by a piston of section
$A$ located to the right of the origin at distance $L$ and acting as a
lid, so that the volume is $V=AL$.  The microscopic model for the
piston will be a potential $\lis\f(L-\x)$ if $x=(\x,\h,\z)$ are the
coordinates of a particle. The function $\lis\f(r)$ will vanish for
$r>r_0$, for some $r_0\ll L$, and diverge to $+\io$ at $r=0$. Thus
$r_0$ is the width of the layer near the piston where the force of the
wall is felt by the particles that happen to be roaming there.

The contribution to the total potential energy $\F$ due to
the walls is $W_{wall}=\sum_j
\lis\f(L-\x_j)$ and $\dpr_V \lis\f=A^{-1}\dpr_L\lis \f$; assuming
monocyclicity it is necessary to evaluate the time average of $\dpr_L
\F(x)=\dpr_L W_{wall}\=-\sum_j \lis\f'(L-\x_j)$.
As time evolves the particles $x_j$ with
$\x_j$ in the layer within $r_0$ of the wall will feel the force
exercised by the wall and bounce back. One particle in the layer will
contribute to the average of $\dpr_L
\F(x)$ the amount
$$\fra1{\rm total\ time} 2\ig_{t_0}^{t_1}- \lis\f'(L-\x_j)
dt\Eq(4.1)$$
if $t_0$ is the first instant when the point $j$ enters the layer and
$t_1$ is the instant when the $\x$-component of the velocity vanishes
``against the wall''. Since $-\lis\f'(L-\x_j)$ is the $\x$-component
of the force, the integral is $2m|\dot\x_j|$ (by Newton's law),
provided $\dot\x_j>0$ of course.

Suppose that no collisions between particles occur while the
particles travel within the range of the potential of the wall: \ie
the mean free path is much greater than the range of the potential
$\lis\f$ defining the wall.  The collisions contributions to the
average momentum transfer to the wall per unit time is therefore given
by, see \equ(2.1), $\ig_{v>0} 2mv\, f(v)\, \r_{wall}\, A\, v\, dv$ if
$\r_{wall}, f(v)$ are the average density near the wall and,
respectively, the average fraction of particles with a velocity
component normal to the wall between $v$ and $v+dv$. Here $p,f$ are
supposed to be independent of the point on the wall: this should be
true up to corrections of size $o(A)$.

Thus writing the average kinetic energy per particle and per velocity
component, $\ig
\fra{m}2 v^2 f(v) dv$, as $\fra12\b^{-1}$ (\cfr \equ(2.1)) it follows
that %
$$p\defi -\media{\dpr_V\F}= \r_{wall} \b^{-1}\Eq(4.2)$$
has the physical interpretation of pressure. The $\fra12 \b^{-1}$ is
the average kinetic energy per degree of freedom: hence it is
proportional to the absolute temperature $T$ (\cfr Sect.\sec(2)).

{\it On the other hand} if motion on the energy surface takes place on
a single periodic orbit the quantity $p$ in \equ(4.2) is the right
quantity that would make the heat theorem work, see \equ(3.2). Hence
regarding the trajectory on each energy surface as periodic (\ie the
system as monocyclic) leads to the heat theorem with $p,U,V,T$ having
the {\it right physical interpretation} corresponding to their
appellations.  This shows that monocyclic systems provide natural
models of thermodynamic behavior.

Assuming that a chaotic system like a gas in a container of volume $V$
will satisfy ``for practical purposes'' the above property a quantity
$p$ can be defined such that $dU+{p\, dV}$ admits the inverse of the
average kinetic energy $\media{K}$ as an integrating factor and
furthermore $p,U,V,\media{K}$ have the physical interpretations of
pressure, energy, volume and (up to a proportionality factor) absolute
temperature.

Boltzmann's conception of space (and time) as discrete allowed him to
conceive the property that the energy surface is constituted by
``points'' all of which belong to a single trajectory: a property that
would be impossible if phase space was really a continuum.  Regarding
phase space as consisting of a finite number of ``cells'' of finite
volume $h^{dN}$, for some $h>0$ (rather than of a continuum of points)
allowed him to think, without logical contradiction, that the energy
surface consisted of a single trajectory and, hence, that motion was a
cyclic permutation of its points (actually cells).

Furthermore it implied that the time average of an observable $F(\V
P,\V Q)$ {\it had to be} identified with its average on the energy
surface computed via the Liouville distribution $C^{-1}\,\ig F(\V P,\V
Q) \d(H(\V P,\V Q) -U)\,d\V Pd\V Q$ with $C=\ig\d(H(\V P,\V Q)
-U)\,d\V Pd\V Q$ (the appropriate normalization factor): a property
that was written symbolically $\fra{dt}T=\fra{d\V Pd\V Q}{\ig d\V Pd\V
Q}$ or

$$\lim_{T\to\io}\fra1T\ig_0^T F(S_t(\V P,\V Q)) dt=
\fra{\ig  F(\V P',\V Q') \d(H(\V P',\V Q') -U)
\,d\V P'd\V Q'}{\ig  \d(H(\V P',\V Q') -U)\,d\V P'd\V Q'}\Eq(4.3)$$
The validity of \equ(4.3) for all (piecewise smooth) observables $F$
and for all points of the energy surface, with the exception of a set
of zero area, is called the {\it ergodic hypothesis}
\*

\0{\it Bibliography:} [Bo84], [Ga99].
\*

\Section(5, Ensembles)
\*

Eventually (1884) Boltzmann realized that the validity of the heat
theorem for averages computed via the \rhs of \equ(4.3) held {\it
independently} of the ergodic hypothesis, \ie the \equ(4.3) was not
necessary because the heat theorem (\ie \equ(3.1)) could also be
derived under the only assumption that the averages involved in its
formulation were computed as averages over phase space with respect to
the probability distribution in the \rhs of \equ(4.3).

Furthermore, if $T$ was identified with the average kinetic energy,
$U$ with the average energy, $p$ with the average force per unit
surface on the walls of the container $\O$ with volume $V$, the
relation \equ(3.1) held {\it for a variety of families of probability
distributions on phase space} besides the \equ(4.3).  Among which
\*

\0(a) the {\it microcanonical ensemble}, which is the collection
of probability distributions in the \rhs of \equ(4.3) parameterized by
$u=\fra{U}N,v=\fra{V}N$ (energy and volume per particle)

$$\m^{mc}_{u,v}(d\V Pd\V Q)= \fra{1}{Z_{mc}(U,N,V)}\d(H(\V P,\V Q) -U)
\,\fra{d\V Pd\V Q}{N!\,
h^{d N}},\Eq(5.1)$$
where $h$ is a constant with the dimensions of an action which, in the
discrete representation of phase space mentioned in Sect.\sec(4), can
be taken such that $h^{dN}$ equals the cells volume and, therefore,
the integrals with respect to \equ(5.1) can be interpreted as
an (approximate) sum over the cells conceived as microscopic
configurations of $N$ {\it indistinguishable} particles (whence the $N!$).

\0(b) the {\it canonical ensemble} which is the collection of probability
distributions parameterized by $\b,v=\fra{V}N$

$$\m^{c}_{\b,v}(d\V Pd\V Q)= \fra1{Z_c(\b,N,V)}
e^{-\b H(\V P,\V Q)}\fra{d\V Pd\V Q}{N!\,
h^{d N}} \Eq(5.2)$$
to which more ensembles can be added. For instance {\cs Gibbs} introduced
\*

\0(c) the {\it grand canonical ensemble} which is the
collection of probability distributions parameterized by $\b,\l$ and
defined over the space $\FF_{gc}=\cup_{N=0}^\io \FF(N)$

$$\m^{gc}_{\b,\l}(d\V Pd\V Q)=
\fra1{Z_{gc}(\b,\l,V)} {e^{ \b\l N-\b H(\V P,\V Q)}}
\,\fra{d\V Pd\V Q}{N!\,
h^{d N}}\Eq(5.3)$$
Hence there are {\it several} different models of Thermodynamics:
a key test for accepting them as real microscopic
descriptions of macroscopic Thermodynamics is that

\0(1) a correspondence between the macroscopic states of thermodynamic
equilibrium and the elements of a collection of probability
distributions on phase space can be established by identifying on one side
macroscopic thermodynamic states with given values of the
thermodynamic functions and, on the other side, probability
distributions attributing  the {\it same average values}
to the corresponding  microscopic observables (\ie whose averages have the
interpretation of thermodynamic functions).

\0(2) once the correct correspondence between the elements of the different
ensembles is established, \ie once the pairs $(u,v)$, $(\b,v)$,
$(\b,\l)$ are so related to produce the same values for the averages
$U,\, V,\, k_B T\defi\b^{-1}, \, p\,|\dpr\O|$ of

$$ H(\V P,\V Q),\ V,\ \fra{2K(\V P)}{3N},\ \ig \d_{\dpr\O}(\V q_1)\,2m (\V
v_1\cdot\V n)^2\,d\V q_1, \Eq(5.4)$$
($\d_{\dpr\O}(\V q_1)$ is a delta function pinning $\V q_1$ to the
surface $\dpr\O$) then the averages of all physically interesting
observables {\it should coincide at least in the thermodynamic limit},
$\O\to\io$. In this way the elements $\m$ of the considered collection
of probability distributions can be identified with the states of
macroscopic equilibrium of the system. The $\m$'s depend on parameters
and therefore they form an {\it ensemble}: each of them corresponds to
a macroscopic equilibrium state whose thermodynamic functions are
appropriate averages of microscopic observables and therefore are
functions of the parameters identifying $\m$.
\vskip1mm

{\it Remark:} The word {\it ensemble} is often used to indicate the
individual probability distributions of what is called here
ensemble. The meaning used here seems closer to the original sense in
the 1884 paper of Boltzmann (in other words often by ``ensemble'' one
means that collection of the phase space points on which a given
probability distribution is considered, and this does not seem to be
the original sense).
\vskip1mm

For instance in the case of the microcanonical distributions this
means interpreting energy, volume, temperature and pressure of the
equilibrium state with specific energy $u$ and specific volume $v$ as
proportional, through appropriate universal proportionality constants, to
the integrals with respect to $\m^{mc}_{u,v}(d\V P\,d\V Q)$ of the
mechanical quantities in \equ(5.4); and the averages of other
thermodynamic observables in the state with specific energy $u$ and
specific volume $v$ should be given by their integrals with respect to
$\m^{mc}_{u,v}$.

Likewise one can interpret energy, volume, temperature and pressure of
the equilibrium state with specific energy $u$ and specific volume $v$
as the averages of the mechanical quantities \equ(5.4) with respect to
the canonical distribution $\m^{c}_{\b,v}(d\V P\,d\V Q)$ which has
average specific energy precisely $u$; and the averages of other
thermodynamic observables in the state with specific energy and volume
$u$ and $v$ are given by their integrals with respect to
$\m^{c}_{\b,v}$. And a similar definition can be given for the
description of thermodynamic equilibria via the grand canonical
distributions.
\*

\0{\it Bibliography:} [Gi81], [Ga99].
\*

\Section(6, Equivalence of ensembles)
\*

{\cs Boltzmann} proved that, computing averages via the microcanonical
or canonical distributions, the essential property
\equ(3.1) was satisfied when changes in their parameters (\ie $u,v$ or
$\b,v$ respectively) induced changes $du,dv$ on energy and
volume. He also proved that the function $s$, whose existence is
implied by \equ(3.1), was {\it the same} function once expressed as a
function of $u,v$ (or of any pair of thermodynamic parameters, \eg of
$T,v$ or of $p,u$).  A close examination of Boltzmann's proof
shows that the \equ(3.1) holds {\it exactly} in the canonical ensemble
and up to corrections tending to $0$ as $\O\to\io$ in the
microcanonical ensemble. Identity of thermodynamic functions
evaluated in the two ensembles holds, as a consequence, up to
corrections of this order. And Gibbs added that the same held for the
grand canonical ensemble.

Of course {\it not every collection of stationary probability
distributions on phase space} would provide a model for
Thermodynamics: Boltzmann called {\it orthodic} the collections of
stationary distributions which generated models of Thermodynamics
through the above mentioned identification of its elements with
macroscopic equilibrium states. The microcanonical, canonical and the
later grand canonical ensembles are the chief examples of orthodic
ensembles: which Boltzmann and Gibbs {\it proved to be not only
orthodic but to generate the same thermodynamic functions}, \ie to
generate the same Thermodynamics

This freed from the analysis of the truth of the doubtful ergodic
hypothesis (still unproved in any generality) or of the monocyclicity
(manifestly false if understood literally rather than regarding phase
space as consisting of finitely many small, discrete, cells), and
allowed Gibbs to formulate the problem of Statistical Mechanics of
equilibrium as

\vskip1mm
\0{\it Problem: study the properties of the collection of
probability distributions constituting (any) one of the above ensembles}.
\vskip1mm

However by no means the three ensembles just introduces exhaust the
class of orthodic ensembles producing the same models of
Thermodynamics in the limit of infinitely large systems. The wealth of
ensembles with the orthodicity property, hence leading to equivalent
mechanical models of Thermodynamics, can be naturally interpreted in
connection with the phase transitions phenomenon, see Sect.\sec(9).

Clearly the quoted results do not ``prove'' that thermodynamic
equilibria ``are'' described by the microcanonical or canonical or
grand canonical ensembles. However they certainly show that for most
systems, independently of the number of degrees of freedom, one can
define {\it quite unambiguously a  mechanical model of Thermodynamics}
establishing {\it parameter free, system independent, physically
important} relations between Thermodynamic quantities (\eg $\dpr_u
\fra{p(u,v)}{T(u,v)}\=\dpr_v\fra{1}{T(u,v)}$, from \equ(3.1)).

The ergodic hypothesis which was at the root of the mechanical
theorems on heat and entropy cannot be taken as a justification of
their validity. {\it Naively} one would expect
that the time scale necessary to see an equilibrium attained, called
{\it recurrence time scale}, would have to be at least the time that a
phase space point takes to visit all possible microscopic states of
given energy: hence an explanation of why the necessarily enormous size of
the recurrence time is not a problem becomes necessary..

In fact the recurrence time can be estimated once the phase space is
regarded as discrete: for the purpose of countering mounting criticism
Boltzmann assumed that momentum was discretized in units of $(2m k_B
T)^{\fra12}$ (\ie the average momentum size) and space was discretized
in units of $\r^{-\fra13}$ (\ie the average spacing), implying a
volume of cells $h^{3N}$ with $h\defi \r^{-\fra13}(2m k_B T)^{\fra12}$;
then he calculated that, even with such a {\it gross} discretization,
a cell representing a microscopic state of $1cm^3$ of hydrogen at
normal condition would require a time (called ``recurrence time'') of
the order of $\sim 10^{10^{19}}$ times the age of the Universe (!) to
visit the entire energy surface. [In fact the phase space volume is
$\G=(\r^{-3}N(2m k_B T)^{\fra32})^N\=h^{3N}$ and the number of cells of
volume $h^{3N}$ is $\G/(N! h^{3N})\simeq e^{3N}$: and the time to
visit all will be $e^{3N}\t_0$ with $\t_0$ a typical atomic unit, \eg
$10^{-12} sec\,$: but $N=10^{19}$]. In this sense the
statement boldly made by the young Boltzmann that ``aperiodic motions
can be regarded as periodic with infinite period'' was even made
quantitative.

The time is clearly so long to be {\it irrelevant for all purposes}:
nevertheless the correctness of the microscopic theory of
Thermodynamics can still rely on microscopic dynamics once it is
understood, as stressed by Boltzmann, that the reason why we observe
approach to equilibrium, and equilibrium itself, over ``human'' time
scales, {\it far shorter than the recurrence times}, is due to the
property that on most of the energy surface the (very few) observables
whose averages yield macroscopic thermodynamic functions (namely
pressure, temperature, energy $\ldots$) {\it assume the same value}
even if $N$ is only very moderately large (of the order of $10^3$
rather than $10^{19}$). This implies that this value coincides with
the average and therefore satisfies the {\it heat theorem} without any
contradiction with the length of the recurrence time. The latter
rather concerns the time needed to the {\it generic observable} to
{\it thermalize}, \ie to reach its time average: the generic
observable will indeed take a very long time to ``thermalize'' {\it
but no one will ever notice because the generic observable} (\eg the
position of a pre-identified particle) {\it is not relevant for
Thermodynamics}.

The word ``proof'' used so far in this paper is not in the
mathematical sense: the relevance of a {\it mathematically rigorous}
analysis was widely realized only around the
1960's at the same time when the first numerical studies of the
thermodynamic functions became possible and rigorous results became
needed to check the correctness of various numerical simulations.

\*

\0{\it Bibliography:} [Bo66], [Bo84], [Ga99].

\*
\Section(7, Thermodynamic limit)
\*

Adopting Gibbs axiomatic point of view it is interesting to see the
path to be followed to achieve an equivalence proof of three ensembles
introduced in Sect \sec(4).

A preliminary step is to consider, given a cubic box $\O$ of
volume $V=L^d$, the normalization factors $Z^{gc}(\b,\l,V),
Z^c(\b,N,V),Z^{mc}(U,N,V) $ in \equ(5.1),\equ(5.2),\equ(5.3), and to
check that the following {\it thermodynamic limits} exist:

$$\eqalign{
\b p_{gc}(\b,\l)&\defi \lim_{V\to\io}
\fra1V\log Z_{gc}(\b,\l,V)\cr
-\b
f_c(\b,\r)&\defi\lim_{V\to\io, \fra{N}V=\r} \fra1N\log
Z^c(\b,N,V)\cr
k_B^{-1}s_{mc}(u,\r)&\defi\lim_{V\to\io,
\fra{N}V=\r,\fra{U}N=u} \fra{1}N \log Z^{mc}(U,N,V)\cr}\Eq(7.1)
$$
where the density $\r\defi v^{-1}\=\fra{N}V$ is used instead of $v$
for later reference.  The normalization factors play an important role
because they have simple thermodynamic interpretation, see
Sect. \sec(8): they are called grand canonical, canonical and
microcanonical {\it partition functions}, respectively.

Not surprisingly assumptions on the interparticle potential $\f(\V
q-\V q')$ are necessary to achieve an existence proof of the limits in
\equ(7.1). The assumptions on $\f$ are not only quite general but also
have a clear physical meaning. They are
\vskip1truemm

\0(1) {\it stability}: \ie existence of a constant $B\ge0$ such that
$\sum_{i<j}^N\f(\V q_i-\V q_j)\ge -B N$ for all $N\ge0, \V q_1,\ldots,\V
q_N\in \RRR^d$, and
\\
(2) {\it temperedness}: \ie existence of constants $\e_0,R>0$ such that
    $|\f(\V q-\V q')|< B |\V q-\V q'|^{-d-\e_0}$ for $|\V q-\V q'|>R$.
\vskip1truemm

The assumptions are satisfied by essentially all microscopic
interactions {\it with the notable exceptions of the gravitational and
Coulombic interactions}, which require a separate treatment (and lead
to somewhat different results on the Thermodynamic behavior).

For instance the (1), (2) are satisfied if $\f(\V q)$ is $+\io$ for
$|\V q|< r_0$ and smooth for $|\V q|>r_0$, for some $r_0\ge0$, and
furthermore $\f(\V q)> B_0 |\V q|^{-(d+\e_0)}$ if $r_0<|\V q|\le R$,
while for $|\V q|>R$ it is $|\f(\V q)|< B_1 |\V q|^{-(d+\e_0)}$ with
$B_0,B_1,\e_0>0, R>r_0$ suitably chosen. Briefly {\it $\f$ is fast
diverging at contact and fast approaching $0$ at large distance}. This
is called a (generalized) {\it Lennard--Jones potential}. If $r_0>0$
the $\f$ is called a {\it hard core potential}. If $B_1=0$ the
potential is said to have {\it finite range}. See Appendix A1 for
physical implications of violations of the above stability and
temperedness properties. {\it However} in the following it will
necessary, both for simplicity and to contain the length of the
exposition, to restrict consideration to the case $B_1=0$, \ie to

$$\f(\V q)> B_0 |\V q|^{-(d+\e_0)},\quad r_0<|\V q|\le R\quad {\rm and}\quad
|\f(\V q)|\=0,\quad |\V q|>R\Eq(7.2)$$
unless explicitly stated.

Assuming (1) and (2) the existence of the limits in \equ(7.1) can be
mathematically proved: in Appendix A2 the proof of the first is
analyzed to provide the simplest example of the technique.  A
remarkable property of the functions $\b p_{gc}(\b,\l), -\b\r
f_c(\b,\r), \r s_{mc}(u, \r)$ is that they are {\it convex} functions:
hence they are continuous in the interior of their domains of
definition and at one variable fixed are, with at most countably many exceptions,
differentiable in the other.

In the case of a potential without hard core ($\r_{\max}= \io$) the
$-\r f_c(\b,\r)$ can be checked to tend to $0$ slower than $\r$ as
$\r\to0$ and to $-\io$ faster than $-\r$ as $\r\to\io$ (essentially
proportionally to $-\r\log\r$ in both cases).  Likewise in the same
case $s_{mc}(u,\r)$ can be shown to tend to $0$ slower than
$u-u_{min}$ as $u\to u_{min}$ and to $-\io$ faster than $-u$ as
$u\to\io$. The latter asymptotic properties can be exploited to derive
from the relations between the partition functions in
\equ(7.1)

$$\eqalign{&
\txt Z^{gc}(\b,\l,V)=\sum_{N=0}^\io e^{\b\l N} Z^c(\b,N,V) \quad
{\rm and}\cr
&\txt Z^c(\b,N,V)= \ig_{-B}^\io e^{-\b U} Z^{mc}(U,N,V)\, dU \cr}\Eq(7.3)$$
and from the mentioned convexity the consequences

$$\eqalign{
\b p_{mc}(\b,\l)=&
\max_v (\b \l v^{-1}  -\b v^{-1}f_c(\b,v^{-1}))\cr
-\b f_c(\b,v^{-1})=&\max_u (-\b u+ k_B^{-1} s_{mc}(u,v^{-1}))\cr}
\Eq(7.4)$$
and that the maxima are attained in points, {\it or intervals},
internal to the intervals of definition.  Let $v_{gc},u_c$ be points
where the maxima are respectively attained in \equ(7.4).

Remark that the quantity $\fra{e^{\b\l N} Z^c(\b,N,V)}{
Z^{gc}(\b,\l,V)}$ has the interpretation of probability of a density
$v^{-1}=N/V$ evaluated in the grand canonical distribution. It follows that {\it
if the maximum in the first of
\equ(7.4) is strict}, \ie it is reached at a single point, the values
of $v^{-1}$ in closed intervals {\it not containing} the maximum point
$v_{gc}^{-1}$ have a probability behaving as $< e^{-c\,V}, c>0$, as
$V\to\io$, compared to the probability of $v^{-1}$'s {\it in any
interval containing} $v_{gc}^{-1}$.  Hence $v_{gc}$ has the
interpretation of average value of $v$ in the grand canonical
distribution, in the limit $V\to\io$.

Likewise the interpretation of $\fra{e^{-\b u N}
Z^{mc}(u N,N,V)}{Z^c(\b,N,V)}$ as probability in the canonical
distribution of an energy density $u$ shows that if the maximum in the
second of \equ(7.4) is strict the values of $u$ in
closed intervals {\it not containing} the maximum point $u_c$ have a
probability behaving as $< e^{-c\,V}, c>0$, as $V\to\io$, compared to the
probability of $u$'s in any interval containing $u_c$. Hence in the
limit $\O \to \io$ the average value of $u$ in the canonical
distribution is $u_c$.

If the maxima are {\it strict} the \equ(7.4) also establishes a relation
between the grand canonical density, the canonical free energy and the
grand canonical parameter $\l$, or between the canonical energy, the
microcanonical entropy and the canonical parameter $\b$:

$$\l=\dpr_{v^{-1}} (v_{gc}^{-1} f_c(\b,v_{gc}^{-1})),\qquad k_B
\b=\dpr_u s_{mc}(u_c,v^{-1})\Eq(7.5)$$
where convexity and strictness of the maxima imply the derivatives
existence.

\vskip1mm
\0{\it Remark:}
Therefore in the equivalence between canonical and microcanonical
ensembles the canonical distribution with parameters $(\b, v)$ should
correspond with the microcanonical with parameters $(u_c,v)$. And the
grand canonical distribution with parameters $(\b,\l)$ should
correspond with the canonical with parameters $(\b, v_{gc})$ .
\*
\0{\it Bibliography:} [Ru69], [Ga99].

\*
\Section(8, Physical interpretation of thermodynamic functions)
\*

The existence of the limits \equ(7.1) implies several properties of
interest. The first is the possibility of finding the physical meaning
of the functions $p_{gc},f_c,s_{mc}$ and of the parameters $\l,\b$

Note first that, for all $V$ the grand canonical average
$\media{K}_{\b,\l}$ is $\fra{d}2
\b^{-1}\media{N}_{\b,\l}$ so that $\b^{-1}$ is proportional
to the temperature $T_{gc}=T(\b,\l)$ in the gand canonical distribution:
$\b^{-1}=k_B T(\b,\l)$.
Proceeding {\it heuristically} the physical meaning of $p(\b,\l)$ and of
$\l$ can be found by the following remarks.
\vskip1mm

Consider the microcanonical distribution $\m^{mc}_{u,v}$ and denote by
$\ig^*$ the integral over $(\V P,\V Q)$ extended to the domain of the
$(\V P,\V Q)$ such that $H(\V P,\V Q)=U$ {\it and}, at the same time,
$\V q_1\in d V$ where $dV$ is an infinitesimal volume surrounding the
region $\O$.  Then by the microscopic definition of the pressure $p$,
see Sect. \sec(1), it is

$$p dV=\fra{N}{Z(U,N,V)}
\ig^*\d\,{2\over 3}{p_1^2\over 2m}\fra{d\V P\,d\V Q}{N! h^{d N}}\=
{2\over 3 Z(U,N,V)}\ig^*\d\, K(\V P){d\V P\,d\V Q\over N!h^{d N}}\Eq(8.1)$$
where $\d\=\d(H(\V P,\V Q)-U)$.
The \rhs of \equ(8.1) can be compared with ${\dpr_V Z(U,N,V) d
V\over{Z(U,N,V)}} ={N\over{Z(U,N,V)}} \ig^*{d\V P\,d\V Q\over N! h^{dN}}$
to give ${{\dpr_V Z}\, d V\over Z}=N{p\,d V\over {2\over 3}
\media{K}^*}=\b p dV$
because $\media{K}^*$, which denotes the average $\ig^*K/\ig^* 1$,
{\it should be essentially the same as the microcanonical average
$\media{K}_{mc}$} (\ie insensitive to the fact that one
particle is constrained to the volume $dV$) if
$N$ is large. In the limit $V\to\io,\fra{V}N=v$ the latter remark
together with the second of \equ(7.5) yields

$$k_B^{-1} \dpr_v s_{mc}(u,v^{-1})=\b\,p(u,v),\qquad k_B^{-1} \dpr_u
s_{mc}(u,v)=\b\Eq(8.2)$$
respectively. Note that $p\ge0$ and it is {\it not increasing in $v$}
because $s_{mc}(\r)$ is {\it concave} as a function of $v=\r^{-1}$ (in
fact by the remark following \equ(7.2) $\r s_{mc}(u,\r)$ is convex in
$\r$ and, in general, if $\r g(\r)$ is convex in $\r$ then $g(v^{-1})$ is,
always, concave in $v=\r^{-1}$).

Hence $d s_{mc}(u,v)= \fra{du+p dv}T$ so that taking into account the
physical meaning of $p,T$ (as pressure and temperature, see
Sec. \sec(2)), $s_{mc}$ is, in Thermodynamics, the {\it
entropy}. Therefore, see the second of \equ(7.4), $-\b f_c(\b,\r)= -\b
u_c +k_B^{-1} s_{mc}(u_c,\r)$ becomes

$$ f_c(\b,\r)=u_c-T_c s_{mc}(u_c,\r),\qquad df_c=-p\, dv-s_{mc}\, dT\Eq(8.3)$$
and since $u_c$ has the interpretation, as mentioned in Sect. \sec(7),
of average energy in the canonical distribution $\m^c_{\b,v}$ it follows
that $f_c$ has the thermodynamic interpretation of {\it free
energy}  (once compared with the free energy definition
$F=U-TS$ in Thermodynamics).

By \equ(7.5), \equ(8.3) $\l= \dpr_{v^{-1}} (v_{gc}^{-1}
f_c(\b,v_{gc}^{-1}))\=u_c-T_c s_{mc}+p\, v_{gc}$ and $v_{gc}$ has the
meaning of specific volume $v$. Hence, after comparison with the
chemical potential definition $\l V=U-TS+pV$ in Thermodynamics, it
follows that the thermodynamic interpretation of $\l$ is the {\it
chemical potential} and, see \equ(7.4),\equ(7.5), the grand canonical relation
$\b p(\b,\l)= \b \l v_{gc}^{-1} -\b v_{gc}^{-1}(-\b u_c+k_B^{-1}
s_{mc}(u_c,v^{-1}))$ shows that $p(\b,\l)\=p$ implying that $p(\b,\l)$ is
the pressure expressed, however, as a function of temperature
and chemical potential.
\vskip1mm

To go beyond the heuristic derivations above it should be remarked
that convexity and the property that the maxima in \equ(7.4),\equ(7.5)
are reached in the interior of the intervals of variability of $v$ or
$u$ are sufficient to turn the above arguments into rigorous
mathematical deductions: this means that given \equ(8.2) as
definitions of $p(u,v),\b(u,v)$ the second of
\equ(8.3) follows as well as $p(\b,\l)\=p(u_v,v_{gc}^{-1})$.
But the values $v_{gc}$ and $u_c$ in
\equ(7.4) {\it are not necessarily unique}: convex functions can
contain horizontal segments and therefore the general conclusion is
that the maxima may possibly be attained in
intervals. Hence instead of a single $v_{gc}$ there might be a whole
interval $[v_-,v_+]$ where the
\rhs of \equ(7.4) reaches the maximum and instead of a single $u_c$
there might be a whole interval $[u_-,u_+]$ where the
\rhs of \equ(7.5) reaches the maximum.

Convexity implies that the values of $\l$ or of $\b$ for which the
maxima in \equ(7.4) or \equ(7.5) are attained in intervals rather than
in single points are {\it rare} (\ie at most denumerably many): the
interpretation is, in such cases, that the thermodynamic functions
show discontinuities: and the corresponding phenomena are called {\it
phase transitions}, see Sect.\sec(9).
\*

\0{\it Bibliography:} [Ru69], [Ga99].
\*

\Section(9, Phase transitions and boundary conditions)
\*

The analysis in Sect.\sec(7),\sec(8) of the relations between
elements of ensembles of distributions describing
macroscopic equilibrium states not only allows us to obtain mechanical
models of Thermodynamics but it also shows that the models, for a
given system, coincide at least as $\O\to\io$.
Furthermore the equivalence between the thermodynamic
functions computed via corresponding distributions in different
ensembles can be extended to a full equivalence of the
distributions.

If the maxima in \equ(7.4) are attained in single points $v_{gc}$ or
$u_c$ the equivalence should take place in the sense that a
correspondence between $\m^{gc}_{\b,\l},\m^{c}_{\b,v},\m^{mc}_{u,v}$
can be established so that given {\it any local observable} $F(\V P,\V
Q)$, defined as an observable {\it depending on $(\V P,\V Q)$ only
through the $\V p_i,\V q_i$ with $\V q_i\in\L$} where $\L\subset\O$ is
a finite region, has the {\it same average} with respect to
corresponding distributions in the limit $\O\to\io$.

The correspondence is established by considering $(\l,\b)\otto
(\b,v_{gc})\otto (u_{mc},v)$ where $v_{gc}$ is where the maximum in
\equ(7.4) is attained, $u_{mc}\=u_c$ is where the maximum in
\equ(7.5) is attained and $v_{gc}\=v$, (\cfr also
\equ(8.2),\equ(8.3)). This means that the limits

$$\lim_{V\to\io} \ig F(\V P,\V Q)\m^{a}(d\V P\,d\V Q)\defi
\media{F}_a=a-{\rm independent},\qquad a=gc,c,mc\Eq(9.1)$$
coincide if the averages are evaluated by the distributions
$\m^{gc}_{\b,\l},\m^{c}_{\b,v_c},\m^{mc}_{u_{mc},v_{mc}}$

Exceptions to the \equ(9.1) are possible: and are certainly likely to
occur at values of $u,v$ where the maxima in \equ(7.4) or \equ(7.5)
are attained in intervals rather than in isolated points; but this does
not exhaust, in general, the cases in which \equ(9.1) may not hold.

However no case in which \equ(9.1) fails {\it has to be
regarded as an exception}. It rather signals that an interesting and
important phenomenon occurs. To understand it properly it is necessary
to realize that the grand canonical, canonical and microcanonical
families of probability distributions are {\it by far} not the only
ensembles of probability distributions whose elements can be
considered to generate models of Thermodynamics, \ie which are
orthodic in the sense of  Sect. \sec(6).
More general families of orthodic statistical ensembles of probability
distributions can be very easily conceived.
In particular
\vskip1mm

\0{\bf Definition:} {\it
consider the grand canonical, canonical and
microcanonical distributions associated with an energy function in
which the potential energy contains, besides the interaction $\F$
between particles located inside the container, {\it also} the
interaction energy $\F_{in,out}$ between particles inside the
container and external particles, identical to the ones in the
container but {\it not allowed to move} and fixed in positions such
that in every unit cube $\D$ external to $\O$ there is a finite number
of them bounded independently of $\D$.
Such configurations of external particles will be called ``boundary
conditions of fixed external particles''.}
\vskip1mm

The thermodynamic limit with such boundary conditions is obtained by
considering the grand canonical, canonical and microcanonical
distributions constructed with potential energy function $\F
+\F_{in,out}$ in containers $\O$ of increasing size taking care that,
while the size increases, the fixed particles that would become
internal to $\O$ are eliminated. The argument used in Sec.\sec(7) to
show that the three models of thermodynamics, considered there, did
define the same thermodynamic functions can be repeated to reach the
conclusion that the also the (infinitely many) ``new'' models of
Thermodynamics in fact give rise to the same thermodynamic functions
and averages of local observables. Furthermore the values of the
limits corresponding to \equ(7.1) can be computed using the new
partition functions and {\it coincide} with the ones in \equ(7.1) (\ie
are boundary conditions independent).

However it may happen, and in general it really happens, for
many models and for particular values of the state parameters, that the
limits in \equ(9.1) do not coincide with the analogous limits
computed in the new ensembles: \ie the averages of some
local observables are {\it unstable} with respect to changes of fixed
particles boundary conditions.

There is a very natural interpretation of such apparent
ambiguity of the various models of Thermodynamics: namely at the
values of the parameters that are selected to describe the macroscopic
states under consideration {\it there may correspond different
equilibrium states with the same parameters}. When the maximum in
\equ(7.4) is reached on an interval of densities one should not think
of any failure of the microscopic models for Thermodynamics: rather
one has to think that there are several states possible with the same
$\b,\l$ and that they can be identified with the probability
distributions obtained by forming the grand canonical, canonical or
microcanonical distributions with {\it different kinds of boundary
conditions}.

For instance a boundary condition with high density may produce an
equilibrium state with parameters $\b,\l$ which has high density, \ie
the density $v^{-1}_+$ at the right extreme of the interval in which the
maximum in \equ(7.4) is attained, while using a low density boundary
condition the limit in \equ(9.1) may describe the averages taken in a
state with density $v^{-1}_-$ at the left extreme of the interval or,
perhaps, with a density intermediate between the two
extremes. Therefore the following definition emerges.
\*

\0{\bf Definition:} {\it If the grand canonical distributions with
parameters $(\b,\l)$ and different choices of fixed external particles
boundary conditions generate for some local observable $F$ average values
which are different by more than a quantity $\d>0$ for all large
enough volumes $\O$ then one says that the system has a phase
transition at $(\b,\l)$. This implies that the limits in \equ(9.1), when
existing, will depend on the boundary condition and their values will
represent averages of the observables in ``different
phases''.
A corresponding definition is given in the case of the canonical and
microcanonical distributions when, given $(\b,v)$ or $(u,v)$, the limit
in \equ(9.1) depends on the boundary conditions for some $F$.}
\vskip2mm

\0{\it Remarks:} (1) The idea is that by fixing one of the
thermodynamic ensembles and by varying the boundary conditions one can
realize all possible states of equilibrium of the system that can
exist with the given values of the parameters determining the state
in the chosen ensemble (\ie $(\b,\l)$, $(\b,v)$ or $(u,v)$ in the
grand canonical, canonical or microcanonical cases respectively).
\\
(2) The impression that in order to define a phase transition the
thermodynamic limit is necessary is {\it incorrect}: the definition
{\it does not require} considering the limit $\O\to\io$.  The
phenomenon is that by changing boundary conditions the average of a
local observable can change {\it at least by amounts
independent of the system size}. Hence occurrence of a phase transition
is perfectly observable in finite volume: it suffices to check that by
changing boundary conditions the average of some observable changes by
an amount whose minimal size is volume {\it independent}.  It is a
manifestation of an instability of the averages with respect to
boundary conditions changes: an instability which does not fade
away when the boundary recedes to infinity, \ie boundary perturbations
produce bulk effects and at a phase transitions the averages of the
local observable, if existing at all, will exhibit a nontrivial
boundary conditions dependence. This is also called ``{\it long
range order}''.
\\
(3) It is possible to show that when this happens then some
thermodynamic function whose value is boundary condition independent
(like the free energy in the canonical distributions) has
discontinuous derivatives in terms of the parameters of the ensemble
defining them. This is in fact one of the frequently used alternative
definitions of phase transitions: the latter two natural definitions
of first order phase transition are equivalent. However it is very
difficult to prove that a given system shows a phase transition. For
instance existence of a liquid-gas phase transition is still an open
problem in systems of the type considered until Sect.\sec(15).
\\
(4) A remarkable unification of the theory of the equilibrium
ensembles emerges: all distributions of any ensemble describe equilibrium
states. If a boundary condition is fixed once and for all then some
equilibrium states might fail to be described by an element of an
ensemble. However if all boundary conditions are allowed then all
equilibrium states should be realizable in a given ensemble by varying
the boundary conditions.
\\
(5) The analysis leads us to consider as completely equivalent {\it
without exceptions} grand canonical, canonical or microcanonical
ensembles enlarged by adding to them the distributions with potential
energy augmented by the interaction with fixed external particles.
\\
(6) The above picture is {\it really proved} only for special classes
of models (typically in models in which particles are constrained to
occupy points of a lattice and in systems with hard core interactions,
$r_0>0$ in \equ(7.2)) but it is believed to be correct in general: at
least it is consistent with all what is so far known in classical
statistical mechanics. The difficulty is that conceivably one might
even need boundary conditions more complicated than the fixed
particles boundary conditions (like putting different particles
outside interaction with the system with an arbitrary potential,
rather than via $\f$).

The discussion of the equivalence of the ensembles and the question of
boundary conditions importance has already imposed the consideration
of several limits as $\O\to\io$. Occasionally it will again come up.
For conciseness purposes it is useful to
set up a formal definition of equilibrium states of an
``infinite volume''  system: although infinite volume is an idealization void
of physical reality it is nevertheless useful to define such states
because certain notions, like that of {\it pure state} can be sharply
defined, with few words and avoiding wide circumvolutions, in terms of
them. Therefore let:
\vskip1mm

\0{\bf Definition:} {\it An {\it infinite
volume} state with parameters $(\b,\l)$, or $(\b,v)$ is a collection
of {\it average values} $F\to \media{F}$ obtained, respectively, as
limits of finite volume averages $\media{F}_{\O_n}$ defined from
canonical, microcanonical or grand canonical distributions in $\O_n$
with fixed parameters $(\b,\l)$, $(\b,v)$, or $(u,v)$ and with general
boundary condition of fixed external particles, on sequences
$\O_n\to\io$ for which such limits exist simultaneously for all local
observables $F$.}
\vskip1mm

Having set the definition of infinite volume state consider a local
observable $G(X)$ and let $\t_\x G(X)=G(X+\x),\, \x\in\RRR^d$, with
$X+\x$ denoting the configuration $X$ in which all particles are
translated by $\x$: then an infinite volume state is a {\it pure
state} if for any pair of local observables $F,G$ it is

\vskip-5mm

$$\media{F\t_\x G}-\media{F}\media{\t_\x G}\tende{\x\to\io}0\Eq(9.2)$$
which is called a {\it cluster property} of the pair $F,G$.

The result alluded to in remark (6) is that at least in the case of
hard core systems (or of the simple lattice systems discussed in
Sect.\sec(15)) the infinite volume equilibrium states in the above
sense exhaust at least the totality of the infinite volume {\it pure
states}. And furthermore the other states that can be obtained in the
same way are convex combinations of the pure states, \ie they are
``statistical mixtures'' of pure phases.  Note that $\media{\t_\x G}$
cannot be replaced, in general, by $\media{G}$ because not all
infinite volume states are necessarily translation invariant and in
simple cases (crystals) it is even possible that no translations
invariant state is a pure state.
\vskip1mm

\0{\it Remarks:} (1) This means that, in the latter models,
generalizing the boundary conditions, \eg considering external
particles not identical to the ones in the system, or using periodic
or partially periodic boundary conditions or {\it the widely used
alternative device of introducing a small auxiliary potential and
first taking the infinite volume states in presence of it and then
letting the potential vanish} does not enlarge further the set of
states (but may sometimes be useful: an example of a study of a phase
transition by using the latter method of small fields will be given in
Sect.\sec(13)).
\\
(2) If $\ch$ is the indicator function of a local
event it will make sense to consider the probability of occurrence of
the event in a infinite volume state defining it as $\media{\ch}$. In
particular the probability density for finding $p$ particles in $\V
x_1,\V x_2,\ldots,\V x_p$, called the {\it $p$-point correlation
function}, will thus be defined in an infinite volume state. For
instance if the state is obtained as a limit of canonical states
$\media{\cdot}_{\O_n}$ with parameters $\b,\r$, $\r=\fra{N_n}{V_n}$,
in a sequence of containers $\O_n$, then $\r(\V x)=\lim_n
\media{\sum_{j=1}^{N_n}\d(\V x-\V q_j)}_{\O_n}$, and $\r(\V x_1,\V
x_2,\ldots,\V x_p)=\lim_n\media{\sum^{N_n}_{i_1,\ldots,i_p}
\prod_{j=1}^p\d(\V x_j-\V q_{i_j})}_{\O_n}$, where the sum is over
the {\it ordered} $p$-ples $(j_1,\ldots,j_p)$.
Thus the pair correlation $\r(\V q,\V q')$ and its possible
cluster property are

$$\eqalign{
&\r(\V q,\V q')\defi\lim_n \fra{\ig_{\O_n} e^{-\b U(\V q,\V q',\V
q_1,\ldots,\V q_{N_n-2})}d\V q_1\ldots d\V q_{N_n-2}}
{(N_n-2)!\,Z^c_0(\b,\r,V_n) }\cr
&\r(\V q,(\V q'+\Bx))-\r(\V q)\r(\V q'+\Bx)\tende{\Bx\to\io} 0\cr}\Eq(9.3)$$
where $Z^c_0\defi \ig e^{-\b U(\V Q)}d\V Q$ is the
``configurational'' partition function.
\*

\0{\it Bibliography:} [Ru69], [Do68], [LR69], [Ga99].
\*

\Section(10, Virial theorem and atomic dimensions)
\*

For a long time it has been doubted that ``just changing boundary
conditions'' could produce such dramatic changes as macroscopically
different states (\ie phase transitions in the sense of the definition
in Sect.\sec(9)).
The first evidence that by taking the thermodynamic limit very regular
analytic functions like $N^{-1}\log Z^{c}(\b,N,V)$ as a function of
$\b,v=\fra{V}N$ could develop in the limit $\O\to\io$ singularities
like discontinuous derivatives (corresponding to the maximum in
\equ(7.4) being reached on a plateau and to a {\it consequent}
existence of several pure phases) arose in the {\it van der Waals'
theory} of liquid gas transition.

Consider a real gas with $N$ identical particles with mass $m$ in a
container $\O$ with volume $V$.  Let the force acting on the $i$-th
particle be $\V f_i$; multiplying both sides of the equations of
motion $m\ddot{\V q}_i=\V f_i$ by $-\fra12\V q_i$ and summing over $i$
it follows $-\fra12\sum_{i=1}^N m \V q_i\cdot\ddot{\V q}_i=
-\fra12\sum_{i=1}^N\V q_i\cdot\V f_i\defi \fra12 C(\V q)$ and the
quantity $C(\V q)$ defines the {\it virial of the forces} in the
configuration $\V q$; note that $C(\V q)$ is {\it not} translation
invariant because of the presence of the forces due to the walls.

Writing the force $\V f_i$ as a sum of the internal and of the
external forces (due to the walls) the virial $C$ can be expressed
naturally as sum of the virial $C_{int}$ of the internal forces
(translation invariant) and of the virial $C_{ext}$ of the external
forces.

By dividing both sides of the definition of the virial by $\t$ and
integrating over the time interval $[0,\t]$ one finds in the limit
$\t\to+\io$, \ie up to quantities relatively infinitesimal as
$\t\to\io$, that $\media{K}=\fra12 \media{C}$, {\it and}
$\media{C_{ext}}=3 p V$ where $p$ is the pressure
and $V$ the volume. Hence $\media{K}=\fra32pV+\fra12\media{C_{int}}$ or
$$
\fra1\b=pv+\fra{\media{ C_{int}}}{3N}\Eq(10.1)$$
Equation \equ(10.1) is the {\it Clausius' virial theorem}: in the case
of no internal forces it yields $\b p v =1$, the ideal gas equation.

The internal virial
$C_{int}$ can be written, if $\V f_{j\to i}=-\V\dpr_{\V q_i}\f(\V
q_i-\V q_j)$, as $C_{int}=-\sum_{i=1}^N$ $\sum_{i\ne j}\V f_{j\to i}\cdot\V q_i\=
-\sum_{i<j}\V \dpr_{q_i}\f(\V q_i-\V q_j)\,\cdot\,(\V q_i-\V
q_j)$
which shows that the contribution to the virial by the internal
repulsive forces is {\it negative} while that of the attractive forces
is {\it positive}. The average of $C_{int}$ can be computed by the
canonical distribution, which is convenient for the purpose.  van der
Waals first used the virial theorem to perform an actual computation
of the corrections to the perfect gas laws.  Simply neglect the third
order term in the density and use the approximation $\r(\V q_1,\V
q_2)=\r^2 e^{-\b \f(\V q_1-\V q_2)}$ for the pair correlation
function, \equ(9.3), then
$$\fra12\media{C_{int}}=V \fra3{2\b}\r^2 I(\b)+VO(\r^3)\Eq(10.2)$$
where $I(\b)=\fra12\int \big(e^{-\b \f(\V q)}-1\big) \,d^3 \V q$ and
the equation of state \equ(10.1) becomes $pv+ \fra{I(\b)}{\b
v}+O(v^{-2})=\b^{-1}$.

For the purpose of illustration the calculation of $I$ can be
performed approximately at ``high temperature'' ($\b$ small) in the
case $\f(r)=4\e\big((\fra{r_0}{r})^{12}- (\fra{r_0}{r})^6\big)$ (the
classical Lennard--Jones potential), $\e,r_0>0$. The result is $I\cong
-(b-\b a)$, $b=4 v_0,\ a=\fra{32}3\e v_0,\
v_0=\fra{4\p}3(\fra{r_0}2)^3,\ $: hence $p v+\fra{a}v-\fra{b}{\b
v}=\fra1\b+O(\fra1{\b v^2})$,
$(p+\fra{a}{v^2})\,v=(1+\fra{b}v)\fra1\b=\fra1{1-\fra{b}v}\fra1\b+O(\fra1{\b
v^2})$ or:
$$(p+\fra{a}{v^2})(v-b)\b=1+O(v^{-2}),\Eq(10.3)$$
which gives the equation of state for $\b\e\ll1$.  Equation \equ(10.3)
can be compared with the well-known empirical {\it van der Waals
equation} of state:
$$\b(p+ a/v^2)(v -b)= 1\quad\hbox{or}\quad (p +An^2/V^2)(V- nB)=
nRT\Eq(10.4)$$
where, if $N_A$ is Avogadro's number, $A= a N^2_A,\, B = b N_A,\, R =
k_B N_A,\, n=N/N_A$. It shows the possibility of accessing the
microscopic parameters $\e$ and $r_0$ of the potential $\f$ via
measurements detecting deviations from the Boyle-Mariotte law $\b p v=
1$ of the rarefied gases: $\e= 3a/8b = 3A/8BN_A,\quad r_0 =
(3b/2\pi)^{1/3}=(3B/2\pi N_A)^{1/3}.  $

As a final comment it is worth stressing that {\it the virial theorem
gives in principle the exact corrections to the equation of state, in a
rather direct and simple form, as time averages of the virial of the
internal forces}. Since the virial of the internal forces is easy to
calculate from the positions of the particles as a function of
time the theorem {\it provides a method for computing the
equation of state} in numerical simulations. In fact this idea has
been exploited in many numerical experiments, in which the
\equ(10.1) plays a key role.
\*

\0{\it Bibliography:} [Ga99].
\*

\Section(11, van der Waals theory)
\*

Equation \equ(10.4) is empirically used {\it beyond} its
validity region (small density and small $\b$) by regarding
$A,B$ as phenomenological parameters to be experimentally determined
by measuring them near generic values of $p,V,T$. The measured values
of $A,B$ do not ``usually vary too much'' as functions of $v,T$ and,
apart from this small variability, the
predictions of \equ(10.4) have reasonably agreed with
experience until, as experimental precision increased over
the years, serious inadequacies eventually emerged.
\*
\*\*

\eqfig{160pt}{45pt}{
\ins{18.pt}{-7.00pt}{$v_l$}
\ins{90pt}{-7pt}{$v_g$}
\ins{-7.0pt}{63.pt}{$p$}
\ins{125pt}{-7.pt}{$v$}
}{fig1}{(1)}
\*
\0{\nota The van der Waals equation of state at a temperature $T<T_c$
where the pressure is not monotonic. The horizontal line illustrates the
``Maxwell rule''.}

\vskip1mm

Certain consequences of \equ(10.4) are appealing: \eg Fig.1 shows that
it does not give a $p$ monotonic non increasing in
$v$ if the temperature is small enough.  A {\it critical temperature}
can be defined as the largest value $T_c$ of the temperature below
which the graph of $p$ as a function of $v$ is not monotonic
decreasing; the critical volume $V_c$ is the value of $v$ at the
horizontal inflection point occurring for $T=T_c$.

For $T<T_c$ the van der Waals interpretation of the equation of state
is that the function $p(v)$ may describe {\it metastable states} while
the actual equilibrium states would follow an equation with a
monotonic dependence on $v$ and $p(v)$ becoming horizontal in the
coexistence region of specific volumes. The precise value of $p$ where
to draw the plateau (see Fig.1) would then be fixed by experiment or
theoretically predicted via the simple rule that the plateau
associated with the represented isotherm is drawn at a height such
that the area of the two cycles in the resulting loop are equal.

This is {\it Maxwell's rule}: obtained by assuming that the isotherm
curve joining the extreme points of the plateau and the plateau itself
defines a cycle, see Fig.1, representing a sequence of possible
macroscopic equilibrium states (the ones corresponding to the plateau)
or states with extremely long time of stability (``metastable'')
represented by the curved part. This would be an isothermal Carnot
cycle which therefore could not produce work: since the work produced
in the cycle (\ie $\oint pdv$) is the signed area enclosed by the
cycle the rule just means that the area is zero. The argument
is doubtful at least because it is not clear that the {\it
intermediate} states with $p$ increasing with $v$ could be realized
experimentally or could even be theoretically possible.

A striking prediction of \equ(10.4), taken literally, is that
the gas undergoes a ``gas-liquid'' phase
transition with a critical point at a temperature $T_c$, volume $v_c$
and pressure $p_c$ that can be computed via \equ(10.4) and are given
by $R T_c= 8 A/27 B, V_c= 3 B,\  (n=1)$.

At the same time this is interesting as it shows that there are
simple relations between the critical parameters and
the microscopic interaction constants  $(\e\simeq k_B T_c$ and
$r_0\simeq(V_c/N_A))^{1/3}$:
$\e= 81 k_BT_c/64,\ r_0=(V_c/2\pi N_A)^{1/3}$
if a classical Lennard--Jones potential (\ie $\f=4\e\big(
(\fra{r_0}{|\V q|})^{12}-(\fra{r_0}{|\V q|})^{6}\big)$, see
Sect.\sec(10)) is used for the interaction potential $\f$.

However \equ(10.4) {\it cannot be accepted acritically} not only
because of the approximations (essentially the
neglecting of  $O(v^{-1})$ in the equation of state), but mainly because,
as remarked above, for $T<T_c$ the function $p$ is {\it no longer
monotonic} in $v$ as it {\it must be}, see comment following
\equ(8.2).

The van der Waals' equation, refined and complemented by Maxwell's
rule, predicts the following behavior:
$$\eqalign{ (p -p_c)\propto(v - v_c)^\d&\qquad\d= 3,\,T =T_c\cr
(v_g -v_l)\propto (T_c - T)^\b&\qquad\b=1/2,\,\hbox{for } T\to
T^-_c\cr}\Eq(11.1)$$
which are in sharp contrast with the experimental data gathered in the
twentieth century. For the simplest substances one finds {\it instead}
$\d\cong5,\,\b\cong 1/3$.

Finally blind faith in the equation of state \equ(10.4)
is untenable, last but not least, also because nothing in the
analysis would change if the space dimension was $d=2$ or $d=1$: but
in the last case, $d=1$, it is easily {\it proved} that the system, if
the interaction decays rapidly at infinity, does not undergo {\it
phase transitions}, see Sect. \sec(12).

In fact it is now understood that van der Waals' equation {\it
represents rigorously only a limiting situation}, in which
particles have a hard core interaction (or a strongly repulsive one at
close distance) and a further smooth interaction $\f$ with very
long range. More precisely suppose that the part of the potential
outside a hard core radius $r_0>0$ is attractive (\ie non negative)
and has the form $\g^d\f_1(\g^{-1}|\V q|)\le0$ and call $ P_0(v)$
the ($\b$-independent) product of $\b$ times the pressure of the
hard core system {\it without any attractive tail}
($P_0(v)$ is not explicitly known except if $d=1$ in which case it is
$P_0(v) (v-b)=1, \, b=r_0$), and let $a=-\fra12\ig_{|\V q|>r_0}|\f_1(\V q)|d\V
q$. If $p(\b,v;\g)$ is the pressure when $\g>0$ then it can be proved

$$\b\, p(\b,v){\buildrel def \over =}\lim_{\g\to0}\b\, p(\b,v;\g)
=\Big[-\fra{\b\,a}{v^2}+P_0(v)\Big]_{Maxwell\ rule}\Eq(11.2)$$
where the subscript means that the graph of $p(\b,v)$ as a function of
$v$ is obtained from the function in square bracket by applying to it
Maxwell's rule, described above in the case of the van der Waals'
equation. The \equ(11.2) reduces exactly to the van der Waals equation
in dimension $d=1$ and for $d>1$ it leads to an equations with
identical critical behavior (even though $P_0(v)$ cannot be explicitly
computed).

\*
\0{\it Bibliography:} [LP79], [Ga99].
\*

\Section(12, Absence of Phase Transitions: $d=1$)
\*

One of the most quoted ``no go'' theorems in Statistical Mechanics is
that $1$--dimensio\-nal systems of particles interacting via short
range forces do not exhibit phase transitions (\cfr Sect. \sec(13))
unless the somewhat unphysical situation of having zero absolute
temperature is considered.  This is particularly easy to check in the
case of ``nearest neighbor hard core interactions'': call $r_0$ the
hard core size, so that the interaction potential $\f(r)=+\io$ if
$r\le r_0$, and suppose also that $\f(r)\=0$ if $f\ge 2 r_0$. In this
case the thermodynamic functions can be exactly computed and checked
to be analytic: hence the equation of state cannot have any phase
transition plateau.  This is a special case of {\it van Hove's
theorem} establishing smoothness of the equation of state for
interactions extending beyond the nearest neighbor and rapidly
decreasing at infinity.

If the definition of phase transition based on sensitivity of the
thermodynamic limit to variations of boundary conditions is adopted
then a more general, conceptually simple, argument can be given to
show that in one-dimensional systems there cannot be any phase
transition if the potential energy of mutual interaction between a
configuration $\V Q$ of particles to the left of a reference particle
(located at the origin $O$, say) and a configuration $\V Q'$ to the
right of the particle (with $\V Q\cup O\cup \V Q'$ compatible with the
hard cores) is {\it uniformly bounded below}. Then a mathematical
proof can be devised showing that boundary conditions influence
disappears as the boundaries recede to infinity.  One also says that
{\it no long-range order can be established in $1$-dimensional}, in
the sense that one loses .

The analysis fails if the space dimension is $\ge 2$: in this case,
{\it even if the interaction is short ranged}, the energy of
interaction between two regions of space separated by a boundary is
{\it of the order of the boundary area}. Hence one cannot bound above
and below the probability of any two configurations in two half-spaces
by the product of the probabilities of the two configurations, each
computed as if the other was not there: because such bound would be
proportional to the exponential of the surface of separation, which
tends to $\io$ when the surface grows large.  This means that we
cannot consider, at least not in general, the configurations in the
two half spaces as independently distributed.

Analytically a condition on the potential sufficient to imply that the
energy between a configuration to the left and one to the right of the
origin is bounded below, if the dimension $d$ is $d=1$, is simply
expressed by $\ig_{r'}^\io r\,\,|\,\f(r)\,|\,\, dr<+\io$ for $r'>r_0$.

Therefore {\it in order to have phase transitions in $d=1$ a potential
is needed that is ``so long range'' that it has a divergent first
moment}. It can be shown by counterexamples that if the latter
condition fails there can be phase transitions even in $d=1$
systems.

The results just quoted apply also to discrete models like {\it
lattice gases} or {\it lattice spin} models that will be considered
later, see Sect. \sec(15).
\*
\0{\it Bibliography:} [LL67], [Dy69],  [Ga99], [GBG04].
\*

\Section(13, Continuous symmetries: ``no $d=2$ crystal'' theorem.)
\*

A second case in which it is possible to rule out existence of phase
transitions or at least of certain kinds of transitions arises when
the system under analysis enjoys large symmetry.
{\it By symmetry it is meant a group of transformations acting on the
configurations of a system and transforming each of them into
a configuration which, at least for one boundary condition} (\eg
periodic or open), {\it has the same energy}.

A symmetry is said to be {\it continuous} if the group of
transformations is a continuous group. For instance continuous systems
have translational symmetry if considered in a container $\O$ with
periodic boundary conditions. Systems with ``too much symmetry''
sometimes cannot show phase transitions. For instance the continuous
translation symmetry of a gas in a container $\O$ with periodic
boundary conditions is sufficient to exclude the possibility of
crystallization in dimension $d=2$ .

To discuss this, which is a prototype of a proof which can be used to
infer absence of many transitions in systems with continuous
symmetries, consider the {\it translational symmetry} and a potential
satisfying, besides the usual \equ(7.2) and with the symbols used in
\equ(7.2), the further property that $|\V q|^2|\dpr^2_{ij} \f(\V
q)|<\lis B|\V q|^{-(d+\e_0)} $, with $\e_0>0$, for some $\lis B$ holds
for $r_0<|\V q|\le R $. This is a very mild extra requirement (and it
allows for a hard core interaction).

Consider an ``ideal crystal'' on a square lattice (for simplicity) of
spacing $a$, {\it exactly fitting in its container $\O$ of side $L$}
assumed with periodic boundary conditions: so that $N=(\fra{L}a)^d$ is
the number of particles and $a^{-d}$ is the density, which is {\it
supposed smaller than the close packing density} if the interaction
$\f$ has a hard core. The probability distribution of the particles is
rather trivial: $\lis\m=\sum_{p} \prod_{\V n} \d(\V q_{p(\V n)}- a\,\V n)
\fra{d\V Q}{N!}$, the sum running over the permutations $\V m\to p(\V
m)$ of the sites $\V m\in \O$, $\V m\in \ZZZ^d, 0<m_i\le L a^{-1}$.
The density at $\V q$ is $\widehat\r(\V q)= \sum_\V n\d(\V q-a\,\V
n)\=\media{\sum_{j=1}^N \d(\V q-\V q_j)}$ and its Fourier transform is
proportional to $\r(\V k)\defi \fra1N\media{\sum_j e^{-i\V k\cdot\V
q_j}}$ with $\V k=\fra{2\p}{L}\V n$ with $\V n\in \ZZZ^d$. The $\r(\V
k)$ has value $1$ for all $\V k$ of the form $\V K=\fra{2\p}{a}\V n$
and $\fra1N O(\max_{c=1,2}|e^{ik_c a}-1|^{-2})$ otherwise.  In presence of
interaction it has to be expected that, in a crystal state,
$\r(\V k)$ has peaks near the values $\V K$: but the value of
$\r(\V k)$ can depend on the boundary conditions.

Since the system is translation invariant a crystal state defined as a
state with a distribution ``close'' to $\lis\m$, \ie with $\hat\r(\V
q)$ with peaks at the ideal lattice points $\V q= \V n a$, cannot be
realized under periodic boundary conditions, even when the system
state is crystalline. To realize such a state a symmetry breaking term
in the interaction is needed.

This can be done in several ways: \eg by changing the boundary
condition. Such a choice implies a discussion of how much the boundary
conditions influence the positions of the peaks of $\r(\V k)$: for
instance it is not obvious that a boundary condition will not generate
a state with a period different from the one that \ap has been
selected to disprove (a possibility which would imply a reciprocal
lattice of $\V K$'s different from the one considered to begin
with). Therefore here the choice will be to imagine that an external
weak force with potential $\e W(\V q)$ acts forcing a symmetry
breaking that favors the occupation of regions around the points of
the ideal lattice (which would mark the average positions of the
particles in the crystal state that is being sought). The proof ({\it
Mermin's theorem}) that no equilibrium state with particles
distribution ``close'' to $\lis \m$, \ie with peaks in place of the
delta functions (see below), is essentially reproduced below.

Take $W(\V q)=\sum_{\V n a\in\O} \ch(\V q-\V n a)$ where $\ch(\V q)\le
0$ is smooth and $0$ everywhere except in a small vicinity of the
lattice points around which it decreases to some negative minimum
keeping a rotation symmetry around them: the potential $W$ is assumed
invariant under the translations by the lattice steps.  By the choice
of the boundary condition and of $\e W$ the density
$\widetilde\r_\e(\V q)$ {\it will be periodic} with period $a$ so that
$\r_\e(\V k)$ will, possibly, not have a vanishing limit as $N\to\io$
only if $\V k$ is a reciprocal vector $\V K=\fra{2 \p}a\V n$. If the
potential is $\f+\e W$ and {\it if a crystal state in which particles
have higher probability of being near the lattice points $\V n a$
exists}, it should be expected that for $\e>0$ small the system will
be found in a state with Fourier transform $\r_\e(\V k)$ of the
density satisfying, for some vector $\V K\ne\V0$ in the reciprocal lattice,

$$\kern1cm\lim_{\e\to0}\lim_{N\to\io} |\r_\e(\V K)|=r>0,
\Eq(13.1)$$
\ie {\it the requirement is that uniformly in $\e\to0$ the Fourier transform
of the density has a peak at some $\V K\ne\V0$}: note that if $\V k$
is not in the reciprocal lattice $\r_\e(\V k)\tende{N\to\io}0$, being
bounded above by $\fra1N O(\max_{c=1,2}|e^{ik_j a}-1|^{-2})$ because
$\fra1N \widetilde\r_\e$ is periodic and has integral over $\V q$ equal to
$1$. Hence excluding crystal existence will be identified with the
impossibility of the
\equ(13.1). Other criteria can be imagined, \eg considering
crystals with a lattice different from simple cubic: they lead to the
same result by following the same technique. Nevertheless it is not
mathematically excluded (but unlikely) that with some weaker
definition of crystal state existence could be possible even in $2$
dimensions.

The following inequalities hold under the present assumptions on the
potential and in the canonical distribution with periodic boundary
conditions and parameters $(\b,\r),\,\r=a^{-3}$ in a box $\O$ with
side multiple of $a$ (so that $N=(L a^{-1})^d$) and potential of
interaction $\f+\e W$. The further assumption that the lattice $\V n
a$ is not a close packed lattice is (of course) necessary when the
interaction potential has a hard core. Then for suitable
$B_0,B,B_1,B_2>0$, independent of $N$, 
and $\e$ and for $|\Bk|<\fra\p{a}$ and for all $\O$ ({\it if}
$\V K\ne\V0$)

\vskip-4mm
$$\eqalign{
&\fra1N\media{|\sum_{j=1}^N  e^{-i(\Bk+\V K)\cdot\V q_j}|^2}\ge
 B\fra{ (\r_\e(\V K)+\r_\e(\V K+2\Bk))^2}{B_1\,\Bk^2+ \e\,B_2}
\cr
&\fra1N\sum_{\Bk}
\g(\Bk)\,\fra{d\Bk}N\,\media{|\sum_{j=1}^N e^{-i(\Bk+\V K)\cdot\V
q_j}|^2}\le B_0<\io
\cr
}\Eq(13.2)$$
where the averages are in the canonical distribution $(\b,\r)$ with
periodic boundary conditions and a symmetry breaking potential $\e
W(\V q)$; $\g(\V k)\ge0$ is an (arbitrary) smooth a function vanishing
for $2|\Bk|\ge\d$ with $\d<\fra{2\p}a$. See Appendix A3 for the
derivation of \equ(13.2).

Multiplying both sides of the first in \equ(13.2) by $N^{-1}
\g(\Bk)$ and summing over $\Bk$ the cristallinity condition in the form
\equ(13.1) implies that the \rhs is $\ge
B r^2 a^d\ig_{|\Bk|<\d}\fra{\g(\Bk)\,d\Bk}{\Bk^2
B_1+\e\,B_2}$. For $d=1,2$ the integral diverges, as
$\e^{-\fra12}$ or $\log\e^{-1}$ respectively, implying $|\r_\e(\V
K)|\tende{\e\to0}r=0$: the criterion of crystallinity \equ(13.1) {\it
cannot be satisfied if $d=1,2$}.

The above inequality is an example of a general class of inequalities
called {\it infrared inequalities} stemming from another inequality
called {\it Bogoliubov's inequality}, see Appendix \sec(A3), which
lead to the proof that certain kinds of ordered phases cannot exist
{\it if the dimension of the ambient space is $d=2$} when a system in
suitable boundary conditions (\eg periodic) shows a {\it continuous
symmetry}. The excluded phenomenon is, more precisely, the non
existence of equilibrium states exhibiting, {\it in the thermodynamic
limit}, a symmetry {\it lower} than the continuous symmetry holding in
periodic boundary conditions.

In general existence of thermodynamic equilibrium states with symmetry
lower than the symmetry enjoyed by the system in finite volume and
under suitable boundary conditions is called a {\it spontaneous
symmetry breaking}. It is yet another manifestation of instability
with respect to changes in boundary conditions, hence its occurrence
reveals a phase transition. There is a large class of systems for
which an infrared inequality implies absence of spontaneous symmetry
breaking: in most one or two dimensional systems a continuous symmetry
cannot be {\it spontaneously broken}.

The limitation to dimension $d\le2$ is a strong limitation
to the generality of the applicability of infrared theorems to exclude
phase transitions. More precisely systems can be divided into classes
each of which has a ``critical dimension'' below which too much
symmetry implies absence of phase transitions (or of certain kinds of
phase transitions).

It should be stressed that, at the critical dimension, the symmetry
breaking is usually so weakly forbidden that one might need
astronomically large containers to destroy small effects (due to
boundary conditions or to very small fields) which break the symmetry:
\eg in the crystallization just discussed the Fourier transform peaks
are only bounded by $O(\fra1{\sqrt{\log\e^{-1}}})$: hence from a
practical point of view it might still be possible to have some kind
of order even in large containers.
\*

\0{\it Bibliography:} [Me68], [Ho67], [Ru69].
\*

\Section(14, High Temperature and small density)
\*

There is {\it another} class of systems in which no phase transitions
take place.  These are the systems with stable and tempered
interactions $\f$, \eg satisfying \equ(7.2), in the high temperature
and low density region.  The property is obtained by showing that the
equation of state is analytic in the variables $(\b,\r)$ near the
origin $(0,0)$.

A simple algorithm ({\it Mayer's series}) yields the coefficients of
the virial series $\b p(\b,\r)=\r+\sum_{k=2}^\io c_k(\b) \r^k$. It has
the drawback that the $k$-th order coefficient $c_k(\b)$ is expressed
as a sum of many terms (a number growing {\it more} than exponentially
fast in the order $k$) and it is not so easy (but possible) to show
combinatorially that their sum is bounded exponentially in $k$ if $\b$
is small enough.  A more common approach leads quickly to the desired
solution.  Denoting $\F(\V q_1,\ldots,\V q_n)\defi\sum_{i<j}\f(\V
q_i-\V q_j)$ consider the (``spatial or configurational'') correlation
functions defined, in the grand canonical distribution with parameters
$\b,\l$ (and empty boundary conditions), by
$$\r_\O(\V q_1,\ldots, \V q_n)\defi\fra1{Z^{gc}(\b,\l,V)} \sum_{m=0}^\io
z^{n+m}\ig_\O e^{-\b \F(\V q_1,\ldots, \V q_n,\V y_1,\ldots,\V y_m)}
\fra{d\V y_1\ldots d\V y_m}{m!}\Eq(14.1)$$
This is the probability density for finding particles with any
momentum in the volume element $d \V q_1\ldots d \V q_n$ (irrespective
of where other particles are), and $z=e^{\b\l}(\sqrt{2\p
m\b^{-1}h^{-2}})^d$ accounts for the integration over the momenta
variables and is called the {\it activity}: it has the dimension of a
density (\cfr \equ(9.3)).

Assuming that the potential has a hard core (for simplicity) of radius
$R$ the interaction energy $\F_{\V q_1}(\V q_2,\ldots,\V q_n)$
of a particle in $\V q_1$ with any
number of other particles in $\V q_2,\ldots,\V q_m$ with $|\V q_i-\V
q_j|>R$ is bounded below by $-B$ for some $B\ge0$ (related but not
equal to the $B$ in \equ(7.2)).
The functions $\r_\O$ will be regarded as a {\it sequence} of
functions ``of one, two,$\ldots$ particle positions'':
$\r_\O=\{\r_\O(\V q_1,\ldots,\V q_n)\}_{n=1}^\io$ vanishing for $\V
q_j\not\in \O$. Then one checks that

$$\eqalignno{
&\r_\O(\V q_1,\ldots, \V q_n)= z \d_{n,1}\ch_\O(\V
q_1)+K\,\r_\O(\V q_1,\ldots,
\V q_n)\qquad {\rm with}\cr
&K\,\r_\O(\V q_1,\ldots, \V q_n)\defi e^{-\b \F_{\V q_1}(\V q_2,\ldots,
\V q_n)}\Big(\r_\O(\V q_2,\ldots,\V  q_n)\,\d_{n>1}+ &\eq(14.2)\cr
&+\sum_{s=1}^\io \ig_\O \fra{d\V y_1\ldots d\V y_s}{s!}  \prod_{k=1}^s
\big(e^{-\b \f(\V q_1-\V y_{k})}-1\big)\r_\O(\V q_2,\ldots, \V q_n,\V y_1,\ldots,
\V y_s)\Big)\cr}$$
where $\d_{n,1},\d_{n>1}$ are Kronecker deltas and $\ch_\O(\V q)$ is
the indicator function of $\O$. The \equ(14.2) is called the {\it
Kirkwood-Salzburg equation} for the family of correlation functions in
$\O$.  The kernel $K$ of the equations is {\it independent of $\O$},
but the domain of integration is $\O$.

Calling $\a_\O$ the sequence of functions $\a_\O(\V q_1,\ldots, \V
q_n)\=0$ if $n\ne1$ and $\a_\O(\V q)=z \ch_\O(\V q)$, a
recursive expansion arises, namely
$$ \r_\O=z\a_\O+z^2  K\a_\O+z^3  K^2\a_\O+z^4
K^3\a_\O+\ldots\Eq(14.3)$$
It gives the correlation functions, {\it
provided the series converges}. The remark
$$|  K^p\a_\O(q_1,\ldots,q_n)|\le e^{(2\b B+1)p}\big(\ig |e^{-\b \f(\V
q)}-1| d\V q\big)^p\defi e^{(2\b B+1)p}\, r(\b)^{3p}\Eq(14.4)$$
shows that the series \equ(14.3), called {\it Mayer's series}
converges if $|z|< e^{-(2\b B+1)} r(\b)^{-3}$. Convergence is uniform
(as $\O\to\io$) and $( K^p)\a_\O(\V q_1,\ldots,\V q_n)$ tends to a
limit as $V\to\io$ at fixed $\V q_1,\ldots,\V q_n$ and the limit is
simply $( K^p\a)(\V q_1,\ldots,\V q_n)$, if $\a(\V q_1,\ldots,\V
q_n)\=0$ for $n\ne1$, and $\a(\V q_1)\=1$. This is because the kernel
$K$ contains the factors $(e^{-\b \f(\V q_1-\V y)}-1)$ which decay
rapidly or will even be eventually $0$, if $\f$ has finite range. It
is also clear that $( K^p\a)(q_1,\ldots,q_n)$ is translation
invariant.

Hence if $|z| e^{2\b B+1}\,r(\b)^3<1$ the limits as $\O\to\io$ of the
correlation functions exist and can be computed by a convergent
power series in $z$; the correlation functions will be translation
invariant (in the thermodynamic limit).

In particular the one-point correlation function $\r=\r(q)$ is $\r=z\,
\big(1+O(zr(\b)^3)\big)$, which to lowest order in $z$ just shows that
activity and density essentially coincide when they are small enough.
Furthermore $\b\, p_\O\=\fra1V\log Z^{gc}(\b,\l,V)$ is such that
$z\,\dpr_z\,\b p_\O=\fra1V\ig\r_\O(q)dq$ (from the definition of
$\r_\O$ in \equ(14.1)). Therefore
$$\b p(\b,z)=\lim_{V\to\io}\fra1V\log Z^{gc}(\b,\l,V)=
\ig_0^z \fra{d z'}{z'} \r(\b, z')\Eq(14.5)$$
showing that the density $\r$ is analytic in $z$ as well and $\r\simeq
z$ for $z$ small: therefore the grand canonical {\it pressure is
analytic in the density} and $\b\, p=\r\, (1+O(\r^2))$, at small
density. In other words the equation of state is, to lowest order,
essentially the equation of a perfect gas. All quantities that are
conceivably of some interest turn out to be analytic functions of
temperature and density.  The system is essentially a free gas and it
has no phase transitions in the sense of a {\it discontinuity or of a
singularity} in the dependence of a thermodynamic function in terms of
others.  Furthermore the system cannot show phase transitions in the
sense of sensitive dependence on boundary conditions of fixed external
particles: this also follows, with some extra work, from the
Kirkwood-Salzburg equations.
\*

\0{\it Bibliography:} [Ru69], [Ga99].
\*

\Section(15, Lattice models)
\*

The problem of proving existence of phase transitions in models of
homogeneous gases with pair interactions is still open.  Therefore it
makes sense to study phase transitions problem in simpler models,
tractable to some extent but nontrivial and which even have an
interest in their own from the point of view of applications.

The simplest models are the so-called {\it lattice models} in which
particles are constrained to points of a lattice: they cannot move in
the ordinary sense of the word (but of course they could jump) and
therefore their configurations do not contain momentum variables.

Interaction energy is just potential energy and ensembles are defined
as collections of probability distributions on the position
coordinates of the particle configurations. Usually the potential is a
{\it pair potential} decaying fast at $\io$ and, often, with a hard
core forbidding double or higher occupancy of the same lattice site.
For instance the {\it lattice gas with potential $\f$}, in a cubic box
$\O$ with $|\O|=V=L^d$ sites of a square lattice with mesh $a>0$, is defined
by the potential energy attributed to the configuration $X$ of
occupied {\it distinct} sites, \ie subsets $X\subset\O$:

$$H(X)=-\sum_{(x,y)\in X} \f(x-y).\Eq(15.1)$$
where the sum is over pairs of distinct points in $X$.  The {\it
canonical ensemble} and the {\it grand canonical ensemble} are the
collections of distributions, parameterized by $(\b,\r)$
($\r=\fra{N}V$) or, respectively, by $(\b,\l)$, attributing to $X$
the probability:

$$p_{\b,\r}(X)=\fra{e^{-\b H(X)}}{Z^c_p(\b,N,\O)}\d_{|X|,N},\qquad{\rm
or}\qquad p_{\b,\l} (X)=\fra{e^{\b\l|X|}e^{-\b
H(X)}}{Z^{gc}_p(\b,\l,\O)},\Eq(15.2)$$
where the denominators are normalization factors that can be called,
in analogy with the theory of continuous systems, {\it canonical} and
{\it grand canonical} partition functions; the subscript $p$ stands
for particles.

A lattice gas in which in each site there can be at most one particle
can be regarded as a model for the distribution of a family of {\it
spins} on a lattice.  Such models are quite common and useful: for
instance they arise in studying systems with magnetic properties.
Simply identify an ``occupied'' site with a ``spin up'' or $+$ and an
``empty'' site with a ``spin down'' or $-$ (say).  If
$\Bs=\{\s_x\}_{x\in \O}$ is a spin configuration, the energy of the
configuration ``for potential $\f$ and magnetic field $h$'' will
be

$$H(\Bs)=-\sum_{(x,y)\in\O}\f(x-y)\s_x\s_y-h\sum_x\s_x\Eq(15.3)$$
with the sum running over pairs $(x,y)\in\O$ of distinct sites. If
$\f(x-y)\=J_{xy}\ge0$ the model is called a {\it ferromagnetic
Ising model}. As in the case of continuous systems it will assumed
finite range for $\f$: \ie $\f(x)=0$ for $|x|>R$, for some $R$, unless
explicitly stated otherwise.

The canonical and grand canonical ensembles in the box $\O$ with
respective parameters $(\b,m)$ or $(\b,h)$ will be defined as the
probability distributions on the spin configurations
$\Bs=\{\s_x\}_{x\in\O}$ with $\sum_{x\in \O}\s_x=M=m V$ or without
constraint on $M$, respectively; hence
$$
p_{\b,m}(\Bs)=\fra{e^{-\b\sum_{(x, y)}\f(x-y)\s_x\s_y}}
{Z^c_s(\b,M,\O)},\qquad
p_{\b,h}(\Bs)=\fra{e^{-\b h\sum\s_x-\b\sum_{(x,y)}\f(x-y)\s_x\s_y}}
{Z^{gc}_s(\b,h,\O)}
\Eq(15.4)$$
where the denominators are normalization factors again called the {\it
canonical} and {\it grand canonical} partition functions.  As in the
study of the previous continuous systems canonical and grand canonical
ensembles with ``external fixed particle configurations'' can be
defined together with the corresponding ensembles with ``external
fixed spin configurations''; the subscript $s$ stands for spins.

For each configuration $X\subset \O$ of a lattice gas let $\{n_x\}$
be $n_x=1$ if $x\in X$ and $n_x=0$ if $x\not\in X$.  Then the
transformation $\s_x={2n_x-1}$ establishes a correspondence between
lattice gas and spin distributions.  In the correspondence the
potential $\f(x-y)$ of the lattice gas generates a potential
$\fra14\f(x-y)$ for the corresponding spin system and the chemical
potential $\l$ for the lattice gas is associated with a magnetic field
$h$ for the spin system with $h=\fra12(\l+\sum_{x\ne0}\f(x))$.

The correspondence between boundary conditions is natural: for
instance a boundary condition for the lattice gas in which all external
sites are occupied becomes a boundary condition in which
external sites contain a spin $+$. The relation between lattice gas
and spin systems is so close to permit switching from one to the
other with little discussion.

In the case of spin systems empty boundary conditions are often
considered (no spins outside $\O$). In lattice gases and spin systems
(as well as in continuum systems) often periodic and semiperiodic
boundary conditions are considered (\ie periodic in one or more
directions and with empty or fixed external particles or spins in the
others).

Thermodynamic limits for the partition functions
$$\eqalign{
-\b\, f(\b,v)=&
\lim_{\O\to\io\atop \fra{V}N=v}\,\fra1N\log
Z^c_p(\b,N,\O), \qquad
\b\, p(\b,\l)=\lim_{\O\to\io}\fra1V\log
Z^{gc}_p(\b,\l,\O)\cr
-\b\, g(\b,m)=&\lim_{\O\to\io,\atop\fra{M}{V}\to m}
\fra1V\log
Z^c_s(\b,M,\O),\qquad
\b\, f(\b,h)=\lim_{\O\to\io}\fra1V\log
Z^{gc}_s(\b,\l,\O)\,
\cr}\Eq(15.5)$$
can be shown to exist by a method similar to the one discussed in
Appendix A2. They have convexity and continuity properties as in the
cases of the continuum systems. In the lattice gas case the $f,p$
functions are still interpreted as free energy and pressure. In the
spin case $f(\b,h)$ has the interpretation of {\it magnetic free
energy} while $g(\b,m)$ does not have a special name in the
thermodynamics of magnetic systems. As in the continuum systems it is
occasionally useful to define infinite volume equilibrium states:
\vskip1mm

{\bf Definition:} {\it: An {\it infinite
volume} state with parameters $(\b,h)$, or $(\b,m)$ is a collection
of {\it average values} $F\to \media{F}$ obtained, respectively, as
limits of finite volume averages $\media{F}_{\O_n}$ defined from
canonical, or grand canonical distributions in $\O_n$
with fixed parameters $(\b,h)$ or $(\b,m)$, or $(u,v)$ and with general
boundary condition of fixed external spins or empty sites, on sequences
$\O_n\to\io$ for which such limits exist simultaneously for all local
observables $F$.}
\vskip1mm

This is taken {\it verbatim} from the definition in Sect.\sec(9). In
this way it makes sense to define the {\it spin correlation functions}
for $X=(\Bx_1,\ldots,\Bx_n)$ as $\media{\s_X}$ if
$\s_X=\prod_j\s_{\Bx_j}$. For instance we shall call
$\r(\Bx_1,\Bx_2)\defi\media{\s_{\Bx_1}\s_{\Bx_2}}$ and a pure phase
can be defined as an infinite volume state such that

$$\media{\s_X\s_{Y+\Bx}}-\media{\s_X}\media{\s_{Y+\Bx}}\tende{\Bx\to\io}0
\Eq(15.6)$$

\0{\it Bibliography:} [Ru69], [Ga99].
\*

\Section(16, Thermodynamic limits and inequalities)
\*

An interesting property of lattice systems is that it is possible to
study delicate questions like existence of infinite volume
states in some (moderate) generality. A typical tool is the use of
{\it inequalities}. As the simplest example of a vast class of
inequalities consider the {\it ferromagnetic Ising model} with some
finite (but arbitrary) range interaction $J_{xy}\ge0$ in a field
$h_x\ge0$: $J,h$ may even be not translationally invariant. Then the
average of $\s_X\defi \s_{x_1}\s_{x_2}\ldots\s_{x_n}$,
$X=(x_1,\ldots,x_n)$, in a state with ``empty boundary conditions''
(\ie no external spins) satisfies the inequalities $\media {\s_X},\
\dpr_{h_x}\media {\s_X},\ \dpr_{J_{xy}} \media {\s_X} \ge0$
$X=(x_1,\ldots,x_n)$. More generally let $H(\Bs)$ in \equ(15.3) be
replaced by $H(\Bs)=-\sum_X J_X \s_X$ with $J_X\ge0$ and $X$ can be
any finite set; then, if $Y=(y_1,\ldots,y_n),X=(x_1,\ldots,x_n)$, the
following {\it Griffiths inequalities} hold

$$\media
{\s_X}\ge0,\quad
\dpr_{J_Y}\media{\s_X}\=\media{\s_X\s_Y}-\media{\s_X}\media{\s_Y}\
\ge0\Eq(16.1)$$

The inequalities can be used to check, in ferromagnetic Ising models,
\equ(15.3), existence of infinite volume states (\cfr
Sect.\sec(9),\sec(15)) obtained by fixing the boundary condition $\BB$
to be either ``all external spins $+$'' or ``all external sites
empty''. If $\media{F}_{\BB,\O}$ denotes the grand canonical average
with boundary condition $\BB$ and any fixed $\b,h>0$, this means that
{\it for all local observables} $F(\Bs_\L)$, \ie for all $F$ depending
on the spin configuration in any fixed region $\L$, {\it all} the
following limits exist

$$\lim_{\O\to\io} \media{F}_{\BB,\O}= \media{F}_{\BB}\Eq(16.2)$$
The reason is that the
inequalities \equ(16.1) imply that all averages
$\media{\s_X}_{\BB,\O}$ are monotonic in $\O$ for all fixed $X\subset
\O$: so the limit \equ(16.2) exists for $F(\Bs)=\s_X$.  Hence it
exists for all $F$ depending only on finitely many spins, because any
local function $F$ ``measurable in $\L$'' can be expressed (uniquely) as a
linear combination of functions $\s_X$ with $X\subseteq\L$.

Monotonicity with empty boundary conditions is seen by considering
the sites outside $\O$ and in a region $\O'$ with side $1$ unit
larger than that of $\O$ and imagining that the couplings $J_X$ with
$X\subset\O'$ but $X\not\subset\O$ {\it vanish}. Then
$\media{\s_X}_{\O'}\ge\media{\s_X}_{\O}$ because $\media{\s_X}_{\O'}$
is an average computed with a distribution corresponding to an energy
with the couplings $J_X$ with $X\not\subset\O$ but $X\subset\O'$
changed from $0$ to $J_X\ge0$.

Likewise if the boundary condition is $+$ then enlarging the box from $\O$
to $\O'$ corresponds to decreasing an external field acting $h$ on the
external spins from $+\io$ (which would force all external spins to be
$+$) to a finite value $h\ge0$: so that increasing the box $\O$ causes
$\media{\s_X}_{+,\O}$ to decrease. Therefore as $\O$
increases Ising ferromagnets spin correlations increase if the
boundary condition is empty and decrease if it is $+$.

The inequalities can be used in similar ways to prove that the infinite
volume states obtained from $+$ or empty boundary conditions are
translation invariant; and that in zero external field, $h=0$, the $+$
and $-$ boundary conditions generate pure states if the interaction
potential is only a pair ferromagnetic interaction.

There are many other important inequalities which can be used to prove
several existence theorems along very simple paths. Unfortunately
their use is mostly restricted to lattice systems and requires very
special assumptions on the energy (\eg ferromagnetic interactions in
the above example). The quoted examples were among the first
discovered and provide a way to exhibit nontrivial thermodynamic
limits and pure states.

\*
\0{\it Bibliography:} [Ru69], [Le74], [Ga99], [LL01], [Li02].
\*

\Section(17, Symmetry breaking phase transitions)
\*

The simplest phase transitions, see Sect. \sec(9), are {\it symmetry
breaking} transitions in lattice systems: they take place when the
energy of the system in a container $\O$ and with some {\it special}
boundary condition (\eg periodic, or antiperiodic or empty) is
invariant with respect to the action of a group $\GG$ on phase
space. This means that on the points $x$ of phase space acts a group
of transformations $\GG$ so that with each $\g\in\GG$ a map $x\to x\g$ is
associated which transforms $x$ into $x\g$ respecting the composition
law in $\GG$, \ie $(x\g)\g'\=x(\g\g')$. If $F$ is an observable the
action of the group on phase space induces an action on the observable
$F$ changing $F(x)$ into $F_\g(x)\defi F(x\g^{-1})$.

A {\it symmetry breaking transition} occurs when by fixing suitable
boundary conditions and taking the thermodynamic limit a state
$F\to\media{F}$ is obtained in which some local observable shows a non
symmetric average $\media{F}\ne \media{F_\g}$ for some $\g$.

An example is provided by the ``nearest neighbor ferromagnetic Ising
model'' on a $d$--dimensional lattice with energy function given by
\equ(15.2) with $\f(x-y)\=0$ unless $|x-y|=1$, \ie unless $x,y$ are
nearest neighbors, in which case $\f(x-y)=J>0$. With periodic or empty
boundary conditions it exhibits a {\it discrete} ``up-down'' symmetry
$\Bs\to -\Bs$.

Instability with respect to boundary conditions can be revealed by
considering the two boundary conditions, denoted $+$ or $-$, in which
the lattice sites outside the container $\O$ are either occupied by
spins $+$ or by spins $-$. Consider also, for later reference, the
boundary conditions in which the boundary spins in the upper half of
the boundary are $+$ and the ones in the lower part are $-$: call this
the $\pm$-boundary condition; see Fig.2. Or the boundary conditions in which some
of the opposite sides of $\O$ are identified while $+$ or $-$
conditions are assigned on the remaining sides: call these
``cylindrical or semiperiodic boundary conditions''.

A new description of the spin configurations is useful: given $\Bs$
draw a unit segment perpendicular to the center of each bond $b$
having opposite spins at its extremes. An example of this construction
is provided by Fig.2 for the boundary condition $\pm$.

\eqfig{120pt}{112pt}{
\ins{2.pt}{58.0pt}{$\st\bf A$}
\ins{54.pt}{58.0pt}{$\st\bf O$}
\ins{106.pt}{58.0pt}{$\st\bf B$}
}{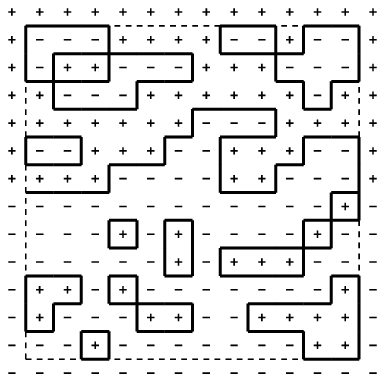}{}

\0{Fig.2:\nota\ The dashed line is the boundary of $\O$; the outer
spins correspond to the $\pm$ boundary condition.  The points $A,B$
are points where an open ``line'' $\l$ ends.}

\*
The set of segments can be grouped into lines separating regions where
the spins are positive from regions where they are negative.  If the
boundary condition is $+$ or $-$ the lines form "closed polygons"
while if the condition is $\pm$ there is also a single polygon $\l_1$
which is not closed (as in Fig. 2). If the boundary condition is
periodic or cylindrical all polygons are closed but some may ``go
around'' $\O$.

The correspondence $(\g_1,\g_2,\ldots,\g_n,\l_1)\otto\Bs$, {\it for
the boundary condition $\pm$} or, {\it for
the boundary condition $+$ (or $-$), $\Bs\otto(\g_1,\ldots,\g_n)$}
is one-to-one and, if $h=0$, the
energy $H_\O(\Bs)$ of a configuration is higher than
$-J\cdot(\hbox{number of bonds in $\O$})$ by an amount $2 J(|\l_1|+
\sum_i|\g_i|) $ or, respectively, $2J\sum_i|\g_i|$. The grand
canonical probability of each spin configuration is therefore
proportional, if $h=0$, respectively to
$$e^{-2\b J(|\l_1|+\sum_i|\g_i|)}\qquad{\rm or}\qquad
e^{-2\b J\sum_i|\g_i|}\Eq(17.1)$$
and the ``up-down'' symmetry is clearly reflected by \equ(17.1).

The average $\media{\s_x}_{\O,+}$ of $\s_+$ {\it with $+$ boundary
conditions} is given by $\media{\s_x}_{\O,+}=1-2
\,P_{\O,+}(-)$, where $P_{\O,+}(-)$ is the probability that the spin $\s_x$
is $-1$. If the site $x$ is occupied by a negative spin then the point
$x$ is {\it inside} some contour $\g$ associated with the spin
configuration $\Bs$ under consideration.  Hence if $\r(\g)$ is the
probability that a given contour belongs to the set of contours
describing a configuration $\Bs$, it is $P_{\O,+}(-)\le \sum_{\g o
x}\r(\g)$ where $\g o x$ means that $\g$ ``surrounds'' $x$.

If $\G=(\g_1,\ldots,\g_n)$ is a spin
configuration and if the symbol $\G\,{comp}\,\g$ means that the
contour $\g$ is ``disjoint'' from
$\g_1,\ldots,\g_n$ (\ie $\{\g\cup\G\}$ is a new spin configuration),
then
$$\r(\g)=\fra{\sum_{\G\ni \g} e^{-2\b J\sum_{\g'\in \G}|\g'|}}
{\sum_{\G} e^{-2\b J\sum_{\g'\in \G}|\g'|}}\=
e^{-2\b J|\g|}
\fra{\sum_{\G\,{comp}\, \g} e^{-2\b J\sum_{\g'\in \G}|\g'|}}
{\sum_{\G} e^{-2\b J\sum_{\g'\in \G}|\g'|}}\le e^{-2\b J|\g|}\Eq(17.2)$$
because the last ratio in \equ(17.2) does not
exceed $1$. Remark that there are at most $3^p$ different
shapes of $\g$ with perimeter $p$ and at most $p^2$ congruent $\g$'s
containing $x$; therefore the probability that the
spin at $x$ is $-$ when the boundary condition is $+$ satisfies the inequality
$P_{\O,+}(-)\le\sum_{p=4}^\io p^2 3^p e^{-2\b J p}\,\tende{\b\to\io}0$.

This probability can be made as small as wished so that
$\media{\s_x}_{\O,+}$ is estimated by a quantity which is as close to
$1$ as desired provided $\b$ is large enough and {\it the closeness of
$\media{\s_x}_{\O,+}$ to $1$ is both $x$ and $\O$ independent}.

A similar argument for the $(-)$-boundary condition, or the remark
that for $h=0$ it is $\media{\s_x}_{\O,-}=-\media{\s_x}_{\O,+}$, leads
to conclude that, at large $\b$,
$\media{\s_x}_{\O,-}\ne\media{\s_x}_{\O,+}$ and the difference between
the two quantities is positive uniformly in $\O$. This is the proof
({\it Peierls' theorem}) of the fact that there is, if $\b$ is large,
a strong instability with respect to the boundary conditions of the
magnetization: \ie the nearest neighbor Ising model in dimension $2$
(or greater, by an identical argument) has a phase transition.  If the
dimension is $1$ the argument clearly fails and no phase transition
occurs, see Sect. \sec(12).
\*
\0{\it Bibliography:} [Ga99].
\*
\Section(18, Finite volume effects)
\*

The description in Sect.\sec(17) of the phase transition in the
nearest neighbor Ising model can be made more precise from the physical
point of view as well as from the mathematical point of view giving
insights into the nature of the phase transitions.  Assume that the
boundary condition is the $(+)$-boundary condition and describe a spin
configuration $\Bs$ by means of the associated closed
disjoint polygons $(\g_1,\ldots,\g_n)$.  Attribute to
$\Bs=(\g_1,\ldots,\g_n)$ a probability proportional to
\equ(17.1). Then the following {\it Minlos--Sinai's theorem}
holds

\vskip1mm
\0{\it If $\b$ is large enough there exist $C>0, \r(\g)>0$
with $\r(\g)\le e^{-2\b J|\g|}$ and such that a spin configuration
$\Bs$ randomly chosen out of the grand canonical distribution with $+$
boundary conditions and $h=0$ will contain, with probability
approaching $1$ as $\O\to\io$, a number $K_{(\g)}(\Bs)$ of contours
congruent to $\g$ such that
$$|K_{(\g)}(\Bs)-\r(\g)\,|\O||\le C\sqrt{|\O|}\, e^{-\b
J|\g|}\Eq(18.1)$$
\0and this relation holds simultaneously for all $\g$'s.}

\vskip1mm
Thus there are very few contours (and the larger they are the smaller
is, in absolute and relative value, their number): a typical spin
configuration in the grand canonical ensemble with $(+)$-boundary
conditions is such that the large majority of the spins is
``positive'' and, in the ``sea'' of positive spins, there are a few
negative spins distributed in {\it small and rare regions} (in a number,
however, still of order of $|\O|$).

Another consequence of the analysis in Sect.\sec(17) concerns the
the approximate equation of state near the phase
transition region at low
temperatures and finite $\O$.

\kern-1.cm

\eqfig{210pt}{80pt}{
\ins{25pt}{13pt}{$\st O(|\O|^{-1/2})$}
\ins{-60pt}{15pt}{$\st-O(|\O|^{-1/2})$}
\ins{-35pt}{25pt}{$ m^*(\b)$}
\ins{5pt}{-18pt}{$-m^*(\b)$}
\ins{70pt}{-7pt}{$h$}
\ins{-60pt}{30pt}{$1$}
\ins{5pt}{47pt}{$m_\O(\b,h)$}
\ins{180pt}{15pt}{$ \st O(|\L|^{-1/2})$}
\ins{110pt}{15pt}{$\st -O(|\L|^{-1/2})$}
\ins{137pt}{25pt}{$ m^*(\b)$}
\ins{175pt}{-18pt}{$-m^*(\b)$}
\ins{240pt}{-7pt}{$h$}
\ins{110pt}{30pt}{$1$}
\ins{175pt}{47pt}{$m_\O(\b,h)$}
}{fig3}{\lower40pt\hbox{(3)}}

\kern1.5cm

\0If $\O$ is finite the graph of $h\to m_\O(\b,h)$ will have a rather
different behavior depending on the possible boundary conditions; \eg
if the boundary condition is $(+)$ or $(-)$ one gets respectively, the
results depicted in Fig.3, where $m^*(\b)$ denotes the {\it
spontaneous magnetization}, \ie $m^*(\b)\defi
\lim_{h\to0^+}\lim_{\O\to\io} m_\O(\b,h)$.

With periodic or empty boundary conditions the diagram changes as in the first
of Fig.4: the thermodynamic limit $m(\b,h)=\lim_{\O\to\io}
m_\O(\b,h)$ exists for all $h\ne0$ and the resulting graph is
in the second of Fig.4,  which shows that at $h=0$ the limit is
discontinuous. It can be proved, if $\b$ is large enough,
that $\io>\lim_{h\to 0^+}\dpr_h m(\b,h)=\chi(\b)>0$ (\ie the angle
between the vertical part of the graph and the rest is sharp).

\eqfig{210pt}{45pt}{
\ins{10pt}{15pt}{$\st O(|\O|^{-1/2})$}
\ins{-50pt}{15pt}{$\st -O(|\O|^{-1/2})$}
\ins{-35pt}{28pt}{$m^*(\b)$}
\ins{5pt}{-15pt}{$-m^*(\b)$}
\ins{70pt}{-10pt}{$h$}
\ins{-60pt}{30pt}{$1$}
\ins{7pt}{45pt}{$m_\O(\b,h)$}
\ins{135pt}{25pt}{$m^*(\b)$}
\ins{174pt}{-15pt}{$-m^*(\b)$}
\ins{235pt}{-10pt}{$h$}
\ins{110pt}{30pt}{$1$}
\ins{175pt}{45pt}{$m(\b,h)$}
}{fig4}{\lower23pt\hbox{(4)}}

\kern1.3cm

\0Furthermore it can be proved that $m(\b,h)$ is analytic in $h$ for
$h\ne0$. If $\b$ is small enough analyticity holds at all $h$. For
$\b$ large the function $f(\b,h)$ has an {\it essential singularity
at} $h=0$: a result that can be interpreted as excluding a naive
theory of metastability as a description of states governed by an
equation of state obtained from an analytic continuation to negative values
of $h$ of $f(\b,h)$.

The above considerations and results further clarify what a
phase transition for a finite system means.

\*
\0{\it Bibliography:} [Ga99], [FP04].
\*

\Section(19, Beyond low temperature {(ferromagnetic Ising model)})
\*

A limitation of the results discussed above is the condition of low
temperature (``$\b$ large enough'').  A natural problem is to go
beyond the low-temperature region and to describe fully the phenomena
in the region where boundary condition instability takes place and
first develops. A number of interesting partial results are known,
which considerably improve the picture emerging from the previous
analysis. A striking list, but far from exhaustive, of such results
follows and focuses on {\it properties of ferromagnetic Ising spin
systems}. The reason for restricting to such cases is that they are
simple enough to allow a rather fine analysis which sheds considerable
light on the structure of statistical mechanics suggesting precise
formulation of the problems that it would be desirable to understand
in more general systems.
\vskip1mm

\0(1) Let $z\defi e^{\b h}$ and consider the the product of $z^V$ ($V$
is the number of sites $|\O|$ of $\O$) times the partition function
with periodic or perfect-wall boundary conditions with finite range
feromagnetic interaction, not neceesarily nearest neighbor; a
polynomial in $z$ (of degree ${2V}$) is thus obtained. Its zeros
{\it lie on the unit circle} $|z|=1$: this is {\it Lee-Yang's theorem}.
It implies that the only singularities of $f(\b,h)$ in the region
$0<\b<\io$, $-\io<h<+\io$ can be found at $h=0$.

A singularity can appear only if the point $z=1$ is an accumulation
point of the limiting distribution (as $\O\to\io$) of the zeros on the
unit circle: if the zeros are $z_1,\ldots,z_{2V}$ then
$\fra1{V}\log z^{V} Z(\b,h,\O,{\rm periodic})=2\b J+\b
h+\fra1{V}\sum_{i=1}^{2V} \log (z-z_i)$ and if
$V^{-1}\cdot\big($number of zeros of the form $z_j=e^{i\th_j}\ {\rm
with}\ \th\le\th_j\le \th+d\th$)$\tende{\O\to\io}\fra{d\r_\b(\th)}{2\p}$ it
is

$$\b f(\b,h)=2\b J +
\fra1{2\p}\ig_{-\p}^\p \log (z-e^{i\th})\, d\r_\b(\th) \Eq(19.1)$$
The existence of the measure $d\r_\b(\th)$ follows from the existence
of the thermodynamic limit: but $d\r_\b(\th)$ is not necessarily
$d\th$-continuous, \ie proportional to $d\th$.

\0(2) It can be shown
that, with a not necessarily nearest neighbor interaction, the zeros
of the partition function do not move too much under small
perturbations of the potential even if one perturbs the
energy (at perfect-wall or periodic boundary conditions) into
$$ H'_\O(\Bs)=H_\O(\Bs)+(\d H_\O)(\Bs),\qquad (\d
H_\O)(\Bs)=\sum_{X\subset\O} J'(X)\,\s_X\Eq(19.2)$$
where $J'(X)$ is very general and defined on subsets
$X=(x_1,\ldots,x_k)\subset\O$ such that the quantity
$||J'||=\sup_{y\in Z^d}\sum_{y\in X} |J'(X)|$ is small enough.  More
precisely fixed a ferromagnetic pair potential $J$ suppose that one
knows that, when $J'=0$, the partition function zeros in the variable
$z=e^{\b h}$ lie in a certain closed set $N$ (of the unit circle) in
the $z$-plane. Then if $J'\ne0$ they lie in a closed set $N^1$,
$\O$-independent and contained in a neighborhood of $N$ of width
shrinking to $0$ when $||J'||\to0$.  This leads to establish various
relations between analyticity properties and boundary condition
instability as described in (3) below.

\0(3) In the ferromagnetic Ising model, with a not necessarily nearest
neighbor interaction, one says that there is a {\it gap} around $0$ if
$d\r_\b(\th)=0$ near $\th=0$. It can be shown that if $\b$ is small
enough there is a gap for all $h$ of width uniform in $h$.

\0(4) Another question is whether the boundary condition instability is
always revealed by the one-spin correlation function (\ie by the
magnetization) or whether it might be shown only by some correlation
functions of higher order. It can be proved that no boundary condition
instability takes place for $h\ne0$; at $h=0$ it is possible only if
$$\lim_{h\to0^-} m(\b,h)\ne \lim_{h\to0^+} m(\b,h)\Eq(19.3)$$

\0(5) A consequence of the Griffiths' inequalities, \cfr Sect.\sec(16),
is that if \equ(19.3) is true for a given $\b_{0}$ then it is true for
all $\b>\b_0$.  Therefore item (4) leads to a natural definition of
the {\it critical temperature} $T_c$ as the least upper bound of the
$T$'s such that \equ(19.3) holds ($k_BT=\b^{-1}$).

\0(6) If $d=2$ the free energy of the nearest neighbor ferromagnetic
Ising model has a singularity at $\b_c$ and the value of $\b_c$ is
known exactly from the exact solutions of the model:
$m(\b,0)\defi m^*(\b)\=(1-\sinh^42\b J)^{\fra18}$.
The location and nature of the singularities of $f(\b,0)$ as a function of
$\b$ remains an open question for $d=3$.  In particular the question
of whether there is a singularity of $f(\b,0)$ at $\b=\b_c$ is open.

\0(7) For $\b<\b_c$ there is instability with respect to boundary
conditions (see (6) above) and a natural question is: how many
``pure'' phases can exist in the ferromagnetic Ising model? (\cfr
\equ(9.2) and Sect.\sec(9)).  Intuition suggests that there should be
only two phases: the positively magnetized and the negatively
magnetized ones.

One has to distinguish between translation invariant pure phases and
non translation invariant ones.  It can be proved that, in the case of
the $2$--dimensional nearest neighbor ferromagnetic Ising models, all
infinite volume states (\cfr Sect.\sec(15)) are translationally
invariant. Furthermore they can be obtained by considering just the
two boundary conditions $+$ and $-$: the latter states are pure states
also for models with non nearest neighbor ferromagnetic interaction.
The solution of this problem has led to the introduction of many new
ideas and techniques in statistical mechanics and probability theory.

\0(8) In any dimension $d\ge2$, for $\b$ large enough it can be proved that
the nearest neigbor Ising model has only two translation invariant
phases.
If the dimension is $\ge 3$ and $\b$ is large the $+$ and $-$
phases exhaust the set of translationally invariant pure phases but
{\it there exist non translationally invariant phases}.  For $\b$
close to $\b_c$, however, the question is much more difficult.

\*
\0{\it Bibliography:} [On44], [LY52], [Ru71], [Si91],
[Ga99], [Ai80], [Hi81], [FP04]

\*

\Section(20, Geometry of phase coexistence.)
\*

Intuition about the phenomena connected with the classical phase
transitions is usually based on the properties of the liquid-gas phase
transition; this transition is experimentally investigated in
situations in which the total number of particles is fixed (canonical
ensemble) and in presence of an external field (gravity).

The importance of such experimental conditions is obvious; the
external field produces a nontranslationally invariant situation and
the corresponding separation of the two phases. The fact that the
number of particles is fixed determines, on the other hand, the
fraction of volume occupied by each of the two phases.

Once more consider the nearest neighbor ferromagnetic Ising model: the
results available for it can be used to obtain a clear picture of the
solution to problems that one would like to solve but which in most
other models are intractable with present day techniques.

It will be convenient to discuss phase coexistence in the canonical
ensemble distributions on configurations of fixed total magnetization
$M=m V$, see Sect.\sec(15), \equ(15.4).  Let $\b$ be large enough to
be in the two phases region and, for a fixed $\a\in(0,1)$, let

$$m=\a\,m^*(\b)+(1-\a)\,(-m^*(\b))=\,(1-2\a)\,m^*(\b)\Eq(20.1)$$
\ie $m$ is in the vertical part of the diagram $m=m(\b,h)$ at $\b$
fixed (see Fig.4).

Fixing $m$ as in \equ(20.1) does not yet determine the separation of
the phases in two different regions; for this effect it will be
necessary to introduce some external cause favoring the occupation of
a part of the volume by a single phase.  Such an asymmetry can be
obtained in at least two ways: through a weak uniform external field
(in complete analogy with the gravitational field in the liquid-vapor
transition) or through an asymmetric field acting only on boundary
spins.  This second way should have the same qualitative effect as the
former, because in a phase transition region a boundary perturbation
produces volume effects: see Sect. \sec(9),\sec(17).  From a
mathematical point of view it is simpler to use a boundary asymmetry
to produce phase separations and the simplest geometry is obtained by
considering {\it $\pm$-cylindrical or $++$-cylindrical boundary
conditions}: this means $++$ or $\pm$ boundary conditions periodic in
one direction (\eg in Fig.2 imagine the right and left boundary
identified after removing the boundary spins on them).

Spins adjacent to the bases of $\O$ act as symmetry-breaking
external fields. The $++$-cylindrical boundary condition should
favor the formation inside $\O$ of the positively magnetized phase;
therefore it will be natural to consider, in the canonical distribution,
this boundary condition only when the total magnetization is fixed to
be the spontaneous magnetization $m^*(\b)$.

On the other hand, the $\pm$-boundary condition favors the
separation of phases (positively magnetized phase near the top of $\O$
and negatively magnetized phase near the bottom).  Therefore it will
be natural to consider the latter boundary condition in the case of a
canonical distribution with magnetization $m=(1-2\a)\,m^*(\b)$ with
$0<\a<1$, \equ(20.1).  In the latter case the positive phase can be
expected to adhere to the top of $\O$ and to extend, in some sense to
be discussed, up to a distance $O(L)$ from it; and then to change into
the negatively magnetized pure phase.

To make precise the phenomenological description  consider
the spin configurations $\Bs$ through the associated sets
of disjoint polygons (\cfr Sect.\sec(17)).
Fix the boundary conditions to be $++$ or $\pm$-cylindrical
boundary conditions and note that polygons associated with a spin
configuration $\Bs$ are {\it all closed} and of two types:
the ones of the first type, denoted $\g_1,\ldots,\g_n$, are polygons
which do not encircle $\O$, the second type of polygons, denoted by the
symbols $\l_\a$, are the ones which wind up, at least once, around $\O$.

So a spin configuration $\Bs$ will be described by a set of
po\-ly\-gons the statistical weight of a configuration $\Bs=(\g_1
,\ldots,\g_n,\l_1,\ldots,\l_h)$ is (\cfr
\equ(17.1)):

$$ e^{-2\b J(\sum_i|\g_i|+\sum_j|\l_j|)}\,.\Eq(20.2)$$
The reason why the contours $\l$ that go around the cylinder $\O$ are
denoted by $\l$ (rather than by $\g$) is that they ``look like'' open
contours, see Sect.\sec(17), if one forgets that the opposite sides of
$\O$ have to be identified.  In the case of the $\pm$-boundary
conditions then the number of polygons of $\l$-type {\it must be odd}
(hence $\ne0$), while for the $++$-boundary condition the number of
$\l$-type polygons {\it must be even} (hence it could be $0$).
\*
\0{\it Bibliography:} [Si91], [Ga99]
\*

\Section(21, Separation and Coexistence of Phases)
\*

In the context of the geometric description of the spin configuration
in Sect.\sec(20) consider the canonincal distributions with
$++$-cylindrical or the $\pm$-cylindrical boundary conditions and zero
field: they will be denoted briefly as $\m_{\b,++}$, $\m_{\b,\pm}$.
The following theorem {\it Minlos-Sinai's theorem} provided the
foundations of the microscopic theory of coexistence: it is formulated
in dimension $d=2$ but, modulo obvious changes, it holds for $d\ge2$.
\*

\0{\it For $0<\a<1$ fixed let $m=(1-2\a)\,m^*(\b)$; then
for $\b$ large enough a spin configuration
$\Bs=(\g_1,\ldots,\g_n,\l_1,\ldots,\l_{2h+1})$ randomly chosen
with the distribution $\m_{\b,\pm}$ enjoys the properties (1)-(4)
below with a $\m_{\b,\pm}$--probability approaching $1$ as $\O\to\io$:

\0(1) $\Bs$ contains {\it only one} contour of $\l$-type and
$$|\,|\l|-(1+\e(\b))L|<o(L)\Eq(21.1)$$
where $\e(\b)>0$ is a suitable ($\a$-independent) function of $\b$
tending to zero exponentially fast as $\b\to\io$.

\0(2) If $\O^+_\l,\O^-_\l$ denote the regions above and below $\l$ and
$|\O|\=V,|\O^+|,|\O^-|$ are the volumes of $\O,\O^+,\O^-$ it is
$$|\,|\O^+_\l|-\a\,V\,|<\k(\b)\,V^{3/4}\qquad
|\O^-_\l|-(1-\a)V\,|<\k(\b)\,V^{3/4}\Eq(21.2)$$
where $\k(\b)\tende{\b\to\io}0$ exponentially fast; the exponent
$\fra34$, here and below, is not optimal.

\0(3) If $M_\l=\sum_{x\in \O_\l}\s_x$, it is
$$|M_\l-\a\, m^*(\b)\,V|<\k(\b)V^{3/4},\qquad
M'_\l-(1-\a)\, m^*(\b)\,V|<\k(\b)V^{3/4}\Eq(21.3)$$.

\0(4) If $K^\l_{\g}(\Bs)$ denotes the number of contours congruent to a
given $\g$  and lying in $\O^+_\l$ then, simultaneously for all the
shapes of $\g$:
$$|\,K^\l_{\g}(\Bs)-\r(\g)\,\a\, V\,|\le C e^{-\b
J|\g|}{V}^{\fra12}\qquad C>0\Eq(21.4)$$
where $\r(\g)\le e^{-2\b J |\g|}$ is the same quantity already
mentioned in the text of the theorem of \S(6.8). A similar result
holds for the contours below $\l$ (\cfr the comments on \equ(18.1)).}
\*

The theorem not only provides a detailed and rather satisfactory
description of the phase separation phenomenon, but it also furnishes
a precise microscopic definition of the line of separation between the
two phases, which should be naturally identified with the (random)
line $\l$.

A similar result holds in the canonical distribution
$\m_{\b,++,m^*(\b)}$ where (1) is replaced by: {\it no} $\l$-type
polygon is present, while (2), (3) become superfluous and (4) is
modified in the obvious way. In other words a typical configuration
for the distribution the $\m_{\b,++,m^*(\b)}$ has the same appearance
as a typical configuration of the corresponding grand canonical
ensemble with $(+)$-boundary condition (whose
properties are described by the theorem of Sect.\sec(19)).
\*
\0{\it Bibliography:} [Si91], [Ga99].
\*

\Section(22, Phase separation line and surface tension)
\*

Continuing to refer to the nearest neighbor Ising ferromagnet, the
theorem of Sect.\sec(21) means that, if $\b$ is large enough, then the
microscopic line $\l$, separating the two phases, is almost straight
(since $\e(\b)$ is small). The deviations of $\l$ from a straight line
are more conveniently studied in the grand canonical distributions
$\m^0_\pm$ with boundary condition set to $+1$ in the upper half of
$\dpr\O$, {\it vertical sites included}, and to $-1$ in the lower
half: this is illustrated in Fig.2, Sect.\sec(17). The results can be
converted into very similar results for grand canonical distributions
with $\pm$-cylindrical boundary conditions of Sect.\sec(21),\sec(22).

Define $\l$ to be {\it rigid} if the probability that $\l$ passes
through the center of the box $\O$ (\ie $0$) {\it does not} tend to
$0$ as $\O\to\io$; otherwise it is {\it not rigid}.

The notion of rigidity distinguishes between the possibilities for the
line $\l$ to be ``straight'': the ``excess'' length $\e(\b)L$, see
\equ(21.1), can be obtained in two ways: either the line $\l$
is essentially straight (in the geometric sense) with a few ''bumps''
distributed with a density of order $\e(\b)$ or, otherwise, it is only
locally straight and with an important part of the excess length being
gained through a small bending on a large length scale.  In three
dimensions a similar phenomenon is possible.  Rigidity of $\l$, or its
failure, can in principle be investigated by optical means; there can
be interference of coherent light scattered by macroscopically
separated surface elements of $\l$ only if $\l$ is rigid
in the above sense.

It has been rigorously proved that, the line $\l$ is
{\it not rigid} in dimension $2$.  And, at least at low temperature,
the fluctuation of the middle point is of the order $O(\sqrt{L})$. In
dimension $3$ however it has been shown
that the surface $\l$ is rigid at low enough temperature.

A deeper analysis is needed to study the shape of the separation
surface under otherconditions, \eg with $+$ boundary conditions and in
a canonical distribution with magnetization intermediate between $\pm
m^*(\b)$. It involves, as a prerequisite, the definition and many
properties of the {\it surface tension} between the two phases. Here
only the definition of surface tension in the case of $\pm$-boundary
conditions in the $2$-dimensional case will be mentioned. If
$Z^{++}(\O,m^*(\b))$ and $Z^{+-}(\O,m)$ are the canonical partition
functions for the $++$ and $\pm$-cylindrical boundary conditions the
tension $\t(\b)$ is defined as
$\b\,\t(\b)=-\lim_{\O\to\io}\fra1L\log
\fra{Z^{+-}(\O,m)}{Z^{++}(\O,m^*(\b))}$.
The limit can be shown to be $\a$-independent for $\b$ large enough:
the definition motivation and justification is based on the
microscopic geometric description in Sect.\sec(20). The definition can
be naturally extended to higher dimension (and to more general non
nearest neighbor models). If $d=2$ the tension $\t$ can be exactly
computed at all temperature below criticality and is
$\b\,\t(\b)=+2\b J+\log\tanh \b J$.

More remarkably the definition can be extended to define the surface
tension $\t(\b,\V n)$ in the ``direction $\V n$'', \ie when the
boundary conditions are such that the line of separation is in the
average orthogonal to the unit vector $\V n$. In this way if $d=2$ and
$\a\in(0,1)$ is fixed it can be proved that at low enough temperature
the canonical distribution with $+$ boundary conditions and
intermediate magnetization $m=(1-2\a)m^*(\b)$ has typical
configurations containing a region of spins $-$ of area $\simeq \a V$;
furthermore if the container is rescaled to size $L=1$ the region will
have a limiting shape filling an area $\a$ bounded by a {\it smooth} curve
whose form is determined by the classical macroscopic {\it Wulff's
theory} of the shape of crystals in terms of the surface tension
$\t(\V n)$.

An interesting question remains open in the three-dimensional case: it
is conceivable that the surface, although rigid at low temperature,
might become ``loose'' at a temperature $\widetilde T_c$ smaller than
the critical temperature $T_c$ (the latter being defined as the
highest temperature below which there are at least two pure
phases). The temperature $\widetilde T_c$, whose existence is rather
well established in numerical experiments, would be called the ``{\it
roughening transition}'' temperature.  The rigidity of $\l$ is
connected with the existence of translationally noninvariant
equilibrium states. The latter exist in dimension $d=3$, but not in
dimension $d=2$ where the discussed nonrigidity of $\l$, established
all the way to $T_c$, provides the intuitive reason for the absence of
non translationally invariant states. It has been shown that in $d=3$
the roughening temperature $\widetilde T_c(\b)$ necessarily {\it
cannot be smaller} than the critical temperature of the
$2$--dimensional Ising model with the same coupling.

Note that existence of translationally noninvariant equilibrium states
is not necessary for the description of coexistence phenomena. The
theory of the nearest neighbor two-dimensional Ising model is a clear
proof of this statement.
\vskip2mm
\0{\it Bibliography:} [On44], [Be75], [Si91], [Mi95], [PV99], [Ga99],

\*

\Section(23, Critical points)
\*

Correlation functions for a system with short range interactions and
in an equilibrium state which is a pure phase have {\it cluster
properties}, see \equ(9.2): their physical meaning is that in a pure
phase there is independence between fluctuations occurring in widely
separated regions.  The simplest cluster property concerns the ``pair
correlation function'', \ie the probability density $\r(\V q_1,\V
q_2)$ of finding particles in points $\V q_1,\V q_2$ independently of
where the other particles may happen to be, see \equ(9.3). In the case
of spin systems the pair correlation $\r(\V q_1,\V q_2)=\media{\s_{\V
q_1}\s_{\V q_2}}$ will be considered. The pair correlation of a
translation invariant equilibrium state has a cluster property,
\equ(9.2),\equ(15.6), if

$$|\r(\V q_1,\V q_2)-\r^2|\,\tende{|\V q_1-\V
q_2|\to\io}\,0 \Eq(23.1)$$
where $\r$ is the probability density for finding a particle at $\V q$
(\ie the physical density of the state) or $\r=\media{\s_{\V q}}$ is
the average of the value of the spin at $\V q$ (\ie the magnetization
of the state).

A general definition of {\it critical point} is a point $c$ in the
space of the parameters characterizing equilibrium states, \eg $\b,\l$
in grand canonical distributions or $\b,v$ in canonical distributions
or $\b,h$ in the case of lattice spin systems in a grand canonical
distribution. In systems with short range interaction (\ie with $\f(\V
r)$ vanishing for $|\V r|$ large enough) the point $c$ is a critical
point if the pair correlation tends to $0$, see \equ(23.1), slower
than exponential, for instance as a power of the distance $|\V r|=|\V
q_1-\V q_2|$.

A typical example is the $2$--dimensional Ising model on a square
lattice and with nearest neighbor ferromagnetic interaction of size
$J$. It has a single critical point at $\b=\b_c$, $h=0$ with $\sinh
2\b_cJ=1$. The cluster property is that
$\media{\s_x\s_y}-\media{\s_x}\media{\s_y}\tende{|x-y|\to\io}0$ as

\vskip-1mm
$$A_+(\b) \fra{e^{-\k(\b)|x-y|}}{\sqrt{|x-y|}},
\qquad A_-(\b)\fra{e^{-\k(\b)|x-y|}}{{|x-y|^2}},\qquad
A_c \fra{1}{{|x-y|^\fra14}} \qquad \b=\b_c
\Eq(23.2)$$
for $\b<\b_c$, $\b>\b_c$ or $\b=\b_c$ respectively, where
$A_\pm(\b),A_c,\k(\b)>0$.  The properties \equ(23.2) stem from the
{\it exact solution} of the model.

At the critical point several interesting phenomena occurr: the lack
of exponential decay indicates lack of a length scale over which
really distinct phenomena can take place and properties of the system
observed at different length scales are likely to be simply related by
suitable scaling transformations. Many efforts have been dedicated at
finding ways of understanding quantitatively the scaling properties
which relate different observables. The result has been the
development of the {\it renormalization group approach} to critical
phenomena, \cfr Sect.\sec(25). The picture that emerges is that the
closer the critical point is the larger becomes the maximal scale of
length {\it below which scaling properties are observed}. For instance
in a lattice spin system in $0$ field the magnetization $M|\L|^{-a}$
in a box $\L\subset\O$ should have essentially the same distribution
for all $\L$'s with side $< l_0(\b)$ and $l_0(\b)\to\io$ as
$\b\to\b_c$, {\it provided} $a$ is suitably chosen. The number $a$ is
called a {\it critical exponent}.

There are several other ``critical exponents'' that can be defined
near a critical point. They can be associated with singularities of
the thermodynamic function or with the behavior of the correlation
functions involving joint densities at two or more than two points. As
an example consider a lattice spin system: then the ``$2n$--spins
correlation'' $\media{\s_0\s_{\x_1}\ldots\s_{\x_{2n-1}}}_c$ could
behave proportionally to $\chi_{2n}(0,\x_1,\ldots,\x_{2n-1})$,
$n=1,2,3,\ldots,$, for a suitable family of {\it homogeneous}
functions $\chi_n$, of some degree $\o_{2n}$, of the coordinates
$(\x_1,\ldots,\x_{2n-1})$ at least when the reciprocal distances are
large but $<l_0(\b)$ and $l_0(\b)=
const\,(\b-\b_c)^{-\n}\tende{\b\to\b_0}\io$. This means that if
$\x_i$ are regarded as points in $\RRR^d$ there are functions
$\ch_{2n}$ such that:

$$\chi_{2n}(0,\fra{\x_1}{\l},\ldots,\fra{\x_{2n-1}}{\l})=
\l^{\o_{2n}}\, \chi_{2n}(0,\x_1,\ldots,\x_{2n-1})\qquad 0<\l\in \RRR\Eq(23.3)$$
and $ \media{\s_0\s_{\x_1}\ldots\s_{\x_{2n-1}}}_c\propto
\chi_{2n}(0,\x_1,\ldots,\x_{2n-1})
$ if $1\ll |x_i-x_j|\ll l_0(\b)$. The numbers $\o_{2n}$ define a
sequence of critical exponents.

Other critical exponents can be associated with approaching the
critical point along other directions: for instance along $h\to0$ at
$\b=\b_c$. In this case the length up to which there are scaling
phenomena is $l_0(h)=\ell_o \, h^{-\lis \n}$. And the magnetization
$m(h)$ tends to $0$ as $h\to0$ at fixed $\b=\b_c$ as $m(h)=m_0
h^{\fra1\d}$ for $\d>0$.

None of the critical exponents is in any generality known rigorously,
including its existence. An exception is the case of the
$2$--dimensional nearest neighbor Ising ferromagnet where some
exponents are known exactly (\eg $\o_2=\fra14$, $\o_{2n}=n \o_2$, or
$\n=1$ while $\d,\lis\n$ are not rigorously known). Nevertheless for
Ising ferromagnets (even not nearest neighbor but, as always here,
finite range) in all dimensions all mentioned exponents are
conjectured to be the same as those of the nearest neighbor Ising
ferromagnet. A further exception is the derivation of rigorous
relations between critical exponents and, in some cases, even their
values under the assumption that they exist.
\*

\0{\it Remark:} Naively it could be expected that in a pure state in
$0$ field with $\media{\s_x}=0$ the quantity
$s=|\L|^{-\fra12}\sum_{x\in\L}\s_x$, if $\L$ is a cubic box of side
$\ell$, should have a probability distribution which is Gaussian with
dispersion $\lim_{\L\to\io}
\media{s^2}$. This is ``usually true'': {\it but not always}. The
\equ(23.2) shows that in the $d=2$ ferromagnetic nearest neighbor
Ising model $\media{s^2}$ diverges proportionally to $\ell^{2-\fra14}$
so that the variable $s$ cannot have a Gaussian distribution. The
variable $S=|\L|^{-\fra78}\sum_{x\in\L}\s_x$ will have a finite
dispersion: however there is no reason that it should be
Gaussian. This makes clear the great interest of a fluctuation theory
and its relevance for the critical point studies, see
Sect.\sec(24),\sec(25).
\*

\0{\it Bibliography:} [On44], [DG72], [MW73], [Ai82].
\*

\pagina
\Section(24, Fluctuations)
\*

As it appears from Sect.\sec(23) fluctuations of observabes around
their averages have interesting properties particularly at critical
points. Of particular interest are observables which are averages over
large volumes $\L$ of local functions $F(x)$ on phase space: this is
so because macroscopic observables often have this form. For instance
given a region $\L$ inside the system container $\O$, $\L\subset\O$,
consider a configuration $x=(\V P,\V Q)$ and the number of particles
$N_\L=\sum_{\V q\in\L} 1$ in $\L$, or the potential energy
$\F_\L=\sum_{(\V q,\V q')\in\L} \f(\V q-\V q')$ or the kinetic energy
$K_\L= \sum_{\V q\in\L}\fra1{2m}\V p^2$. In the case of lattice spin
systems consider a configuration $\Bs$ and, for instance, the
magnetization $M_\L=\sum_{i\in\L}\s_i$ in $\L$.  Label the above four
examples by $\a=1,\ldots,4$.

Let $\m_\a$ be the probability distribution describing the equilibrium
state in which the quantities $X_\L$ are considered; let
$x_\L={\media{\fra{X_\L}{|\L|}}}_{\kern3mm\llap{\hbox{$\st\m_\a$}}}$
and $p\defi\fra{X_\L-x_\L}{|\L|}$. Then typical fluctuations
properties to investigate are ($\a=1\ldots4$),
\vskip2mm

\0(1) {\it for all $\d>0$ it is $\lim_{\L\to\io}
\m_\a(|p|>\d)=0$ (law of large numbers)}

\0(2) {\it there is $D_\a>0$ such that
$\m(p\sqrt{|\L|}\in [a,b]) \tende{\L\to\io} \ig_a^b\fra{dz}{\sqrt{2\p
 D_\a}} e^{-\fra{z^2}{2D_\a}}$ (central limit law)}

\0(3) {\it there is an interval $I_\a=(p_{\a,-}^*,p_{\a,+}^*)$
and a concave function $F_\a(p), \, p\in I$, such that if
$[a,b]\subset I$ then $ \fra1{|\L|}
\log\m\big(p\in [a,b]\big)\tende{\L\to\io}
\max_{p\in [a,b]} F_\a(p)$ (large deviations law)}
\vskip2mm

The law of large numbers provides the {\it certainty} of the macroscopic
values; the central limit law controls the {\it small fluctuations} (of
order $\sqrt{|\L|}$) of $X_\L$ around its average; the large
deviations law concerns the {\it fluctuations of order $|\L|$}.

{\it The relations (1),(2),(3) are not always true}: they can be
proved under further general assumptions if the potential $\f$ satisfies
\equ(7.2) in the case of particle systems or if $\sum_{\V q} |\f(\V
q)|<\io$ in the case of lattice spin systems. The function $F_\a(p)$
is defined in terms of the thermodynamic limits of suitable
thermodynamic functions associated with the equilibrium state $\m_\a$.
The further assumption is, essentially in all cases, that a suitable
thermodynamic function in terms of which the $F_\a(p)$ will be
expressed is smooth and with a nonvanishing second derivative.

For the purpose of a simple concrete example consider a lattice spin
system of Ising type with energy $-\sum_{x,y\in
\O}\f(x-y)\s_x\s_y-\sum_x h \s_x$ and the fluctuations of the
magnetization $M_\L=\sum_{x\in\L}\s_x, \L\subset\O$, in the grand
canonical equilibrium states $\m_{h,\b}$.

Let the free energy be $\b f(\b,h)$, see \equ(15.5), let $
m=m(h)\defi\media{\fra{M_\L}{|\L|}}$ and let $h(m)$ the inverse
function of $m(h)$.  If $p=\fra{M_\L}{|\L|}$ the function $F(p)$ is

$$
F(p)=\b\big(f(\b,h(p))-f(\b,h)-\dpr_h
f(\b,h)(h(p)-h)\big)\Eq(24.1)$$
then a quite general result is
\vskip1mm

\0{\it The relations (1),(2),(3) hold if the potential satisfies $\sum_x
|\f(x)|<\io$ and if $F(p)$, \equ(24.1), is smooth and $F''(p)\ne0$ in
open intervals around those in which $p$ is considered, \ie around
$p=0$ for the law of large numbers and for the central limit law or in
an open interval containing $a,b$ for the case of the large deviations
law.}
\vskip1mm

In the envisaged cases the theory of equivalence of the ensembles
implies that the function $F$ can also be computed via thermodynamic
functions naturally associated with other equilibrium ensembles. For
instance instead of the grand canonical $f(\b,h)$ one could consider
the canonical $\b g(\b,m)$, see \equ(15.5), then

$$
F(p)=-\b\big(g(\b,p)-g(\b,m)-\dpr_m
g(\b,m)(p-m)\big)\Eq(24.2)$$

It has to be remarked that there should a strong relation between the
central limit law and the law of large deviations. Setting aside
stating the conditions for a precise mathematical theorem, the
statement can be rapidly illustrated in the case of a ferromagnetic
lattice spin system and with $\L\=\O$, by showing that the large
deviations law in small intervals, around the average $m(h_0)$, at a
value $h_0$ of the external field is implied by the validity of the central
limit law for all values of $h$ near $h_0$ and viceversa (here $\b$ is
fixed). Taking $h_0=0$ (for simplicity) the heuristic reasons are the
following. Let $\m_{h,\O}$ be the grand canonical distribution in
external field $h$. Then

\0(1) The probability $\m_{h,\O}(p\in dp)$ is proportional, by definition,
to $\m_{0,\O} (p\in
dp) e^{-\b h m |\O|}$. Hence if the central limit law holds for
all $h$ near $h_0=0$ there will exist two functions $m(h)$ and $D(h)>0$,
defined for $h$ near $h_0=0$, with $m(0)=0$ and

$$\m_0(p\in dp) e^{-\b h p}= const\,
 e^{-|\O|\fra{(p-m(h))^2}{2D(h)}+ o(\O)}\,dp\Eq(24.3)$$

\0(2) there is a function
$\z(m)$ such that $\dpr_m\z(m(h))=\b h$ and $\dpr^2_m
\z(m(h))=D(h)^{-1}$. {\bf[}This is obtained by remarking that, given $D(h)$,
the differential equation $\dpr_m \b h=D(h)^{-1}$ with the initial
value $h(0)=0$ determines the function $h(m)$; therefore $\z(m)$ is
determined by a second integration, from $\dpr_m\z(m)=\b h(m)$.

It then follows, heuristically, that the probability of $p$ in zero
field has the form $const\, e^{\z(p) |\O|}\,dp$ so that the
probability that $p\in [a,b]$ will be $const\,
e^{|\O|\max_{p\in[a,b]}\z(p)}$.{\bf]}

Viceversa the large deviations law for $p$ at $h=0$ implies the
validity of the central limit law for the fluctuations of $p$ in
all small enough fields $h$: this simply arises from the function
$F(p)$ having a negative second derivative.

{\it This means that there is a `''{\it duality}'' between central
limit law and large deviation law or that the large deviations law is
a ``global version'' of the central limit law. In the sense that

\0(a) if the central
limit law holds for $h$ in an interval around $h_0$ then the
fluctuations of the magnetization at field $h_0$ satisfy a large
deviation law in a small enough interval $J$ around $m(h_0)$,

\0(b) if a large deviation law is satisfied in an interval around $h_0$ then
the central limit law holds for the fluctuations of magnetization
around its average in all fields $h$ with $h-h_0$ small enough.}

\vskip1mm
Going beyond the heuristic level in establishing the duality amounts to
giving a precise meaning to  ``small enough'' and to discuss
which properties of $m(h)$ and $D(h)$ or of $ F(p)$ are needed to
derive properties (a),(b).

For illustration purposes consider the Ising model with ferromagnetic
interaction $\f$: then the central limit law holds for all $h$ if
$\b$ is small enough and, under the same condition on $\b$, the large
deviations law holds for all $h$ and all intervals $[a,b]\subset
(-1,1)$. If $\b$ is not small then the condition $h\ne0$ has to be
added. Hence the conditions are fairly weak and the apparent
exceptions concern the value $h=0$ and $\b$ not small where the
statements may become invalid because of possible phase transitions.

In presence of phase transitions the law of large numbers, the central
limit law and law of large deviations should be reformulated.
Basically one has to add the requirement that fluctuations are
considered in {\it pure phases} and change in a natural way the
formulation of the laws. For instance the large fluctuations of
magnetization in a pure phase of the Ising model in zero field and
large $\b$ (\ie in a state obtained as limit of finite volume states
with $+$ or $-$ boundary conditions) in intervals $[a,b]$ which {\it
do not contain the average magnetization $m^*$ are not necessarily
exponentially small with the size of $|\L|$: if
$[a,b]\subset[-m^*,m^*]$ they are exponentially small but only with
the size of the surface of $\L$} (\ie with $|\L|^{\fra{d-1}{d}}$)
while they are exponentially small with the volume if
$[a,b]\cap[-m^*,m^*]=\emptyset$.

The discussion of Sect.\sec(23) shows that also at the critical point
the nature of the large fluctuations is expected to change: no central
limit law is expected to hold in general because of the example of
\equ(23.2) with the divergence of the average of the normal
second moment of the magnetization in a box as the side tends to
$\io$.

\*
\0{\it Bibliography:} [Ol87]
\*

\Section(25, Renormalization group)
\*

The theory of fluctuations just discussed concerns only fluctuations
of a single quantity. The problem of {\it joint fluctuations} of
several quantities is also interesting and led to really new
developments in the 1970's. It is necessary to restrict attention to
rather special cases in order to illustrate some ideas and the
philosophy behind the approach. Consider therefore the equilibrium
distribution $\m_0$ associated with one of the classical equilibrium
ensembles. For instance the equilibrium distribution of an Ising
energy function $\b H_0$, having included the temperature factor in
the energy: the inclusion is done because the discussion will deal
with the properties of $\m_0$ as a function of $\b$. It will also be
assumed that the average of each spin is zero (``no magnetic field'',
see \equ(15.3) with $h=0$). Wanting to keep in mind a concrete case
imagine that $\b H_0$ is the energy function of the nearest neighbor
Ising ferromagnet in zero field.

Imagine that the volume $\O$ of the container has periodic boundary
conditions and is very large, ideally infinite. Define the family of
{\it blocks} $k\Bx$, parameterized by $\Bx\in\ZZZ^d$ and $k$ integer,
consisting of the lattice sites $\V x=\{k\x_i\le x_i< (k+1)
\x_i\}$. This is a lattice of cubic blocks with side size $k$ that
will be called the ``$k$-rescaled lattice''. The quantities
$m_{\Bx}=k^{-\a d}\sum_{\V x\in k \Bx} \s_{\V x}$ are called the {\it
block spins}.

Given $\a$ define the map $R^*_{\a,k}\m_0=\m_k$ transforming the
initial distribution on the original spins into the distribution of
the block spins. Note that if the initial spins have only two values
$\s_{\V x}=\pm1$ the block spins take values between $-\fra{k^d}{k^{\a
d}}$ and $\fra{k^d}{k^{\a d}}$ at steps of size $\fra2{k^{\a
d}}$. Furthermore the map $R^*_{\a,k}$ makes sense independently of how
many values the initial spins can assume, and even if they assume a
continuum of values $S_{\V x}\in\RRR$.

Taking $\a=1$ means, for $k$ large, looking at the probability
distribution of the joint large fluctuations in the bloks $k\Bx$.
Taking $\a=\fra12$ corresponds to studying a joint central limit
property for the block variables.

Considering a one parameter family of initial distributions $\m_0$
parameterized by a parameter $\b$ (that will be identified with the
inverse temperature), {\it typically} there will be a {\it unique}
value $\a(\b)$ of $\a$ such that the joint fluctuations of the block
variables admit a limiting distribution,

$${\rm prob}_k(m_\Bx\in [a_\Bx,b_\Bx],\,
 \Bs\in\L)\tende{k\to\io}\ig_{\{a_\Bx\}}^{\{b_\Bx\}} g_\L((S_\Bx)_{\Bx\in\L})\,
 \prod_{\Bx\in\L} d S_{\Bx}\Eq(25.1)$$
for some distribution $g_\L(\V z)$ on $\RRR^\L$.

If $\a>\a(\b)$ the limit will then be $\prod_{\Bx\in\L}
\d(S_{\Bx})\,dS_{\Bx}$, or if $\a<\a(\b)$ the limit will not exist
(because the block variables will be too large, with a dispersion
diverging as $k\to\io$).

It is convenient to choose as sequence of $k\to \io$ the sequence
$k=2^n$ with $n=0,1,\ldots$ because in this way it is
$R^*_{\a,k}\=R^{* n}_{\a,1}$ and the limit $k\to\io$ along the
sequence $k=2^n$ can be regarded as a sequence of iterations of a map
$R^*_{\a,1}$ acting on the probability distributions of generic spins
$S_{\V x}$ on the lattice $\ZZZ^d$ (the sequence $3^n$ would
be equally suited).

It is even more convenient to consider probability distributions that
are expressed in terms of energy functions $H$ which generate, in the
thermodynamic limit, a distribution $\m$: then $R^*_{\a,1}$ defines an
action $R_{\a}$ on the energy functions so that $R_\a H=H'$ if $H$
generates $\m$, $H'$ generates $\m'$ and $R^*_{\a,1}\,\m\,=\,\m'$. Of
course the energy function will be more general than \equ(15.3) and a
form like $\d U$ in \equ(19.2) has to be admitted, at least.

In other words $R_\a$ gives the result of the action of $R^*_{\a,1}$
expressed as a map acting on the energy functions. Its iterates define
also a semigroup which is called the {\it block spin renormalization
group}.

{\it While the map $R^*_{\a,1}$ is certainly well defined as a map of
probability distributions into probability distributions, it is by no
means clear that $R_{\a}$ is well defined as a map on the energy
functions.}  Because if $\m$ is given by an energy function it is not
clear that $R^*_{\a,1}\m$ is such.

A remarkable theorem can be (easily) proved when $R^*_{\a,1}$ and its
iterates act on initial $\m_0$'s which are equilibrium states of a
spin system with short range interactions and {\it at high
temperature} ($\b$ small). In this case if $\a=\fra12$ the sequence of
distributions $R^{*n}_{\fra12,1}\m_0(\b)$ admits a limit which is
given by a product of independent Gaussians

$${\rm prob}_k(m_\Bx\in [a_\Bx,b_\Bx],\,
 \Bs\in\L)\tende{k\to\io}\ig_{\{a_\Bx\}}^{\{b_\Bx\}} \prod_{\Bx\in \L}
 e^{-\fra1{2 D(\b)}{S_\Bx^2}} \prod_{\Bx\in\L} \fra{d
 S_{\Bx}}{\sqrt{2\p D(\b)}}\Eq(25.2)$$
Note that this theorem is stated without even mentioning the
renormalization maps $R^n_{\fra12}$: it can nevertheless be interpreted
as saying that

$$R^n_{\fra12} \b H_0\, \tende{n\to\io}\, \sum_{\Bx\in \ZZZ^d} \fra1{2
D(\b)}\,{S_\Bx^2}\Eq(25.3)$$
but the interpretation is not rigorous because \equ(25.2) does not
even say that $R^n_{\fra12} H_0(\b)$ makes sense for $n\ge1$.
The \equ(25.2) says that at high temperature block
spins have {\it normal independent fluctuations: it is therefore an
extension of the central limit law.}

There are a few cases in which the map $R_\a$ can be rigorously shown to be
well defined at least when acting on special equilibrium states like
the high temperature lattice spin systems: but these are exceptional
cases of relatively little interest.

Nevertheless there is a vast literature dealing with approximate
representations of the map $R_\a$. The reason is that assuming not
only its existence but also assuming that it {\it has the properties that
one would normally expect to hold for a map acting on a finite
dimensional space} it follows that a number of consequences can
be drawn, quite non trivial as they led to the first theory of the
critical point that goes beyond the van der Waals theory of
Sect.\sec(11).

The argument proceeds essentially as follows. At the critical point
the fluctuations are expected to be {\it anomalous} (\cfr the last
remark in Sect.\sec(23)) in the sense that $\media{ \big(\fra{\sum_{\V
x\in\L} \s_\Bx}{\sqrt{|\L|}}\big)^2}$ will tend to $\io$, because
$\a=\fra12$ does not correspond to the right fluctuation scale of
$\sum_{\x\in\L}\s_\x$, signaling that $R^{*n}_{\fra12,1}\m(\b_c)$ will
{\it not have a limit} but, possibly, there is $\a_c>\fra12$ such that
$R^{*n}_{\a_c,1}\m_0(\b_c)$ converges to a limit in the sense
\equ(25.1) ($\a_c=\fra78$ in the case of the critical nearest neighbor
Ising ferromagnet, see ending remark in Sect.\sec(23)).

Therefore if the map $R^*_{\a_c,1}$ is considered as acting on
$\m_0(\b)$ it will happen that for all $\b<\b_c$,
$R^{*n}_{\a_c,1}\m_0(\b_c)$ will converge to a trivial limit
$\prod_{\Bx\in\L}
\d(S_{\Bx})\,dS_{\Bx}$ because the value $\a_c$ is greater than
$\fra12$ while normal fluctuations are expected.

If the map $R_{\a_c}$ can be considered as a map on the energy
functions this says that $\prod_{\Bx\in\L}
\d(S_{\Bx})\,dS_{\Bx}$ is a {\it ``(trivial) fixed point of the
renormalization group''} which ``attracts'' the energy functions $\b H_0$
corresponding to the high temperature phases.

The existence of the critical $\b_c$ can be associated with the
existence of a {\it non trivial fixed point} $H^*$ for $R_{\a_c}$
which is hyperbolic with {\it just one} Lyapunov exponent $\l>1$,
hence it has a stable manifold of codimension $1$.  Call $\m^*$ the
probability distribution corresponding to $H^*$.

The migration towards the trivial fixed point for $\b<\b_c$ can be
simply explained by the fact that for such values of $\b$ the initial
energy function $\b H_0$ is {\it outside the stable manifold of the
non trivial fixed point} and under application of the renormalization
transformation $R^n_{\a_c}\,\b H_0$ migrates toward the trivial fxed point,
which is {\it attractive in all directions}.

Increasing $\b$ it may happen that, for $\b=\b_c$, $\b H_0$ {\it
crosses} the stable manifold of the non trivial fixed point $H^*$ for
$R_{\a_c}$.  Then $R^n_{\a_c}\, \b_c H_0$ will no longer tend to the
trivial fixed point but it will tend to $H^*$: this means that the
block spin variables will exhibit a completely different fluctuation
behavior. If $\b$ is close to $\b_c$ the iterations of $R_{\a_c}$ will
bring $R_{\a_c}^n \b H_0$ close to $H^*$, {\it only to be eventually
repelled along the unstable direction} reaching a distance from it
increasing as $\l^n|\b-\b_c|$.

This means that up to a scale length $O(2^{n(\b)})$ lattice units with
$\l^{n(\b)}|\b-\b_c|=1$,
\ie up to a scale $O(|\b-\b_c|^{-\log_2 \l})$
the fluctuations will be close to those of the fixed point
distribution $\m^*$, but beyond that scale they will come close to
those of the trivial fixed point: to see them the block spins would
have to be normalized with index $\a=\fra12$ and they would appear as
uncrorrelated Gaussian fluctuations, \cfr \equ(25.2),\equ(25.3).

The next question is ``how to find the non trivial fixed points?''
this means how to find the energy functions $H^*$ and the
corresponding $\a_c$ which are fixed points of $R_{\a_c}$. If the
above picture is correct the distributions $\m^*$ corresponding to the
$H^*$ would describe the critical fluctuations and if there was only
one choice, or a limited number of choices, of $\a_c$ and $H^*$ this
would open the way to a {\it universality theory} of the critical
point hinted already by the ``primitive'' results of van der Waals'
theory.

The initial hope was, perhaps, that there would be a very small number
of critical values $\a_c$ and of $H^*$ possible: but it rapidly faded
away {\it leaving however the possibility} that the critical
fluctuations could be classified into {\it universality classes}. Each
class would contain many energy functions which upon iterated actions
of $R_{\a_c}$ would evolve under the control of the trivial fixed
point (always existing) for $\b$ small while for $\b=\b_c$ they would
be controlled, instead, by a non trivial fixed point $H^*$ for
$R_{\a_c}$ with the same $\a_c$ and the same $H^*$. For $\b<\b_c$ a
``resolution'' of the approach to the trivial fixed point would be
seen by considering the map $R_{\fra12}$ rather than $R_{\a_c}$ whose
iterates would, however, lead to a Gaussian distribution like
\equ(25.2) (and to a limit energy function like \equ(25.3)).

The picture is highly hypothetical: but it is the first suggestion of
a mechanism leading to critical points with character of universality
and with exponents different from those of the van der Waals' theory
or, for ferromagnets on a lattice, from those of its lattice version
(the {\it Curie-Weiss' theory}): furthermore accepting the
approximations, \eg {\it Wilson--Fisher's $\e$--expansion}, that allow
one to pass from the well defined $R^*_{\a,1}$ to the action of
$R_{\a}$ on the energy functions it is possible to obtain quite
unambiguously values for $\a_c$ and expressions for $H^*$ which are
associated with the action of $R_{\a_c}$ on various classes of models.

For instance it can lead to conclude that the critical behavior of
{\it all} ferromagnetic finite range lattice spin systems (with energy
functions given by \equ(15.3)) have critical points controlled {\it by
the same $\a_c$} and {\it the same} non trivial fixed point: this
property is far from being mathematically proved but it is considered
a major success of the theory. One has to compare it with van der
Waals critical point theory: for the first time an approximation
scheme has led, even though under approximations not fully
controllable, to computable critical exponents which are not equal to
those of the van der Waals theory.

The renormalization group approach to critical phenomena has many
variants, depending on which kind of fluctuations are considered and
on the models to which it is applied. In statistical mechanics there
are a few mathematically complete applications: certain results in
high dimension, theory of dipole gas in $d=2$, hierarchical models,
some problems in condensed matter and in statistical mechanics of
lattice spins and a few others. Its main mathematical successes have
been obtained in various related fields where not only the
philosophy described above can be applied but it leads to
renormalization transformations that can be defined precisely and
studied in detail: \eg constructive field theory, KAM theory of quasi
periodic motions, various problems in dynamical systems.

However the applications always concern special cases and in each of
them the general picture of the {\it trivial-nontrivial} fixed point
dichotomy appears realized but without being accompanied except in
rare cases (like the hierarchical models or the universality theory of
maps of the interval) by the full description of stable manifold,
unstable direction and action of the renormalization transformation on
objects other than the one of immediate interest (a generality which
looks often an intractable problem, but which also turns out
to be not necessary),

In the renormalization group context mathematical physics has played
an important role also by providing clear evidence that universality
classes could not be too few: this was shown by the numerous exact
solutions after Onsager's solution of the nearest neighbor Ising
feromagnet: there are in fact several lattice models in $d=2$ which
exibit critical points with some critical exponents exactly computable
and which depend {\it continuously} on the models parameters.

\*
\0{\it Bibliography:} [MW73], [Ba82], [BS75],
[WF72], [GK83], [GK85], [BG95], [Ma04].
\*

\Section(26, Quantum statistics)
\*

Statistical mechanics is extended to assemblies of quantum particles
rather straightforwardly. In the case of $N$ identical particles the
observables are operators $O$ on the Hilbert space
$\HH_N=L_2(\O)^N_\a$, or $\HH_N=(L_2(\O)\otimes \CC^{2})^N_\a$, where
$\a=+,-$, of the symmetric ($\a=+$, bosonic particles) or
antisymmetric ($\a=-$, fermionic particles) functions $\ps(\V Q)$, $\V
Q=(\V q_1,\ldots, \V q_N)$ of the position coordinates of the
particles or of the positions-spin coordinates $\ps(\V Q,\Bs)$,
$\Bs=(\s_1,\ldots,\s_N)$, normalized so that $\ig |\ps(\V Q)|^2d\V
Q=1$ or $\sum_\Bs\ig |\ps(\V Q,\Bs)|^2d\V
Q=1$, here only $\s_j=\pm1$ is considered. As in classical mechanics a
state is defined by the average values $\media{O}$ that it attributes
to the observables.

Microcanonical, canonical and grand canonical ensembles can be defined
quite easily. For instance consider a system described by the
Hamiltonian ($\hbar=$ Planck's constant)

$$H_N=-\fra{\hbar^2}{2m}\sum_{j=1}^N \D_{\V q_j}+\sum_{j<j'} \f(\V q_j-\V
q_{j'})+ \sum_j w(\V q_j)\defi K+\F\Eq(26.1)$$
where periodic boundary conditions are imagined on $\O$ and $w(\V q)$
is periodic smooth potential (the side of $\O$ is supposed
a multiple of the periodic potential period if $w\ne0$). Then a
{\it canonical equilibrium state} with inverse temperature $\b$ and
specific volume $v=\fra{V}N$ attributes to the observable $O$ the
average value

$$\media{O}\defi\fra{\Tr \, e^{-\b H_N}O}{\Tr \, e^{-\b H_N}}.\Eq(26.2)$$
Similar definitions can be given for the grand canonical equilibrium
states.

Remarkably the ensembles are {\it orthodic} and a ``heat theorem'',
see Sect. \sec(4), can be proved. However ``equipartition'' {\it does
not hold}: \ie $\media{K}\ne\fra{d}2 N \b^{-1}$ although $\b^{-1}$ is
still the integrating factor of $dU+p dV$ in the heat theorem; hence
$\b^{-1}$ keeps being proportional to temperature.

Lack of equipartition is important as it solves paradoxes that arise
in classical statistical mechanics applied to systems with infinitely
many degrees of freedom, like crystals (modeled by lattices of coupled
oscillators) or fields like the electromagnetic field
(black body radiation). However although this has been the first
surprise of quantum statistics (and in fact responsible for the very
discovery of quanta) it is by no means the last.

At low temperatures new unexpected (\ie with no analogue in classical
statistical mechanics) phenomena occurr: Bose-Einstein condensation
(superfluidity), Fermi surface instability (superconductivity) and
appearance of {\it off-diagonal long range order}, with the
acronym {\it ODLRO}, will be selected to illustrate the deeply
different kind of problems of quantum statistical mechanics. Largely
not yet understood, such phenomena pose very interesting problems not
only from the physical point of view but also from the mathematical
point of view and may pose challenges even at a definitory
level. However it should be kept in mind that in the interesting cases
(\ie $3$-dimensional systems and even most $2$ and $1$ dimensional
systems) there is no proof that the objects defined here really exist
for the systems like \equ(26.1), see however the final comment for an
important exception.
\*

\0(1) {\it Bose Einstein condensation:} In a canonical state with
parameters $\b,v$ a definition of the occurrence of Bose condensation
is in terms of the eigenvalues $\n_j(\O,N)$ of the kernel $\r(\V q,\V
q')$ on $L_2(\O)$, called the $1$-particle reduced density matrix,
defined by

$$N\,{\sum_{n=1}^\io \fra{e^{-\b E_n(\O,N)}}{\Tr e^{-\b H_N}}
\ig\lis \ps_n(\V q,\V q_1,\ldots, \V q_{N-1})
\ps_n(\V q',\V q_1,\ldots, \V q_{N-1}) d\V q_1\ldots \V
q_{N-1}}\Eq(26.3)$$
where $E_n(\O,N)$ are the eigenvalues of $H_N$ and $\ps_n(\V
q_1,\ldots,\V q_N)$ are the corresponding eigenfunctions. If $\n_j$
are ordered by increasing value the state with parameters $\b,v$ is
said to contain a Bose-Einstein condensate if $\n_1(\O,N)\ge b\,N
\,>\,0$ for all large $\O$ at $v=\fra{V}{N},\b$ fixed.  This receives
the interpretation that there are more than $bN $ particles with equal
momentum. The {\it free Bose gas} has a Bose condensation phenomenon at
fixed density and small temperature.

\vskip1mm

\0(2) {\it Fermi surface}: the wave functions $\ps_n(\V
q_1,\s_1,\ldots,\V q_N,\s_N)\=\ps_n(\V Q,\Bs)$ are now antisymmetric
in the permutations of the pairs $(\V q_i,\s_i)$. Let $\ps({\V
Q,\Bs;N,n})$ denote the $n$-th eigenfunction of the
$N$ particles energy $H_N$ in \equ(26.1) with eigenvalue $E(N,n)$
(labeled by $n=0,1,\ldots$ and nondecreasingly ordered). Setting $\V
Q''=(\V q_1'',\ldots,\V q_{N-p}'')$, $\Bs''=(\s''_1,\ldots,\s''_{N-p})$,
introduce the kernels $\r_p^{H_N}(\V Q,\Bs;\V Q',\Bs')$ by

$$\eqalign{
\r_p(\V Q,\Bs;\V Q',\Bs')\defi& p!{{N}\choose{p}}\ig \sum_{\Bs''} \,
d^{N-p} \V Q''\,\sum_{n=0}^\io
\fra{e^{-\b E(N,n)}}{\Tr e^{-\b H_N}}\cdot\cr
&\cdot
\lis \ps({\V Q,\Bs; \V Q'',\Bs'';N,n})
\ps({\V Q',\Bs'; \V Q'',\Bs'';N,n})
\cr}\Eq(26.4)
$$
which are called $p$-particles reduced density matrices
(extending the corresponding one particle reduced density matrix
\equ(26.3)). Denote $\r(\V q_1-\V q_2)\defi
\sum_\s\r_1(\V q_1,\s,\V q_2,\s)$.  It is also useful to consider
spinless fermionic systems: the corresponding definitions are obtained
simply by suppressing the spin labels and will not be repeated.

Let $r_1(\V k)$ be the Fourier transform of $\r_1(\V q-\V q')$: the
Fermi surface can be {\it defined} as the locus of the $\V k$'s in the
neighborhood of which $\dpr_{\V k} r_1(\V k)$ is {\it unbounded} as
$\O\to\io,\b\to\io$. The limit as $\b\to\io$ is important because the
notion of a Fermi surface is, possibly, precise only at zero
temperature, \ie $\b=\io$.

So far existence of Fermi surface (\ie the smoothness of $r_{1}(\V
k)$ except on a smooth surface in $\V k$ space) has been proved in
free Fermi systems ($\f=0$) and
\\
(a) certain exactly soluble $1$--dimensional {\it spinless} systems
\\
(b) in rather general one dimensional {\it spinless} systems or
systems with spin and repulsive pair interaction, possibly in an {\it
external periodic potential}.

The spinning case in a periodic potential and dimension $d\ge2$ is the
most interesting case to study for its relevance in the theory of
conduction in crystals. Essentially no mathematical results are
available as the above mentioned ones do not concern any case in
dimension $>1$: this is a rather deceiving aspect of the theory and a
challenge.

In dimension $2$ or higher, for fermionic systems with Hamiltonian
\equ(26.1), not only there are no results available, {\it even without
spin}, but it not even clear that a Fermi surface can exist in
presence of interesting interactions.
\vskip1mm

\0(3) {\it Cooper pairs}: superconductivity theory has been
phenomenologically related to the existence of {\it Cooper pairs}. Consider
the Hamiltonian \equ(26.1) and define, \cfr \equ(26.4), $\r(\V x-\V y,\s; \V
x'-\V y',\s'; \V x -\V x')\defi \r_2(\V x,\s,\V y,-\s; \V
x',\s',\V y',-\s')$.

The system is said to contain Cooper pairs with spins $\s,-\s$ ($\s=+$
or $\s=-$) if there exist functions $g^\a(\V q,\s)\ne0$ with 
$\ig \lis {g^{\a}}(\V
q,\s){g^{\a'}}(\V q,\s)d\V q=0$ if $\a\ne\a'$, such that

$$\lim_{V\to\io} \r(\V x-\V y,\s, \V x'-\V y',\s',
\V x -\V x')\tende{\V x -\V x'\to\io} 
\sum_\a g^\a(\V x -\V y,\s)\lis{g^\a}(\V x'-\V y',\s')\Eq(26.5)$$
In this case $g^\a(\V x-\V y,\s)$ with largest norm can be called the
{\it wave function} of the paired state of {\it lowest energy}: this
is the analogue of the plane wave for a free particle (and, like it,
it is manifestly not normalizable, \ie it is not square integrable as a
function of $\V x,\V y$). If the system contains Cooper pairs and the
non leading terms in the limit \equ(26.5) vanish quickly enough the
$2$-particles reduced density matrix, \equ(26.5), regarded as a kernel
operator has an eigenvalue of order $V$ as $V\to\io$: \ie the state of
lowest energy is macroscopically occupied, quite like the free Bose
condensation on the ground state.

Cooper pairs instability might destroy the Fermi surface in the sense
that $r_1(\V k)$ becomes analytic in $\V k$; but it is also possible
that even in presence of the instability a surface remains which is
the locus of the singularities of the function $r_1(\V k)$. In the
first case there should remain a trace of it as a very steep gradient
of $r_1(\V k)$ of the order of an exponential in the inverse of the
coupling strength; this is what happens in the BCS model for
superconductivity: the model is however a mean field model and this
particular regularity aspect might be one of its peculiarities. In any
event a smooth singularity surface is very likely to exist for some
interesting density matrix (\eg in the BCS model with ``gap parameter
$\g$'' the wave function $g(\V x-\V y,\s)\=
\fra1{(2\p)^d} \ig_{\e(\V k)>0} e^{i\V k\cdot(\V x-\V
y)}\fra{\g}{\sqrt{\e(\V k)^2+\g^2}}\,d\V k$ of the lowest energy level
of the Cooper pairs is singular on a surface coinciding with the Fermi
surface of the free system).
\*

\0(4) {\it ODLRO:} Consider the $k$--fermions reduced density matrix
$\r_k(\V Q,\Bs;\V Q',\Bs')$ as kernel operators $O_k$ on $L_2(\O\times
\CC^{2})^k_-$.  Suppose $k$ even, then if $O_k$ has a (generalized)
eigenvalue of order $N^{\fra{k}2}$ as $N\to\io$, $\fra{N}V=\r$, the
system is said to exhibit {\it off diagonal long range order} of order
$k$. For $k$ odd ODLRO is defined to exist if $O_k$ has an eigenvalue
of order $N^{\fra12(k-1)}$ and $k\ge3$ (if $k=1$ the largest
eigenvalue of $O_1$ is necessarily $\le1$).

For bosons consider the reduced density matrix $\r_k(\V Q;\V Q')$
regarding it as a kernel operator $O_k$ on $L_2(\O)^k_+$ and define
ODLRO of order $k$ to be present if $O(k)$ has a (generalized)
eigenvalue of order $N^{{k}}$ as $N\to\io$, $\fra{N}V=\r$.

The notion of ODLRO can be regarded as a unification of the phenomena
of Bose condensation and of existence of Cooper pairs because Bose
condensation could be said to correspond to the kernel operator
$\r_1(\V q_1-\V q_2)$ in \equ(26.3) having a (generalized) eigenvalue
of order $N$, and to be a case of ODLRO of order $1$. If the state is
pure in the sense that it has a cluster property, see
Sect.\sec(9),\sec(15), then ODLRO, Bose condensation and Cooper pairs
existence imply that the system shows a {\it spontaneously broken
symmetry}: conservation of particle number and clustering imply that
the off-diagonal elements of (all) reduced density matrices vanish at
infinite separation in states obtained as limits of states with
periodic boundary conditions and Hamiltonian \equ(26.1), and this is
incompatible with ODLRO.

The free Fermi gas has no ODLRO, the BCS model of superconductivity
has Cooper pairs and ODLRO with $k=2$, but {\it no Fermi surface} in
the above sense (possibly too strict). Fermionic systems cannot have
ODLRO of order $1$ (because the reduced density matrix of order £$1$
is bounded by $1$).
\*

The contribution of Mathematical Physics has been particularly
effective by providing exactly soluble models: however the soluble
models deal with $1$--dimensional systems and it can be shown that in
dimension $1,2$ no ODLRO can take place. A major advance is the recent
proof of ODLRO and Bose condensation in the case of a {\it lattice}
version of \equ(26.1) at a special density value (and $d\ge3$).

In no case, for the Hamiltonian \equ(26.1) with $\f\ne0$, existence of
Cooper pairs has been proved nor existence of a Fermi surface for
$d>1$. Nevertheless both Bose condensation and Cooper pairs formation
can be proved to occurr rigorously in certain limiting
situations. There are also a variety of phenomena (\eg simple spectral
properties of the Hamiltonians) which are believed to occurr once some
of the above mentioned four do occurr and several of them can be
proved to exist in concrete models.

If $d=1,2$ the ODLRO can be proved to be impossible at $T>0$ through
the use of Bogoliubov's inequality (used in the ``no $d=2$ crystal
theorem'', see Sect.\sec(13)).
\*

\0{\it Bibliography:} [PO56],[Ya62], [Ru69], [Ho67], [Ga99], [ALSSY].
\*

\appendix(A1, The Physical meaning of the stability conditions)
\*

It is therefore useful to see what would happen if the conditions of
stability and temperedness (see \equ(7.2)) are violated.  The analysis
also illustrates some of the typical methods of statistical mechanics.
\*

\0{\it(a) Coalescence catastrophe due to short distance
attraction:} the simplest violation of the first condition in
\equ(7.2) occurs when the potential $\f$ is
smooth and negative at the origin.

Let $\d>0$ be so small that the potential at distances $\le2 \d$ is
$\le -b<0$. Consider the canonical distribution with parameters $\b,N$
in a (cubic) box $\O$ of volume $V$. The probability $P_{collapse}$
that {\it all} the $N$ particles are located in a little sphere of
radius $\d$ around the center of the box (or around any pre-fixed
point of the box) is estimated from below by remarking that $\F\le
-b{N\choose2}\sim -\fra{b}2 N^2$ so that

$$P_{collapse}=\fra{\ig_\CC \fra{d{\V p} d{\V q}}{h^{3N}N!}e^{-\b
(K({\V p})+\F({\V q}))}}
{\ig \fra{d{\V p} d{\V q}}{h^{3N}N!}e^{-\b (K({\V p})+\F({\V q}))}}
\ge
\fra{(\fra{4\p\sqrt{2m\b^{-1}}^3}{3h^3})^N
\fra{\d^{3N}}{N!} e^{\b b\fra12N(N-1)}}{ \ig\fra{
d{\V q}}{h^{3N}N!} e^{-\b \F({\V q})}}\Eqa(A1.1)$$
Phase space is extremely small: nevertheless such configurations are
{\it far} more probable than the configurations which ``look
macroscopically correct'', \ie configurations with particles more or
less spaced by the average particle distance expected in a
macroscopically homogeneous configuration, namely
$(\fra{N}V)^{-1/3}=\r^{-\fra13}$.  Their energy $\F({\V q})$ is of the
order of $u N$ for some $u$, so that their probability will be bounded
above by
$$P_{regular}\le
\fra{\ig \fra{d{\V p} d{\V q}}{h^{3N}N!} e^{-\b (K({\V p})+uN)}}
{\ig \fra{d{\V p} d{\V q}}{h^{3N}N!}e^{-\b (K({\V p})+\F({\V q}))}}
= \fra{ \fra{V^N\sqrt{2m\b^{-1}}^3}{h^{3N}N!}e^{-\b uN}}{
\ig \fra{d{\V q}}{h^{3N}N!}e^{-\b \F({\V q})}}\Eqa(A1.2)$$
{\it However}, no matter how small $\d$ is, the ratio
$\fra{P_{regular}}{P_{collapse}}$ will approach $0$ as
$V\to\io,\,\fra{N}V\to v^{-1}$; extremely fast because $e^{\b bN^2/2}$
eventually dominates over $V^N\sim e^{N\log N}$.

Thus it is far more probable to find the system in a microscopic
volume of size $\d$ rather than in a configuration in which the energy
has some macroscopic value proportional to $N$.  This catastrophe can
be called an {\it ultraviolet catastrophe} (as it is due to the
behavior at very short distances) and it causes the collapse of the
particles into configurations concentrated in regions as small as we
please as $V\to\io$.

\*
\0{\it(b) Coalescence Catastrophe due to Long-Range Attraction.}
It occurs when the potential is too attractive
near $\io$. To simplify suppose that the potential has a {\it hard
core}, \ie it is $+\io$ for $r<r_0$, so that the above discussed
coalescence cannot occur and the system density bounded above by a
certain quantity $\r_{cp}<\io$ ({\it close packing density}).

The catastrophe occurs if $\f({\V q})\sim -g|{\V q}|^{-3+\e}$,
$g,\e>0$, for $|{\V q}|$ large. For instance this is the case of
matter interacting gravitationally; if $k$ is the gravitational
constant, $m$ is the particles mass  then $g =k m^2$ and $\e=2$.

The probability $P_{regular}$ of ``regular configurations'', where
particles are at distances of order $\r^{-1/3}$ from their close
neighbors, is compared with $P_{collapse}$ of ``catastrophic
configurations'', with the particles at distances $r_0$ from their
close neighbors to form a configuration of density $\r_{cp}/(1+\d)^3$
almost in close packing (so that $r_0$ is equal to the hard core
radius times $1+\d$). In the latter case the system {\it does not}
fill the available volume and leaves empty a region whose volume is a
fraction $\sim \fra{\r_{cp}-\r}{\r_{cp}}V$ of $V$.  And it can be
checked thatthe ratio $P_{regular}/P_{collapse}$ tends to $0$ at a rate
$O(e^{g\fra12N (\r_{cp} (1+\d)^{-3}-\r)})$ if $\d$ is small enough
(and $\r<\r_{cp}$).

{\it A system which is too attractive at infinity will not occupy the
available volume but will stay confined in a close packed
configuration even in empty space.}

This is important in the theory of stars: stars cannot be expected to
obey ``regular thermodynamics'' and in particular will not ``evaporate''
because their particles interact via the gravitational force at large
distance. Stars do not occupy the whole volume given to them (\ie the
universe); they do not collapse to a point only because the
interaction has a strongly repulsive core (even when they are burnt out
and the radiation pressure is no longer able to keep them at a
reasonable size, a reasonable size). \*

\0{\it(c) Evaporation Catastrophe:} this is a another {\it
infrared} catastrophe, \ie a catastrophe due to the long-range
structure of the interactions like (b) above; it occurs when the
potential is {\it too repulsive} at $\io$: \ie $\f({\V q})\sim +g
|{\V q}|^{-3+\e}$ as ${\V q}\to\io$ so that the temperedness condition is
again violated.

Also in this case the system does not occupy the whole volume: it will
generate a layer of particles {\it sticking}, in close packed
configuration, to the walls of the container. Therefore if the density
is lower than the close packing density, $\r<\r_{cp}$, the system will
leave a region {\it around the center of the container $\O$} empty;
and the volume of the empty region will still be of the order of the
total volume of the box (\ie its diameter will be a fraction of the
box side $L$).  The proof is completely analogous to the one of the
previous case; except that now the configuration with lowest energy
will be the one sticking to the wall and close packed there, rather
than the one close packed at the center.

Also this catastrophe is important as it is realized in systems
of charged particles bearing the {\it same} charge: the charges adhere
to the boundary in close packing configuration, and dispose themselves
so that the electrostatic potential energy is minimal. Therefore
charges deposited on a metal will {\it not occupy} the whole volume:
they will rather form a surface layer minimizing the potential energy
(\ie so that the Coulomb potential in the interior is constant).  In
general charges in excess of neutrality do not behave
thermodynamically: for instance, besides not occupying the whole
volume given to them, they will not contribute normally to the
specific heat.

Neutral systems of charges behave thermodynamically if they have
hard cores so that the ultraviolet cathstrophe cannot occur or if they
obey quantum mechanical laws and consist of fermionic particles (plus
possibly bosonic particles with charges of only one sign).
\*

\0{\it Bibliography:} [DL67], [LL72], [Li01]
\*

\appendix(A2, The subadditivity method)

A simple consequence of the assumptions is that the
exponential in \equ(5.2) can be bounded above by $e^{\b B N}
e^{-\fra\b{2m}\sum_{i=1}^N \V p_i^2}$ so that

$$\eqalign{
&1\le Z_{gc}(\b,\l,V)\le e^{ V e^{\b \l} e^{\b B} \sqrt{2 m
\b^{-1}}^d}\ \ \tto\cr
& 0\le \fra1V \log Z_{gc}(\b,\l,V)\le e^{\b \l} e^{\b B} \sqrt{2 m
\b^{-1}}^d}
\Eqa(A2.1)$$

Consider for simplicity the case of a hard core interaction with {\it
finite range}, \cfr \equ(7.2). Consider a sequence of boxes $\O_n$
with sides $2^n L_0$ where $L_0>0$ is arbitrarily fixed $>2R$.  The
{\it partition function} $Z_{gc}(\b,z)$ relative to the volume $\O_n$
is $Z_n=\sum_{N=0}^\io
\fra{z^ N}{N!}\ig_{\O_n} {d\V P d\V Q} e^{-\b \F(\V Q)}$ because the
integral over the $\V P$ variables can be explicitly performed and
included in $z^N$ if $z$ is defined as
$z=e^{\b\l}\big({2m\b^{-1}}\big)^{d/2}$.

Then the box $\O_n$ contains $2^d$ boxes $\O_{n-1}$ for $m\ge1$ and

$$1\le Z_n\le Z_{n-1}^{2^d} e^{\b B 2 d (L_{n-1}/R)^{d-1} 2^{2d} }
\Eqa(A2.2)$$
because the corridor of width $2R$ around the boundaries of the $2^d$
cubes $\O_{n-1}$ filling $\O_n$ has volume $2R L_{n-1}2^d$ and
contains at most $(L_{n-1}/R)^{d-1} 2^d$ particles each of which
interacts with at most $2^d$ other particles. Therefore $\b p_n\defi
L_n^{d}\log Z_n\le L_{n-1}^{d}\log Z_{n-1}+\b B \g_d 2^{-n}
(L_0/R)^{d-1}$ for some $\g_d>0$. Hence $0\le\b p_n\le \b p_{n-1} +
\G_d 2^{-n}$ for some $\G_d>0$ and $p_n$ is bounded above and below
uniformly in $n$: so that the limit \equ(7.1) exists on the sequence
$L_n=L_0 2^n$ and defines a function $\b p_\io(\b,\l)$.

A box of arbitrary size $L$ can be filled with about $(L/L_{\lis
n})^d$ boxes of side $L_{\lis n}$ with $\lis n$ so large that,
prefixed $\d>0$, $|p_{\io}-p_{n}|<\d$ for all $n\ge \lis n$. Likewise
a box of size $L_n$ can be filled by about $(L_n/L)^d$ boxes of size
$L$ if $n$ is large. The latter remarks lead to conclude, by standard
inequalities, that the limit in \equ(7.1) exists and coincides with
$p_\io$.

The {\it subadditivity} method just demonstrated for finite range
potentials with hard core can be extended to the potentials satisfying
just stability and temperedness, \cfr Sect.\sec(7).
\*

\0{\it Bibliography:} [Ru69], [Ga99].
\*

\appendix(A3, An infrared inequality)
\*

The infrared inequalities stem from { \it Bogoliubov's
inequality}. Consider as an example the problem of crystallization of
Sect.\sec(13).  Let $\media{\cdot}$ denote average over a canonical
equilibrium state with Hamiltonian $H=\sum^N_{j=1} \fra{\V p_j^2}2+U(\V
Q)+\e W(\V Q)$ with given temperature and density parameters $\b,\r$,
$\r=a^{-3}$. Let
$\{X,Y\}=\sum_j(\dpr_{p_j}X\,\dpr_{q_j}Y-\dpr_{q_j}X\,
 \dpr_{p_j} Y)$ be the Poisson bracket.  Integration by parts, {\it
with periodic boundary conditions}, yields

$$\media{A^*\{C,H\}}\=-
\fra{\ig \,A^*\{C,e^{-\b H}\}d\V P d\V Q}{\b Z_c(\b,\r,N)}\=-
\b^{-1}\media{\{A^*,C\}}
 \Eqa(A3.1)$$
as a general identity.  The latter identity implies, for $A=\{C,H\}$,

$$
\media{\{H,C\}^*\{H,C\}}=-\b^{-1} \media{\{C,\{H,C^*\}\}}\Eqa(A3.2)$$
Hence the Schwartz inequality $\media{A^*A}\media{\{H,C\}^*\{H,C\}}\ge
|\media{\{A^*,C\}}|^2$ combined with the two relations in \equ(A3.1),\equ(A3.2)
yields {\it Bogoliubov's inequality}

$$\media{A^*A}\ge \b^{-1}
  \fra{|\media{\{A^*,C\}}|^2}{
        \media{\{C,\{C^*,H\}\}}}\Eqa(A3.3)$$
Let $g,h$ be arbitrary complex (differentiable) functions and
$\Dpr_j=\Dpr_{\V q_j}$

$$A(\V Q)\defi\sum_{j=1}^N g(\V q_j),\qquad \V C(\V P,\V Q)\defi
\sum_{j=1}^N \V p_j h(\V q_j)\Eqa(A3.4)$$
Then $H=\sum\fra12 \V p_j^2+ \F(\V q_1,\ldots,\V q_N)$, if $\F(\V
q_1,\ldots,\V q_N)=\fra12\sum_{j\ne j'} \f(|\V q_j-\V q_{j'}|)+\e\sum_j W(\V
q_j)$, so that $\{\V C,H\}\=\sum_j (h_j \Dpr_j \F- \V p_j (\V p_j\cdot\Dpr_j)
h_j)$ with $h_j\defi h(\V q_j)$. If $h$ is real valued
$\media{\{C,\{C^*,H\}\}}$ becomes $
\media{\sum_{jj'} h_jh_{j'}\Dpr_j\cdot\Dpr_{j'}
\F(\V Q)}$ $+\media{\e\sum_j h_j^2 \D W(\V q_j) +\fra4{\b}\sum_j(\Dpr_j
h_j)^2}$\\
(integrals on $\V p_j$ just
replace $\V p_j^2$ by $2\b^{-1}$ and $\media{(\V p_j)_i(\V p_j)_{i'}}=
\b^{-1}\d_{i,i'}$). Therefore the average $\media{\{C,\{C^*,H\}\}}$ becomes

$$\media{\fra12\sum_{jj'} (h_j-h_{j'})^2\D \f(|\V q_j-\V
q_{j'}|)+\e\sum_j h_j^2 \D W(\V q_j)+4
\b^{-1}\sum_j(\Dpr_j h_j)^2}\Eqa(A3.5)$$
Choose $g(\V q)\= e^{-i(\Bk+ \V K)\cdot\V q}$, $h(\V q)=\cos \V
q\cdot\Bk$ and bound $(h_j-h_{j'})^2$ by $\Bk^2\,(\V q_j-\V q_{j'})^2$,
$(\Dpr_j h_j)^2$ by $\Bk^2$ and $h_j^2$ by $1$. Hence \equ(A3.5) is
bounded above by $N D(\Bk)$ with

$$ D(\Bk)\defi\, \media{\,\Bk^2\,
\big(4\b^{-1}+\fra1{2N}\sum_{j\ne j'}(\V q_j-\V q_{j'})^2|\D\f(\V q_j-\V
q_{j'})|\big)\,+\, \e \fra1N\sum_j |\D W(\V q_j)|\,}\Eqa(A3.6)$$
This can be used to estimate the denominator in 
\equ(A3.3). For the \lhs remark that $\media{A^*,A}=
|\sum_{j=1}^N e^{-i\V q\cdot (\Bk+\V K)}|^2$ and $|\media{\{A^*,\V
C\}}|^2=|\media{\sum_j h_j \Dpr g_j} |^2=|\V K+\Bk|^2 N^2 (\r_\e(\V
K)+\r_\e(\V K+2\Bk))^2$, hence \equ(A3.3) becomes after multiplying
both sides by the auxiliary function $\g(\Bk)$ (assumed even and
vanishing for $|\Bk|> \fra\p{a}$) and summing over $\Bk$,

$$D_1\defi\fra1N\sum_\Bk\g(\Bk)\media{\fra1N|\sum_{j=1}^N e^{-i(\V K+\Bk)\cdot\V
q_j}|^2}\,\ge\, \fra1N\sum_\Bk\g(\Bk)
\fra{|\V K|^2}{4\b}\fra{(\r_\e(\V K)+\r_\e(\V
K+2\Bk))^2}{D(\Bk)}\Eqa(A3.7)$$
To apply \equ(A3.3) the averages in \equ(A3.6),\equ(A3.7) have to be bounded
above: this is a technical point that is discussed here as it
illustrates a general method of using the results on the thermodynamic
limits and their convexity properties to obtain estimates. 

Note that
$\media{ \fra1N \sum_{\V k}\g(\V k) d^d\V k |\sum_{j=1}^N e^{-i\V
k\cdot\V q_j}|^2}$ is identically $\widetilde \f(0)+$ $\fra2N
\media{\sum_{j<j'}\widetilde \f(\V q_j-\V q_{j'})}$ with
$\widetilde \f(\V q)\defi\fra1N\sum_\Bk \g(\Bk) e^{i \Bk\cdot\V q}$.

Let $\f_{\l,\z}(\V q)\defi\f(\V q)+\l\,
\V q^2|\D\f(\V q)|+\h \widetilde\f(\V q)$ and let $F_{V}(\l,\h,\z)
\defi\fra1N\log Z^c(\l,\h,\z)$
with $Z^c$ the partition function in the volume $\O$ computed with
energy $U'=\sum_{jj'}\f_{\l,\z}(\V q_j-\V q_{j'})+\e\sum_j W(\V q_j)
+\h\,\e\,\sum |\D W(\V q_j)|$. Then $F_{V}(\l,\h,\z)$ is convex in $\l,\h$
and {\it it is uniformly bounded above and below} if $|\h|,|\e|,|\z|\le1$
(say) and $|\l|\le\l_0$: here $\l_0>0$ exists if $\V r^2|\D \f(\V r)|$
satisfies the assumption set at the beginning of Sect.\sec(13) and the
density is smaller than a close packing (this is because the
potential $U'$ will still satisfy conditions similar to \equ(7.2)
uniformly in $|\e|,|\h|<1$ and $|\l|$ small enough).

Convexity and boundedness above and below in an interval imply bounds
on the derivatives in the interior points: in this case on the
derivatives of $F_V$ with respect to $\l,\h,\z$ at $0$. The latter are
identical to the averages in
\equ(A3.6),\equ(A3.7). In this way the constants $B_1,B_2,B_0$ such that
$D(\Bk)\le \Bk^2 B_1+\e B_2$ and $B_0>D_1$ are found.
\*

\0{\it Bibliography:} [Me68].

\*
\0{\bf Bibliography}
\*

{\nota
\0[Ai80] Aizenman, M.: {\it Translation invariance and instability of phase
coexistence in the two dimensional Ising system}, Communications in
Mathematical Physics, {\bf73}, 83--94, 1980.

\0[Ai82] Aizenman, M.: {\it Geometric analysis of $\f^{4}$ fields and
Ising models}, {\bf86}, 1-48, 1982.

\0[ALSSY] Aizenman, M., Lieb, E.H., Seiringer, R., Solovej, J.P.,
Yngvason, J.: {\it Bose--Einstein condensationas a quantum phase
transition in a optical lattice}, cond-math 0412034.

\0[Ba82] Baxter, R.: {\sl Exactly solved models}, Academic Press,
London, 1982.

\0[Be75] van Beyeren, H.: {\it Interphase sharpness in the Ising model},
Communications in Mathematical Physics, {\bf 40}, 1--6, 1975.

\0[BG95] Benfatto, G., Gallavotti, G.: {\it Renormalization group},
Princeton University Press, 1995.

\0[BS75] Bleher, P., and Sinai, Y.:
  {\it Critical indices for Dyson's asymptotically hierarchical models},
  Communications in Mathematical Physics {\bf 45}, 247--278, 1975.

\0[Bo66] Boltzmann, L.: {\it\"Uber die mechanische Bedeutung des
zweiten Haupsatzes der W\"armetheorie}, in
"Wissenschaftliche Abhandlungen", ed. F. Hasen\"ohrl,
vol. I, p. 9--33, Chelsea, New York, 1968.

\0[Bo84] Boltzmann, L.: {\it \"Uber die Eigenshaften monzyklischer
und anderer damit verwandter Systeme}, in "Wissenshafltliche
Abhandlungen", ed. F.P. Hasen\"ohrl, vol. III, p. 122--152,
Chelsea, New York, 1968.

\0[Do68] R.L. Dobrushin, R.L.: {\it Gibbsian random fields for lattice
systems with pairwise interactions}, Functional Analysis and
Applications, {\bf2}, 31--43, 1968;

\0[DG72] Domb, C, Green, M.S.: {Phase transitions and critical
points}, Wiley, New York, 1972.

\0[Dy69] Dyson, F.: {\it Existence of a phase transition in a
one--dimensional Ising ferromagnet}, Communications in Mathematical
Physics, {\bf12}, 91--107, 1969.

\0[FP04] Friedli, S., Pfister, C.: {\it On the Singularity of the Free Energy
at a First Order Phase Transition}, Communications in Mathematical
Physics {\bf 245}, 69--103, 2004

\0[Ga99] Gallavotti, G.: {\it Statitical Mechanics}, Springer Verlag,
1999.

\0[GBG04] Gallavotti, G., Bonetto, F., Gentile, G.:
{\it Aspects of the ergodic, qualitative and statistical properties
of motion}, Springer--Verlag, Berlin, 2004.

\0[GK83]
  Gawedzky, K., and Kupiainen, A. {\it Block spin renormalization gr\-oup for
  dipole gas and $(\dpr\phi)^4$}, Annals of Physics {\bf 147}, 198--243,
  1983.

\0[GK85] Gawedzky, K., and Kupiainen, A.
  {\it Massless lattice $\phi^4_4$ theory: Rigorous control of a
  renormalizable asymptotically free model}, Communications in Mathematical
  Physics {\bf 99}, 197--252, 1985.

\0[Gi81] Gibbs, J.W.: {\it Elementary principles in statistical
mechanics}, Ox Bow Press, Woodbridge (Connecticut), 1981 (reprint of the
1902 edition).

\0[Hi81] Higuchi, Y.: {\it On the absence of non translationally invariant
Gibbs states for the two dimensional Ising system}, in ``Random
fields'', editors J. Fritz, J.L. Lebowitz and D. Szaz, North Holland,
Amsterdam, 1981.

\0[Ho67] Hohenberg, P.C.: {\it Existence of long rabge order in one and
two dimensions}, Physical Review, {\bf 158}, 383--386, 1967.

\0[LL67] Landau, L., Lifschitz, L.E.: {\it Physique Statistique}, MIR,
Moscow, 1967.

\0[Le74] Lebowitz, J.L.: {\it GHS and other inequalities}, Communications
in Mathematical Physics, {\bf28}, 313--321, 1974.

\0[LL72] Lieb, E.H., Lebowitz, J.L.: {\it Lectures on the therodynamic
limit for Coulomb systems}, in {\sl Springer lecture notes in
Physics}, edited by A. Lenard, vol. {\bf20}, p. 135--161, Berlin,
1972.

\0[Li01] Lieb, E.H., Thirring, W.E.: {\it Stability of Matter
from Atoms to Stars}, Springer Verlag, 2001

\0[Li02] Lieb, E.H.: {\it Inequalities}, Springer-Verlag, 2002.

\0[LR69] Lanford, O., Ruelle, D.:  {\it Observables at infinity and
states with short range correlations in statistical mechanics},
Communications in Mathematical Physics, {\bf13}, 194--215, 1969.

\0[LP79] Lebowitz, J.L., Penrose, O.: {\it Towards a rigorous
molecular theory of metastability}, in {\sl Fluctuation phenomena},
ed. E.W. Montroll, J.L. Lebowitz, North Holland, Amsterdam, 1979.

\0[LY52] Lee T.D., Yang C.N.: {\it Statistical theory of equations
of state and phase transitions, II. Lattice gas and Ising model},
Physical Review, {\bf87}, 410--419, 1952.

\0[MW73] McCoy, B.M., Wu, T.T.: {\sl The two dimensional Ising model}, Harvard
University Press, Cambridge, 1973.

\0[Ma04] Mastropietro, V.: {\it Ising models with four spin interaction at
criticality}, Communications in Mathematical Physics, {\bf 244},
595--642, 2004.

\0[Me68] Mermin, N.D.: {\it Crystalline order in two dimensions},
Physical Review, {\bf 176}, 250--254, 1968.

\0[Mi95] Miracle--Sol\'e: {\it Surface tension,
step free energy and facets in the equilibrium crystal shape}, Journal
Statistical Physics, {\bf 79}, 183--214, 1995.

\0[Ol87] Olla, S. {\it Large Deviations for Gibbs Random Fields},
Probability Theory and Related Fields, {\bf 77}, 343--357, 1987.

\0[On44] Onsager L.: {\it Crystal statistics. I. A two dimensional Ising
model with an order--disorder transition}, Physical Review, {\bf65},
117--149, 1944.

\0[PO56] Penrose, O., Onsager, L: {\it Bose-Einstein condensation and
liquid helium}, Physical Review, {\bf 104}, 576-584, 1956.

\0[PV99] Pfister, C., Velenik, Y.: {\it	Interface, Surface Tension and
Reentrant Pinning Transition in the 2D Ising Model},
Communications in Mathematical Physics, {\bf 204}, 269 - 312, 1999.

\0[Ru69] Ruelle, D.: {\it Statistical Mechanics}, Benjamin, New York, 1969.

\0[Ru71] Ruelle, D.: {\it Extension of the Lee--Yang circle theorem},
Physical Review  Letters, {\bf26}, 303--304, 1971.

\0[Si91] Sinai Ya.G.: {\it Mathematical problems of statistical
mechanics}, World Scientific, 1991.

\0{[Ya62]} Yang, C.N.: {\it Concept of off-diagonal long-range order
and the quantum phases of liquid He and of superconductors}, Reviews
of Modern Physics, {\bf34}, 694--704, 1962.

\0[WF72] Wilson, K.G., Fisher, M.E.: {\it Critical exponents in $3.99$
dimensions}, Physical Review Letters, {\bf28}, 240--243, 1972.
}\end

\0[Ru79b] Russo, L.: {\it The infinite cluster
method in the two dimensional Ising model}, Communications in
Mathematical Physics, {\bf67}, 251--266, 1979.

\0[DKS92] Dobrushin, R.L., Kotecky, R., Shlosman, S.: {\it The Wulff
construction}, Tnaslations of Mathematical Monograohs, {\bf104},
American Mathematical Society, Providence, 1992. See also: Dobrushin,
R.L., Kotecky, R., Shlosman, S.: {\it A microscopi justification of the
Wulff construction}, Journal of Statistical Physics, {\bf72}, 1--14,
1993; and Dobrushin, R.L., Hryniv, O.: {\it Fluctuations of the phase
boundary in the $2D$ Ising model}, Communications in Mathematical
Physics, {\bf189}, 395--445, 1997.

\0[DL67] Dyson, F., Lenard, A.: {\it Stability of matter, I}, Journal of
Mathematical Physics, {\bf 8}, 423--434, 1967. And {\it Stability of
matter, II}, Journal of Mathematical Physics, {\bf 9}, 698--711, 1968.

\0[LR69] Lanford, O., Ruelle, D.:  {\it Observables at infinity and
states with short range correlations in statistical mechanics},
Communications in Mathematical Physics, {\bf13}, 194--215, 1969.

\0[MR94] Miracle--Sol\'e, S., Ruiz, J.: {\it On the Wulff construction
as a problem of equivalence of ensembles}, in {\it Micro, Meso and
Macroscopic Approaches in Physics}, ed. M. Fannes, A. Verbeure,
Plenum, New York, 1994.

\0[Mi95] Miracle--Sol\'e, S.: {\it On the microscopic theory of phase
coexistence}, {\it 25 Years of Non-Equilibrium Statistical Mechanics},
ed. J.J. Brey, Lecture Notes in Physics, {\bf 445}, pp. 312--322.
Springer, Berlin, 1995. And {\it Surface tension, step free energy and
facets in the equilibrium crystal shape}, Journal Statistical Physics,
{\bf 79}, 183--214, 1995.

\0[Mo56] Morrey, C.B.: {\it On the derivation of the equations of
hydrodynamics from Statistical Mechancs}, Communications in Pure and
Applied Mathematics, {\bf8}, 279--326, 1955.

\0[Do72] Dobrushin, R.L.: {\it Gibbs state describing coexistence of
phases for a three dimensional Ising model}, Theory of probability and
its applications, {\bf17}, 582--600, 1972.

\0[MS67] Minlos, R.A., Sinai, J.G.: {\it The phenomenon of phase
separation at low temperatures in some lattice models of a gas, I},
Math. USSR Sbornik, {\bf 2}, 335--395, 1967. And: {\it The phenomenon
of phase separation at low temperatures in some lattice models of a
gas, II}, Transactions of the Moscow Mathematical Society, {\bf19},
121--196, 1968. Both reprinted in [Si91].

\0[GK85] Gawedzky, K., and Kupiainen, A.
  {\it Massless lattice $\phi^4_4$ theory: Rigorous control of a
  renormalizable asymptotically free model}, Communications in Mathematical
  Physics {\bf 99}, 197--252, 1985.

\0[GK83]
  Gawedzky, K., and Kupiainen, A. {\it Block spin renormalization gr\-oup for
  dipole gas and $(\dpr\phi)^4$}, Annals of Physics {\bf 147}, 198--243,
  1983.

.Mastropietro, Vieri Ising models with four spin interaction at
criticality.  Comm. Math. Phys.  244  (2004),  no. 3, 595--642